\documentclass[a4paper,epsfig,12pt]{book}
\usepackage[T1]{fontenc}
\usepackage{textcomp}
\usepackage{authblk}
\usepackage{graphicx}\usepackage{epsfig} 
\usepackage{pdfpages}
\usepackage{floatrow}
\usepackage{amsmath}\usepackage{amssymb}\usepackage{amsfonts}
\usepackage{bm}
\usepackage{exscale}\usepackage{rotating}
\usepackage{dcolumn}
\usepackage{color}
\usepackage{enumitem}
\usepackage{soul}
\usepackage{ulem}
\usepackage{hyperref}
\usepackage{fancyhdr}
\fancypagestyle{plain}{
   \fancyhead{}
   
}
\let\oldv\v
\def\diff{\mathrm{d}}

\def\N{\mathcal{N}}
\def\F{\mathcal{F}}

\def\n{\mathrm{n}}
\def\p{\mathrm{p}}
\def\q{\mathrm{q}}
\def\s{\mathrm{s}}
\def\v{\mathrm{v}}

\def\A{\mathrm{A}}
\def\B{\mathrm{B}}
\def\C{\mathrm{C}}
\def\D{\mathrm{D}}

\def\exp{\mathrm{exp}}

\def\Ek{E_{\textrm{k}}}

\def\vecr{\mathbf{r}}
\def\vecp{\mathbf{p}}
\def\veck{\mathbf{k}}
\def\vecn{\mathbf{n}}
\def\Ntest{N_{\textrm{test}}}

\def\eF{\epsilon_{\textrm{F}}}
\def\vF{v_{\textrm{F}}}

\def\vvec{\boldsymbol v}

\def\diff{\mathrm{d}}
\def\beam{{\mathrm{p}}}

\def\vpar{v_{\|}}

\def\bibindent{2em}	
\def\bblab#1#2{\textit{\footnotesize #1$^{\,#2}$}}
\def\bbsty#1#2#3{{\bf #1}, #2 (#3)}	
\begin{document}



%
%
\frontmatter	\pagenumbering{roman}
\thispagestyle{empty}
\title{\vspace{-4.5cm}\begin{center} 
	\scalebox{1.3}[1.8]{Violent Nuclear Reactions:}\\\vspace{1.5ex}
	\scalebox{1.3}[1.8]{Large-Amplitude Nuclear Dynamic}\\\vspace{.2ex}
	\scalebox{1.3}[1.8]{Phenomena in Fermionic Systems}\\
	\vspace{2ex}
	{\includegraphics[angle=0, width=.2\textwidth]{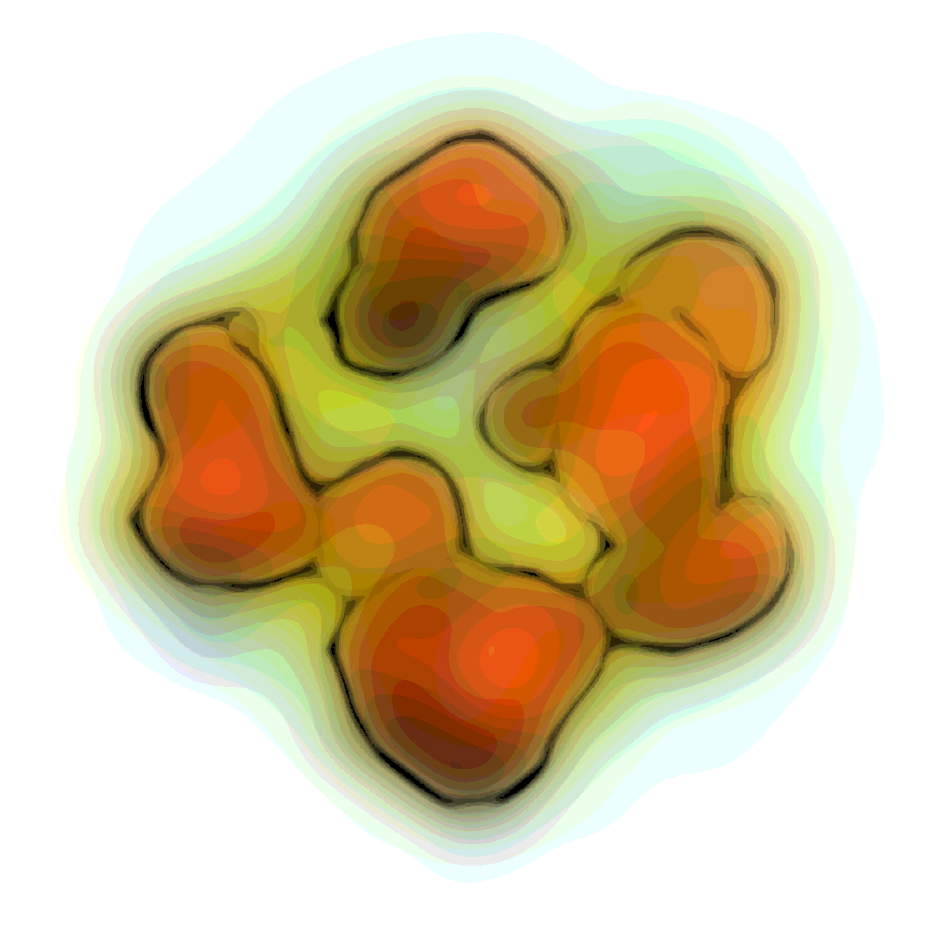}}
	\end{center}
	\vspace{-2ex}
}
\author{\textbf{\Large Paolo Napolitani}}
\affil{Institut de Physique Nucl\'eaire, IN2P3-CNRS, Universit\'e Paris-Sud,
Universit\'e Paris-Saclay, 91406 Orsay Cedex, France}
\date{		
	\vspace{-1.5ex} \textit{HDR report presented 1 February 2017, Orsay, France}\\\vspace{4.5ex}
	\begin{minipage}{1.02\textwidth} 
	\begin{center}\rule{.75\textwidth}{2pt}\end{center} 
Violent nuclear collisions are open systems which require a non--equilibrium description when the process should be followed from the first instants.
The heated system produced in the collision, can no more be treated within an independent-particle picture and additional correlations should be taken into account: they rely to in-medium dissipation and phase-space fluctuations.
Their interplay with the one-body collective behaviour activates the transport dynamics: large-amplitude fluctuations and bifurcations in a variety of mechanisms appear, from fusion to neck formation till eventually freezing out the system into several intermediate-mass clusters.
Starting from fundamental concepts tested on nuclear matter, a microscopic description is built up to address violent processes occurring in heavy-ion collisions at Fermi energies and in spallation reactions, and it is applied to experimental observables. 
	\begin{center}\rule{.75\textwidth}{2pt}\end{center}\end{minipage}
}

\thispagestyle{empty}

\maketitle
\thispagestyle{empty}

\begin{flushleft}
\newpage\thispagestyle{empty}
\textit{
\vspace{5cm}\\
La verit\`a \`e una giovinetta tanto bella quanto pudica e perci\`o va sempre avvolta nel suo mantello.}\\
\bigskip
(The truth is a young maiden as modest as she is beautiful, and therefore she is always seen cloaked.)\\
\bigskip
\bigskip
-- UMBERTO ECO, ``L`isola del giorno prima''
\end{flushleft}
\bigskip
\bigskip
\bigskip

\begin{flushleft}
\textit{Le docteur examina donc l'empreinte deux fois, trois fois, et il fut bien oblig\'e de reconna\^itre son origine extraordinaire.}\\
\bigskip
(Thereupon he looked twice, three times, at the print, and he was obliged to acknowledge its extraordinary origin.)\\
\bigskip
\bigskip
-- Jules Verne, ``Les Aventures du capitaine Hatteras''
\end{flushleft}
\bigskip
\bigskip
\bigskip
\newpage\thispagestyle{empty}

\tableofcontents

\chapter{Preface}

This review was written in order to be conferred the French accreditation to supervise research
(``Habilitation \`a diriger des recherches'').

I wish to start by thanking all the members of the jury, 
\textsc{Maria Colonna}, \textsc{Virginia De La Mota}, \textsc{Eric Suraud},
\textsc{Elias Khan} and \textsc{Oliver Lopez}
for accepting to read and review this document, on a subject they pioneered.
	Thanks to them and to many researchers who prepared the terrain, I could also add some contribution to an already well paved avenue.
	In other words, having them in my jury makes me as happy as a three-year-old kid meeting his favourite team of astronauts!

	As a doctoral student, oscillating between Orsay and Darmstadt, I could finally dedicate fanatically to my favourite amusement, nuclear reactions.
	For a few years, I was among the fission and fusion veterans, experimentalists and visiting theoreticians.
	I will always be grateful to my first research group in GSI (Darmstadt) for teaching me how to distinguish a light- from a heavy-ion beam from the ramping noise of the transformers of the synchrotron; students were betting German superheavy chocolate snacks on the mass number of the ion beam.
	I learnt that when the logbook falls on a keyboard in the control room of a particle accelerator, a relativistic beam of heavy ions can end up on a block of copper, traverse it anyway, and produce a grandiose stream of heavy particles and a blast of neutrons which activate all radiation-safety systems and determines the interruption of the whole machine.	
	In a millisecond scale, this was my electrifying introduction to nuclear physics.
	In better controlled conditions, we discovered several new isotopes, observed new nuclear processes and discoveries were coming daily.
	Too many discoveries (as it used to be in Darmstadt still some years after 2000) inspired new huge projects, huge projects imposed regulations and restrictions, restrictions spoiled the daily-discovery dream.
	Never mind, the theoretical understanding of what had been observed, and how it relates to other fields, should be the natural completion of a successful expedition.

	During my stay in Caen, starting in 2004, I joined a new very exciting research: nuclear pasta phases in compact stars.
	At that time, it was believed that the low-density nuclear matter constituting the outer shells of proto-neutron stars could develop a series of topologies of nucleon arrangements which recall spaghetti, lasagne, maccaroni (and maybe also bucatini, fettuccine, farfalline, gnocchetti, penne, pennette, vermicelli, zitoni, orecchiette...).
	All these configurations were supposed to manifest phase transitions and the overall pasta-phase region was supposed to end-up in a critical point which, eventually, could even be related to a supernova explosion of type II, due to the consequence of a diverging susceptibility on neutrino trapping.
	After two years of statistical modelling, I had the privilege to destroy the field, or at least part of it, because I discovered that stellar matter can not present any second-order phase transition, nor first order.
	Pasta phases survived as a set of crossover structures, but they looked somehow less interesting, at first from a thermodynamic point of view.

	To find back first-order phase transitions in nuclear matter, with all related instabilities, I returned from the stars to my motherland: heavy-ion collisions.
	I have always been thrilled by the fact that in this field, galvanising discussions on the reaction mechanisms, were ending up in a theoretical b\'ehourd among myriads of different interpretations.
	These last years, on the basis of some experience on both statistical and dynamical modelling, and also thanks to the surrounding experimental environment, I could absorb many ideas, recommendations, points of view, indirect experiences from the pioneers of the field in Caen, Catania, Nantes, Orsay and other partner laboratories, and I could also participate to data analyses and discuss new experimental observations.
	Finally, I could crystallise this information in a new comprehensive description of violent nuclear reactions starting from a microscopic approach.

	In this habilitation report, I decided to select only one activity, the dynamical modelling of violent nuclear processes from a one-body point of view, and to present it synthetically in relation to new progress and to the theoretical framework which I contributed to develop, sometimes connecting back to experimental projects where I was, in more than one case, involved as the spokesperson.
	The aim is to produce a useful and readable document which rather briefly draws conclusions on topics that, though being still source of developments, can be very efficiently handled at a high degree of accuracy.
	In particular, neither pure theory, nor experimental analysis is put forward in this report, and a phenomenological line is followed.
	The text is a re-edition of published and to-be-submitted material from my bibliographic list with updates, additional content and connections, all reorganised into a monograph.
	Most of the works I'm reviewing in this report are the results of a long and fruitful collaboration with Maria Colonna, to whom I would like to express my deep gratitude.

	Finally, I'm indebted to all my colleagues, my family and friends for the encouragement on my research.
	When my two very small kids will be able to read, they will learn that I thank them too, for so numerous and inspiring macroscopic analogies of chaotic dynamics involving catastrophic irreversible processes which correspond, in a nuclear context, to the core subject of this report.

\mainmatter									\pagenumbering{arabic}
\chapter[Introduction: context and some history]{Introduction:\\ context and some history 
\label{ch_intro}}
%
%
%
%

\section{Scope and plan}

The splitting of a nuclear complex into fragments constitutes a rich phenomenology in many-body physics which led to a century of discoveries, ranging from nuclear fission to a myriad of different nuclear-reaction channels.
	Correspondingly, a broad spectrum of interpretations have been stimulated.
	The process, when regarding nuclear reactions, can be studied in a laboratory and controlled up to a certain extent, yielding insights on the nuclear interaction.
	Less directly, extrapolations to dense matter in extended astrophysical objects can be searched, connecting microphysics to macrophysics.
	Beside this suggestive bridge between small and large scales, similar behaviours are displayed in other classical and quantum systems in condensed-matter physics.
	The interconnection between these distant fields invited to establish common theoretical frameworks, from energy density functionals to statistical approaches for the equation of state.

	Given the breadth of the subject, rather than undertaking an interdisciplinary overview, we focus on a more specific topic: the fragmentation phenomenology in a nuclear system from nuclear matter to atomic nuclei.
	This topic is nowadays attracting increasing attention due to the recent capability of handling new degrees of freedom, from bosonic states to strangeness but, at least in the nuclear-physics sector, such step forward is still in a developing phase.
	As a second restriction, we focus therefore on aspects of fermionic dynamics in nucleonic systems (and we neglect for instance topics like alpha clustering at very low density, as well as very-large-density effects which characterise relativistic regimes.
	As a third restriction, we focus on ``violent'' processes, and hence an energy range situated around the Fermi regime.
	Due to analogies that will be elucidated in the text, we also include highly excited heavy remnants resulting from ion-ion or nucleon-ion collisions at relativistic energies: these systems are the useful intermediate step for moving from nuclear matter to heavy-ion collisions. 

	Such energy selection corresponds to situations where the formation of fragments in nucleonic matter is induced through a violent external action (mechanical or thermal).
	Without such external action, only a few radioactive isotopes of heavy elements can divide spontaneously into lighter nuclides when subjected to alpha decay or, more seldom, to spontaneous binary fission.
	Even more rarely, some actinides and few other elements may undergo binary and even ternary splits where the lightest products are clusters ranging from He up to the region of Mg.
	The tendency towards cluster decay reflects nuclear-structure properties and it can be enhanced when more states are involved in the decay scheme.
	To access more excited states and channels beyond spontaneous decay, an external action is needed to warm up the system.

	In practice, this situation is achieved in various types of particle-nucleus and nucleus-nucleus collisions, and it is explored in nucleonic matter in very hot astrophysical objects like supernovae and their precursors.
	If the system is excited up to the Fermi regime, the widths of the energy levels fuse into a continuum and footprints of nuclear structure are lost.
In this case, the system can be completely rearranged through a thermodynamic process.
	In particular, the phase-space of the system dilutes, so that nucleonic collisional correlations are no more removed by the Pauli exclusion principle, and they become a continuous seed of dynamical fluctuations and bifurcations.
	Such systems are characterised by a prominent tendency to break into fragments, displaying a huge variety of possible configurations.
	For this reason, we focus on the Fermi regime.
	Indicatively, heavy-ion collisions at incident energies of around 10 to 50 MeV per nucleon correspond to such regime.
	For larger bombarding energies, collective modes are gradually overwhelmed by nucleonic collisional correlations.
	Those latter finally prevail and completely determine the dynamics when reaching relativistic energies.
	However, even at these high energies, collective modes can be restored in the decay sequence, when one large and warm nuclear remnant emerges from the ashes of the reaction.

	The purpose of this report is to analyse the above phenomenology within a dynamical optics which keeps the equivalence to the statistical description at equilibrium: to do this, we start from nuclear matter and from the dispersion relation, which is the foundation of a dynamical approach for large-amplitude perturbations.
Such strategy solves many long-standing questions related to the incomplete picture on the nuclear reaction mechanisms which emerges from statistical or equilibrium approaches alone.

\bigskip
{\bf
This report starts with a discussion on how to handle fluctuations in the dynamics of nuclear matter; a conceptually simple approach based on the Boltzmann-Langevin equation is proposed and taken as a reference throughout the review.
	An analysis of fluctuations in nucleonic Fermi fluids will follow.
	We will then move from nuclear matter to open nuclear systems and study the fluctuation-bifurcation phenomenology in heavy-ion collisions: we focus first of all on spallation mechanisms at relativistic energies.
	We progress to the study of nucleus-nucleus collisions at Fermi energies.
	Anytime it is possible, we investigate connections between the microscopic (transport) description and the macroscopic (statistical) description.
	The conclusions are perspectives.
}

\section{Fragmenting nuclei: a brief history}

	The process of fragment formation in an induced nuclear reaction was already inspiring pioneering research efforts in the middle of the 1930s.
	Bohr might have brought about the challenge in the conclusions of a renowned publication suggesting that~\cite{Bohr1936} when an atomic nucleus is bombarded {\it ``with particles of energies of about a thousand million volts, we must even be prepared for the collision to lead to an explosion of the whole nucleus''}.
	This sentence was so prophetic that it become one of the bestselling citations in prefaces of books and monographs in nuclear physics!

	Indeed, few years later, such scenario started to be outlined in many experiments where light particles (neutrons, protons, deuterons) where directed on a heavy nucleus.
	Already in his doctoral dissertation (1937) and later, in a 1947 publication, Seaborg~\cite{Seaborg1947} coined the term \textit{spallation} as a nuclear process where the entrance channel is a light high-energy projectile hitting a heavy ion, and the exit channel consists of several particles, where heavy residues or fission fragments combine with light ejectiles, including a large neutron fraction.
	The newly introduced reaction was soon associated to technological applications like neutron sources, secondary beams, hadron therapy and accelerator-driven systems for energy production and transmutation; in astrophysics, spallation was associated to the cosmic-ray isotopic spectrum.

	The same year, Serber~\cite{Serber1947} proposed a schematic description where the process essentially translates in a fast stage of excitation followed by a second stage of statistical decay from a fully thermalised compound nucleus~\cite{Thomas68}, thus completing the scenario initiated by Weisskopf in his famous 1937 publication~\cite{Weisskopf1937}.
	This very general description resulted very successful but its shortcoming was soon determined by further new findings.
	The picture become more involved when, firstly Nervik and Seaborg~\cite{Nervik1954}, concentrating on intermediate-energy experiments, and, later on, Robb Grover~\cite{RobbGrover1962}, focusing on relativistic energies, reported that a nucleus undergoing a violent collision does not only disintegrate into light particles but also into intermediate-mass fragments~\footnote{
``intermediate-mass clusters'' or ``intermediate-mass fragments'' or ``IMF'' or ``dynamical clusters'' are equivalent general definitions for reaction products ranging from heavier nuclides than an $\alpha$ particle to about half the system mass. They should not be confused with ``clusters'', which define structure effects like $\alpha$ clustering, Bose condensation or nuclear molecules.
} (IMF), lighter than ordinary fission products but heavier than alpha particles.
	Still exploiting spallation reactions, pioneering experiments with particle accelerators~\cite{Kaufman76,Warwick82,Hirsch84,Andronenko86,Barz86,Korteling90,Kotov95,Hsi97,Avdeyev98,ISIS} further focused on IMF formation.
	Suggestive results on single events were provided by heavy projectiles impinging on photoemulsions~\cite{Jakobsson1982}.

%
%
\begin{figure}[t!]\begin{center}
	\floatbox[{\capbeside\thisfloatsetup{capbesideposition={left,top},capbesidewidth=.45\textwidth}}]{figure}[\FBwidth]
	{\caption{
	A schematic distinction among different energy regimes for heavy-ion collisions, in terms of De Broglie wavelength $\lambda$ as a function of incident energy per nucleon of a projectile.
	For comparison, the radii of $^{208}$Pb and $^{40}$Ca are indicated.
	Two-body nucleon-nucleon collisions are suppressed at low energies, for large $\lambda$, they are recovered at Fermi energy and become dominant at relativistic energies, for small $\lambda$, leading to the participant-spectator picture.
	}
	\label{fig_lambda}}
	{\includegraphics[angle=0, width=.45\textwidth]{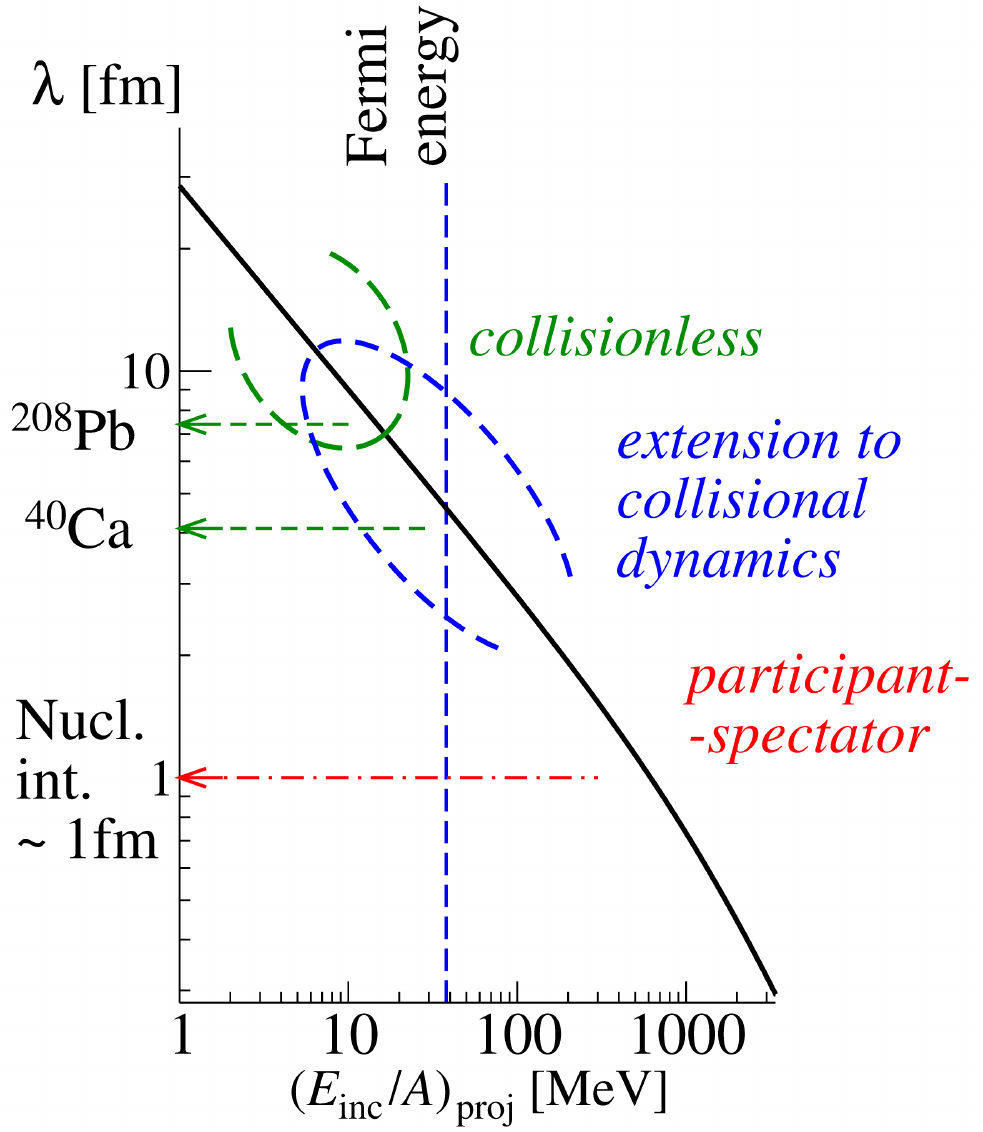}}
\end{center}
\end{figure}
	The phenomenological picture became even richer when, almost contemporaneously, new measurements with heavy-ion beams from Fermi-energy to relativistic domains showed that nuclear systems can produce a sumptuous multifragmentation process in many IMF~\cite{Hufner84,Cole2000,EPJAtopicalWCI2006}.
	Heavy-ion collisions could actually probe a large variety of perturbations acting on a nuclear system as a function of the incident energy.
	At least three regimes could be schematically distinguished, as indicated in Fig.\ref{fig_lambda}.
	Pauli-blocking factors of final orbitals largely suppress two-body nucleon-nucleon collisions at low energy (from the Coulomb barrier to around 15 MeV per nucleon).
	Those latter should on the other hand be included at Fermi energy (from about 15 to 200 MeV per nucleon) and become dominant in the participant-spectator regime (from about 200 MeV per nucleon to relativistic energies).
	Of course, transitions between these domains are smooth. 

	The conditions at Fermi energy are clearly peculiar with respect to higher energies because of a dominant mechanical contribution.
	It's interesting to mention that the earliest models to describe fragment formation in such conditions were in the spirit of fission, or the liquid-drop model in general, and often similarities with normal classical liquids were searched.
	In this respect, Fr\"olich~\cite{Frolich1973} in 1973 and Griffin in 1976~\cite{Griffin1976} proposed a very similar scenario to the fragmentation of liquid jets observed by Rayleigh in 1882~\cite{Rayleigh1882}.
	The picture of the Rayleigh instability could describe efficiently the mechanism of peripheral collisions where a third fragment comes from the neck rupture~\cite{Montoya1994}.
	The Rayleigh instability was proposed again also in more recent speculations about the possibility that nuclei could explode like a bubble~\cite{Moretto1992} and, 
despite such mechanism does not seem to occur, 
it might actually offer insights for some violent out-of-equilibrium processes.
	Also the possibility of successive asymmetric fission events was advocated~\cite{Moretto1988}, or the extension of the saddle-point interpretation to highly excited systems~\cite{Lopez94},
and resulted even successful in describing experimental data, as far as no observables could constrain the chronology of the process.

	Still in parallel to early experiments, possible analogies with phase transitions, which are general phenomena occurring in interacting many-body systems~\cite{Schmidt2001,Hock2009,Kim2004,Steinheimer2012,Moretto2011}, were searched, and inspired improvements on the theoretical description along the avenue of statistical approaches~\cite{Botvina85b,Botvina90,Bondorf95,Raduta1997,Gross2001}, eventually coupled with hydrodynamic descriptions to take into account the entrance channel.
	In particular, due to the analogies between the nuclear forces and the Van-der-Waals interaction, the nuclear matter equation of state (EOS) foresees the possibility~\cite{Jacquaman1983,Muller1995} of experiencing liquid-gas phase transitions connected to the appearing of a mixed vapour phase.
	In this spirit, early experiments on heavy-ion collisions were addressed to tracking signals of caloric curves~\cite{Pochodzalla1995, Schmidt2002, Natowitz2002}, a subject that has been intensively discussed till recently~\cite{Zheng2012, Borderie2013}.

	As soon as more accurate measurements confirmed that the nucleus can disassemble simultaneously (rather than sequentially) into fragments, and that the associated density profile describes dilute nuclear matter, the $1^{\text{st}}$-order phase transition picture was taken as the underlying phenomenology~\cite{Gulminelli1999,Chomaz2002,Gross2011}.

	EOS properties and finite-size effects become the research terrain for several years of experiments with heavy-ion beams in the Fermi-energy range where event-by-event correlation measurements and large angular acceptances were exploited widely~\cite{Souza2006}, leading to new findings on the nuclear multifragmentation phenomenon~\cite{Bowman1991,Dagostino2000,Borderie2008,Moretto2011}.

	In particular, by exploiting event-by-event correlations, more sophisticated thermodynamic analyses could trace first-order phase transitions in finite systems as characterised by negative specific heat and bimodal behaviour of the distribution of the order parameter~\cite{Binder1984,Chomaz2000,Chomaz2003}.
	The latter physically corresponds to the simultaneous presence of different classes of physical states for the same value of the system conditions that trigger the transition (like the temperature, for instance), and is a stronger observable for a first-order phase transition.

	In fact, under suitable conditions, a bimodal character of experimental observables, such as the distribution of the heaviest cluster produced in each collision event~\cite{Bonnet2009}, or the asymmetry between the charges of the two heaviest reaction products~\cite{Pichon2006} has been interpreted as a signal of the finite-size counterpart of the liquid-gas phase transition in nuclear matter, which is associated to a finite latent heat~\cite{Gulminelli2003}: the energy (and density) of the system and the size of the largest fragments are order parameters which rule the evolution of the system from a configuration where a large cluster survives at low excitation energy to a configuration where the system disassembles in small fragments at high-excitation.
	In this framework, liquid-gas coexistence related to density fluctuations trigger the multifragmentation scenario, rather than, like in fission, surface energy.
	For instance, the transition from a compact system to fission is one additional phase transition process triggered by the Coulomb field, without being associated to any latent heat~\cite{Lehaut2009}.


The rich topic outlined above has been historically mostly addressed to from the point of view of statistical models, both on the theory side, and from the corresponding tracing of global experimental observables.
Even though huge progress resulted from such approach, still many questions remained unanswered for long time and are still at the moment topic of research.
We may mention three of them, among many others.

\textit{In spallation}, data on kinematics and correlations were so far sufficient for supporting a general phenomenological framework relying on the fission-evaporation picture, but they were incomplete for also explaining the mechanism of production of light nuclides (see \textsection~\ref{ch_inhomogeneities}).

\textit{In heavy-ion collisions at Fermi and intermediate energies} the bimodal behaviour of fragment-production observables have been related to first-order phase transition in a statistical framework, but it was then argued that also dynamical approaches were producing similar signatures without necessarily requiring the reaching of thermodynamical equilibrium~\cite{LeFevre2009}. In this respect, the thermodynamics of heavy-ion collisions could not be disconnected from microscopic approaches, suited for handling fluctuations, when searching for solid EOS observables.
All conceptual ingredients for microscopic approaches had already been prepared long in advance, but only recently they can be exploited in predictive models (see \textsection~\ref{sec_central}).

\textit{In exotic mechanisms at the threshold between multifragmentation and fission}, like neck fragmentation producing ternary or even quaternary events, approaches which are extension of fission models (like chains of binary fissions) can not be compatible with the kinematics of the mechanism which have been observed to be rather rapid (see \textsection~\ref{sec_peripheral}).

	A fourth natural question would be whether it is possible to answer the three questions above within a common framework.
	On the experimental side, the increasing accuracy of correlation measurements could then bring the study of the origin of IMF formation, limited to a statistical picture of the exit channel, to the study of a process as a function of time, and also extended to out-of-equilibrium intervals of time.
	Those latter characterise the beginning of the process, they entirely characterise some specific channels and they allow to connect different mechanisms to the same entrance channel.
	They also allow to understand how the process of fragment formation evolves microscopically, in presence of instabilities or in presence of a threshold between different exit channels.
	The study of instability growing in the nuclear system, described as a Fermi liquid, progressed on the theoretical side in parallel with the research on the EOS, since the first experimental evidence of phase-transition behaviours.
	It has then been further accelerated by the discovery of the first signals of spinodal decomposition~\cite{Borderie2001}.
	In this spirit, nuclear many-body dynamics and transport approaches emerged as an alternative powerful line of investigation where equilibrium assumptions could be dropped and which keeps the equivalence to the statistical description at equilibrium.
	Transport approaches have been firstly addressed to probe the reaction mechanisms associated to the occurrence of phase transitions~\cite{Aichelin1991,Lipavsk2001,Morawetz2007,Colonna2010}. 

	More recently, isospin effects related to the phase transition phenomenology~\cite{Chomaz2004,Ducoin2007}, as well as the study of currents of neutrons and protons related to transport effects~\cite{Baran2005}, extended the research to the related isospin physics~\cite{EPJAtopicalNSE2014}, further emphasising the need of dynamical approaches to search for features which exceed the statistical picture.

\chapter{Modelling instabilities
\label{ch_instabilities}}

How can we describe the interplay between mean-field and many-body correlation effects in unstable nuclear matter?
In this situation, where clusterisation is observed, the fluctuation of one-body densities determine both mechanical properties of fragments, like excitation, size or density, and how neutrons and protons are shared among emerging fragments and the surrounding environment.
A theory which can handle these conditions can also be adapted to draw a microscopic description of heavy-ion collisions as a function of time. This is the ultimate purpose of this work.
We briefly review in this chapter the reason why a Boltzmann-Langevin approach is well suited for this application.

\textit{Main sources for this chapter}: 
this chapter exploits a publication in preparation~\cite{Napolitani2017}

\section[Large-amplitude fluctuations in fermionic systems]{Large-amplitude fluctuations in \\fermionic systems}

The stochastic microscopic description of non-equilibrium large-amplitude dynamics of many-fermion systems is a long-standing challenge for theoretical modelling.
	Nowadays it is experiencing a strong acceleration due to unprecedented computing resources.
	Even more importantly, these microscopic descriptions are stimulating increasing interest since they have come within the experimental reach in many areas of physics at the same time, from heavy-ion collisions~\cite{EPJAtopicalWCI2006,EPJAtopicalNSE2014}
to solid-state physics (examples are metal clusters~\cite{Calvayrac2000,Fennel2010} or electrons in nanosystems~\cite{ChenG2005}), to ultracold atomic gases~\cite{Dalibard1998,Bloch2008}
and, more indirectly, to some areas of astrophysics~\cite{Horowitz2006,Sebille2011,Schneider2013}.

	Large-amplitude collective motion of {\it fermionic systems}~\cite{Ring1980,Maruhn2010} is efficiently described with the time-dependent Hartree-Fock (TDHF) framework, or time-dependent local density approximation  (TDLDA), in condense-matter applications~\cite{Yabana1999,Reinhard2004}, as far as the variance of the involved observables is small.
	However, when the system experiences a violent dynamics, fluctuations and bifurcations might have to be expected, determining the shortcoming of the TDHF approximation.
	Under these conditions, large fluctuations would tend spontaneously to drive the system far away from the mean-field TDHF evolution along many different directions.
	Depending on the degree of excitation, a large variety of solutions have been developed to include additional correlations into the mean-field representation, ranging from the description of small- to large-amplitude collective motion till addressing chaotic dynamics.

\begin{figure}[b!]
\begin{center}
	\includegraphics[angle=0, width=.85\textwidth]{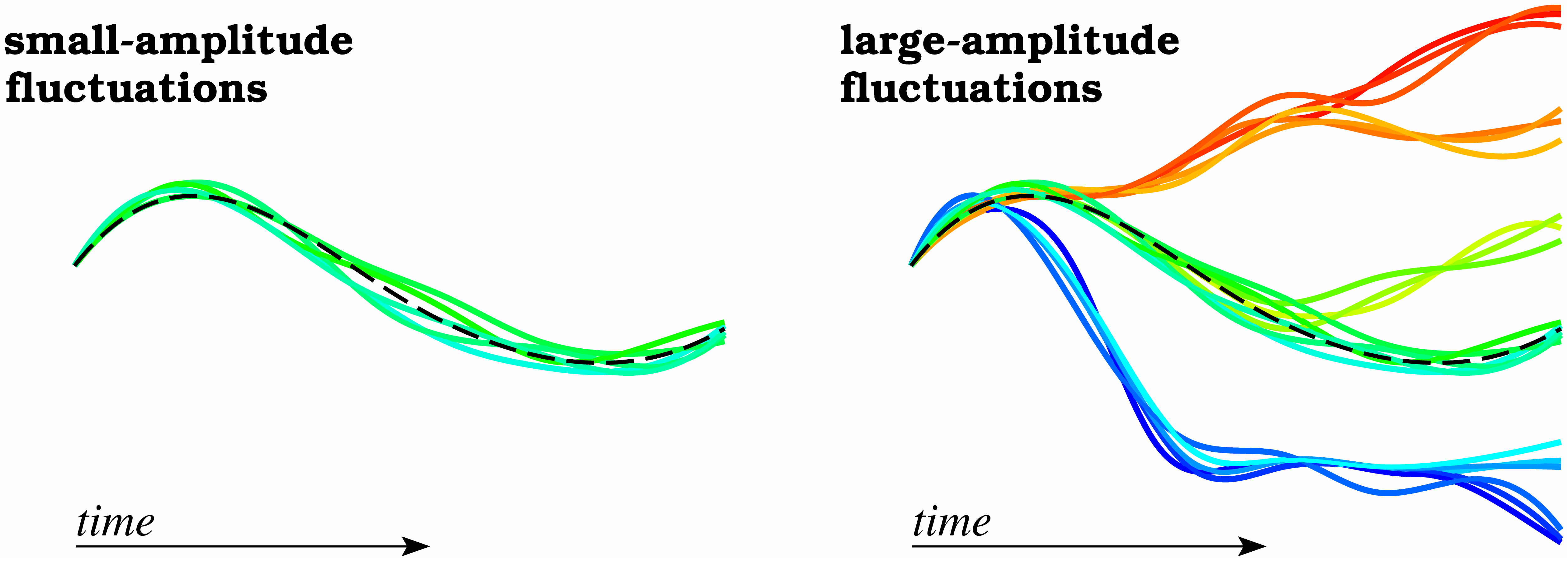}
\end{center}\caption{
	Schematic illustration of mean-field trajectories developing from an initial state
in absence (left) and in presence (right) of large-amplitude fluctuation.
	In the latter case, trajectories split apart in a pattern of bifurcating mean-field subensembles.
	Each subensemble of mean fields is characterised by small fluctuations within an envelope around a corresponding mean trajectory (sketched by a dashed line).
\label{fig_path_small_large} }
\end{figure}

	In this chapter we focus on fermionic systems in nuclear matter, but some concepts could have a larger interdisciplinary interest.
	In the nuclear context, rich discussions on these approaches have been collected in dedicated reviews, e.g. \cite{Simenel2010}.
	In this work we focus on those approaches which can be extended to the description of very-large-amplitude dynamics, and which can eventually handle bifurcating dynamical paths, i.e. the possibility that the system can oscillate between very different exit channels.
	This is for instance the case of threshold conditions between nuclear fusion and multifragmentation which may be encountered in heavy-ion collisions at Fermi energies.
	In this work we focus on stochastic approaches of nuclear dynamics based on mean-field models, and we skip discussions on molecular dynamics (either non-antisimmetrised~\cite{Aichelin1986,Aichelin1991,Peilert1989,Peilert1991,Hartnack1989,Maruyama1998,Chikazumi2001,Papa2001}
or antisimmetrised~\cite{Feldmeier1990,Feldmeier1995,Ono1992}).
	Since early investigations, it was evident that fluctuations could be handled naturally by describing the propagation of several stochastic mean-field trajectories simultaneously, along a bundle of different dynamical paths.
	Intuitively, as schematically illustrated in Fig.~\ref{fig_path_small_large}, each trajectory is a possible ``story'' of the system tracked by the evolution of a given observable as a function of time.
%
%
%
	As far as the bunch of dynamical trajectories could be reduced to correlated channels (as typically in a low-energy framework), it is sufficient to adopt a scheme of coherent-state propagation in the small-amplitude limit as within the time-dependent generator coordinate method (TDGCM) framework~\cite{Reinhard1987,Goutte2005}, or a variational approach \`a la Balian-V\'en\' eroni~\cite{Balian1981,Balian1992,Simenel2012}.
	Beyond the single-particle picture,
	to handle large-amplitude regimes at low energy it is convenient to propagate non-correlated states~\cite{Lacroix2015}; even further beyond this picture, when dealing with a rather excited system, several degrees of freedom are explored in a large-amplitude dynamics.
	In particular, two-body nucleon-nucleon collisions are only partially suppressed by Pauli blocking and they should be taken into account.
	Accordingly, especially in presence of mean-field instabilities,
collective motion may be driven to a chaotic regime; in a nuclear system, this would result in a highly non-linear process, like the formation of nuclear fragments.
	This is the case where it is not sufficient to describe larger spread widths of observables or
dissipative behaviour, like in the extended TDHF (ETDHF) framework~\cite{Wong1978,Wong1979,Lacroix2004}, but it is also required to follow bifurcation paths which deviate from the mean trajectory.	
	In this situation, where the ensemble of mean-field trajectories to be handled is very
large, it was proposed to adopt a	coherent description in a statistical framework, leading to the original formulation of stochastic TDHF (STDHF)~\cite{Reinhard1992}.
	A more recent reformulation of such approach restricts the incoherent correlations to two-fermion collisional correlations~\cite{Suraud2014,Slama2015}.

	Already in the earliest works, it was suggested to further reduce the quantal description to its semiclassical analogue, the Boltzmann Langevin Equation (BLE)~\cite{Ayik1988}, in terms of semiclassical distribution functions.
	Among the various ways followed in deriving the BLE (Density matrix formalism~\cite{Ayik1990}, truncation of the Bogolioubov-Born-Green-Kirkwood-Yvon (BBGKY) hierarchy for density matrices~\cite{Balescu1976}, Green-function techniques~\cite{Reinhard1992bis}), it was also advocated that BLE is a semiclassical analogue of the STDHF approach when mean-field trajectories are reorganised in subensembles~\cite{Reinhard1992,Abe1996}.
	The semiclassical transport treatment defined by the BLE has the minimum requirement of keeping the Fermi statistics for the distribution functions; it is a kinetic equation of Boltzmann-Uehling-Uhlenbeck or Boltzmann-Nordheim-Vlasov type (BUU/BNV)~\cite{Bertsch1988,Cassing1990,Bonasera1994} supplemented by a fluctuation contribution which is incorporated explicitly in the collision term and acts as a continuous source.

	We introduce thereafter approaches to solve the BLE in periodic nuclear matter.
	We test the methods in periodic nuclear matter so as to be able to compare numerical results with analytic expectations and draw conclusions on the range of observables that the model could yield as well as its limitations.
	In particular, we focus on two fluctuation mechanisms which relate the 
properties of the effective interaction 
to the phenomenology observed in several violent nuclear processes.
	These mechanisms are driven by isoscalar fluctuations, which, when mean-field instabilities
occur, lead to the disassembly of the nuclear system (according to the dispersion relation
of the unstable modes) and isovector fluctuations, which should ideally reflect the
strength of the nuclear symmetry energy.

\section{Handling N-body correlations}

\subsection{General framework}

It is usual to describe the evolution of an $N$-body system by replacing the Liouville-von Neumann equation, which involves $6N$ variables, by the equivalent BBGKY hierarchy of equations.  
	Custom approximations can be applied by truncating or reducing the complexity of the hierarchy so as to obtain as many unknown variables as the number of equations: this is the avenue for constructing beyond-mean-field approximations.
	We recall that the kinetic equations for the one-body density matrix, including the BLE, can be obtained from a second-order truncation of the BBGKY hierarchy, for a two-body interaction $V_{ij}$~\cite{Balian1991,Cassing1992,Bonasera1994}, i.e. considering the first two lines of the following chain of coupled equations, which represent the BBGKY hierarchy:
\begin{eqnarray}
i\hbar\frac{\partial\rho_1   }{\partial t} &=& [k_1           ,\rho_1   ] + \textrm{Tr}_2 [V_{12}       ,\rho_{12}] \notag\\
i\hbar\frac{\partial\rho_{12}}{\partial t} &=& [k_1+k_2+V_{12},\rho_{12}] + \textrm{Tr}_3 [V_{13}+V_{23},\rho_{123}] \notag\\
\dots&&\notag\\
i\hbar\frac{\partial\rho_{1^{\dots} k}}{\partial t} &=& \sum_{i=1}^{k}[k_i+\sum_{j<i}^{k}V_{ij},\rho_{1^{\dots} k}]  \notag\\
     &+&\sum_{i=1}^{k}\textrm{Tr}_{k+1}[V_{i k\!+\!1},\rho_{1^{\dots} k\!+\!1}] \notag\\
\dots&&,
	\label{eq:BBGKY}
\end{eqnarray}
where $\textrm{Tr}_{k}$ is a partial trace involving
the many-body density matrix $\rho_{1^{\dots} k}$ 
and $k_i$ are kinetic energy operators.

	The explicit inclusion of correlations beyond order k=2 would be necessary to describe high-coupling regimes~\cite{Lacroix2014}, with also the inclusion of interactions beyond two bodies to include additional nuclear-structure features~\cite{Schuck2016}, or cluster correlations.
	On the other hand, a second-order scheme is well suited for our purpose of studying large-amplitude dynamics of many fermions, where we neglect structure effects.
%

\subsection{Recovering effects beyond the second-order truncation through a stochastic treatment}

	The passages and approximations leading from the second order scheme to the usual form of the BLE are well documented in the literature~\cite{Ayik1990}.
	Let us simply spot the main manipulations which allow to obtain the highly non-linear character of the BLE~\cite{Ayik1988}.
	This scheme imposes to consider that for an interval of time $\tau_{\textrm{BL}}$ there exist a subensemble of mean field trajectories 
$\rho_1^{(m_\lambda)}$ for which fluctuations are small with respect to the mean trajectory $\rho_1^{(m)}$; fluctuations propagate keeping the mean $\rho_1^{(m)}$ unchanged (see Fig.~\ref{fig_path_small_large} for illustration).
	This is equivalent to imposing that $\rho_1^{(m_\lambda)}$ and $\rho_2^{(m_\lambda)}$, i.e. the probability to find two particles at two configuration points, are not decorrelated at all times, while fluctuations act keeping the mean $\rho^{(m)}$ unchanged.
	In this case, the two-body density matrix $\rho_{12}^{(m_\lambda)}$ recovers some correlations of the upper orders of the BBGKY sequence and we can write at a time $t_0=t-\tau_{\textrm{BL}}$:
\begin{eqnarray}
	&\rho_{12}^{(m_\lambda)}(t_0) = \widetilde{\Omega}_{12}\mathcal{A}_{12}(\rho_1^{(m_\lambda)}(t_0)\rho_2^{(m_\lambda)}(t_0))\widetilde{\Omega}_{12}^{+}+\delta\rho_{12}^{(m_\lambda)}(t_0)  \;;\quad&
	\label{eq:fluctuating_term}\\
	&\langle\delta\rho_{12}^{(m_\lambda)}(t_0) \rangle_{\tau_{\textrm{BL}}} = 0 \;;\;&
	\label{eq:fluctuating_mean}\\
	&\langle\delta\rho_{12}^{(m_\lambda)}(t_0) \delta\rho_{12}^{(m_\lambda)}(t') \rangle_{\tau_{\textrm{BL}}} = \textbf{gain} + \textbf{loss} \;.&
	\label{eq:fluctuating_secmom}
\end{eqnarray}
where $\widetilde{\Omega}_{12}$ is the M\o ller wave operator~\cite{Reed1979,Suraud1995} describing the diffusion of a particle with respect to another particle in the nuclear medium, related to a diffusion matrix $G_{12}=V_{12}\widetilde{\Omega}_{12}$, which is on its turn related to the nucleon-nucleon differential cross section $|G_{12}|^2\sim \textrm{d}\sigma/\textrm{d}\Omega$; in other words, the first term of the RHS of eq.\ref{eq:fluctuating_term} contains collision correlations while the second term expresses a fluctuation of vanishing first moment, which may be expressed as a fluctuating collision term, introducing a fluctuation around the collision integral~\cite{Abe1996}.
	We may remark that setting $\delta\rho_{12}^{(m_\lambda)}(t_0)=0$ would impose a full decorrelation between $\rho_1^{(m_\lambda)}$ and $\rho_2^{(m_\lambda)}$, and remove all orders $k>2$.
	This would reduce to the (quantum) Boltzmann kinetic equation, which corresponds to a second-order truncation of the hierarchy without a fluctuation contribution.

	Finally, the description associated with one single mean-field trajectory $n$ in an interval of time $\sim \tau_{\textrm{BL}}$ yields the following form, similar to STDHF: 
\begin{equation}
	i\hbar\frac{\partial\rho_1^{(n)}}{\partial t} \approx [k_1^{(n)}+V_1^{(n)} , \rho_1^{(n)}] 
		+ \bar{I}_{\textrm{coll}}^{(n)} + \delta I_{\textrm{coll}}^{(n)}   ,
\label{eq:STDHF}
\end{equation}
	where the residual terms $\bar{I}_{\textrm{coll}}^{(n)}$ and $\delta I_{\textrm{coll}}^{(n)}$ represent an average collision contribution and a continuous source of fluctuation seeds, respectively. 
	Beyond the interval of time $\tau_{\textrm{BL}}$ the fluctuating term $\delta I_{\textrm{coll}}^{(n)}$ replaces the mean field $n$ by a set of new mean-field subensembles $n_\lambda$, which may eventually yield bifurcations: 
\begin{equation}
	\rho_1^{(n)} \longrightarrow \{ \rho_1^{(n_\lambda)}, \dots \rho_1^{(n_{subens.})} \}   .
\end{equation}
	In the same spirit, in the similar framework of the quantal version of the Stochastic mean field theory in the first BBGKY order, from the analysis of correlation momenta, a demonstration was given that the handling of several mean fields $n$ has the effect of recovering in an approximate form part of the contribution from the orders of the hierarchy which have been cut off, leading to a simplified (rather than a truncated) BBGKY theory~\cite{Lacroix2016}.
	Such argumentation should also apply to Eq.~(\ref{eq:STDHF}).
	Indeed the fluctuations introduced in the theory may account for the unknown and neglected many-body correlations. 
	Moreover, with respect to the quantal version of SMF of ref.~\cite{Lacroix2016}, here we consider two-body effects explicitly, through the inclusion of the collision integral.
	It may be noted that Eq.~(\ref{eq:STDHF}) transforms into an ETDHF theory if the fluctuating term $\delta I_{\textrm{coll}}^{(n)}$ is suppressed.
	Through a Wigner transform we can then replace Eq.~(\ref{eq:STDHF}) by a corresponding set of semiclassical BL mean field trajectories
\begin{equation}
	\frac{\partial f^{(n)}}{\partial t} = \{h^{(n)} , f^{(n)}\} 
		+ I_{\textrm{UU}}^{(n)} + \delta I_{\textrm{UU}}^{(n)}   \;,
\label{eq:SMF}
\end{equation}
where the evolution of a statistical ensemble of Slater determinants is replaced by the evolution of an ensemble of distribution functions $f^{(n)}$, which at equilibrium correspond to a correct Fermi statistics.
	$h^{(n)}$ is the effective Hamiltonian acting on $f^{(n)}$.
	The residual average and fluctuating contributions of Eq.~(\ref{eq:STDHF}) are replaced by Uehling-Uhlenbeck (UU) analogue terms~\cite{Uehling1933}.
	$I_{\textrm{UU}}^{(n)}$ is related to the mean number of transitions within a single phase-space cell $\Delta V_f$.
	While conserving single-particle energies, $\delta I_{\textrm{UU}}^{(n)}$ acts as a Markovian contribution expressed through its correlation~\cite{Colonna1994_a}
\begin{equation}
	\langle \delta I_{\textrm{UU}}^{(n)}(\vecr,\vecp,t) \delta I_{\textrm{UU}}^{(n)}(\vecr'\!,\vecp'\!,t')\rangle 
	= {\bf gain + loss}  
	= 2\mathcal{D}(\vecr,\vecp;\vecr'\!,\vecp'\!,t')\delta(t-t') \;,
\label{eq:correlation}
\end{equation}
where $\mathcal{D}$ is a 
diffusion coefficient.


\section{Obtaining the BLOB description}

From Eq.(2) 
and from the procedure detailed in ref.~\cite{Ayik1990} 
we assume that the fluctuating term $\delta I_{\textrm{UU}}^{(n)}$ in Eq.~(\ref{eq:SMF}) should involve the same contributions composing the average collision term $I_{\textrm{UU}}^{(n)}$, i.e. the transition and the Pauli-blocking terms.
	This implies that also $\delta I_{\textrm{UU}}^{(n)}$ should be expressed in terms of one-body distribution functions. 
This latter possibility can be exploited by replacing the residual terms $(I_{\textrm{UU}}^{(n)} + \delta I_{\textrm{UU}}^{(n)})$ by a similar UU-like term which respects the Fermi statistics both for the occupancy mean value and for the occupancy variance.
	In this case, the occupancy variance at equilibrium should be equal to $f^{(n)}(1-f^{(n)})$ in a phase-space cell $h^3$ and correspond to the movement of extended portions of phase space which have the size of a nucleon, i.e. the residual term $(I_{\textrm{UU}}^{(n)} + \delta I_{\textrm{UU}}^{(n)})$ should carry nucleon-nucleon correlations.
	A natural solution to satisfy such requirement is to rewrite the residual contribution in the form of a similar rescaled UU collision term where a single binary collision involves extended phase-space portions of equal isospin $A$, $B$ to simulate wave packets, and Pauli-blocking factors act on the corresponding final states $C$, $D$, also treated as extended phase-space portions.
	The choice of defining each phase-space portion $A$, $B$, $C$ and $D$ so that its isospin content is either $1$ or $-1$ is necessary to preserve the Fermi statistics for both neutrons and protons, and it imposes that blocking factors are defined accordingly in phase space cells for the given isospin species.
	The above conditions lead to the Boltzmann-Langevin One Body (BLOB) equations~\cite{Napolitani2013}:
\begin{equation}
	\frac{\partial f^{(n)}}{\partial t} - \{h^{(n)} , f^{(n)}\} 
		= I_{\textrm{UU}}^{(n)} + \delta I_{\textrm{UU}}^{(n)}  
	= g\int\frac{\diff\vecp_b}{h^3}\,
	\int
	W({\scriptstyle\A\B\leftrightarrow\C\D})\;
	F({\scriptstyle\A\B\rightarrow\C\D})\;
	\diff\Omega\;,
\label{eq:BLOB}
\end{equation}
where $g$ is the degeneracy factor, integrations are over momenta $\vecp_b$ and scattering angles $\Omega$, and $W$ is the transition rate, in terms of relative velocity between the two colliding phase-space portions and differential nucleon-nucleon cross section
\begin{equation}
	W({\scriptstyle\A\B\leftrightarrow\C\D}) = |v_\A\!-\!v_\B| \frac{\diff\sigma}{\diff\Omega}\;,
\label{eq:transition_rate}
\end{equation}
and $F$ contains the products of occupancies and vacancies of initial and final states over their full phase-space extensions.
\begin{equation}
	F({\scriptstyle\A\B\rightarrow\C\D}) =
	\Big[(1\!\!-\!\!{f}_\A)(1\!\!-\!\!{f}_\B) f_\C f_\D - f_\A f_\B (1\!\!-\!\!{f}_\C)(1\!\!-\!\!{f}_\D)\Big]\;.
\label{eq:Pauli_scaled}
\end{equation}

Fluctuations and bifurcations, due to higher order correlations than the $k$=2 truncation are created through a stochastic approach which exploits the correlations carried in Eq.~(\ref{eq:BLOB}). 

\section{Implementation of Boltzmann-Langevin \\approaches}
%
%
	The stochastic term in Eq.~(\ref{eq:SMF}) can be kept separate and treated as a stochastic force related to an external potential $U'$. 
	Such approach was followed in an early implementation by Suraud and Belkacem~\cite{Suraud1992, Belkacem1993}, where a fluctuating term is obtained from exploiting the quadrupole and octupole momenta of the local momentum distribution in configuration space.
	Again, this strategy is used in the Brownian One Body (BOB) model~\cite{Guarnera1996}, where a fluctuating term is prepared by associating a Brownian force to a stochastic one-body potential, or in the stochastic mean field (SMF) model~\cite{Colonna1998}, where the fluctuating term corresponds to kinetic equilibrium fluctuations of a Fermi gas.
	While in the first approach fluctuation seeds were injected at the beginning of the dynamical process, as undulations in the spacial density landscape, in the second approach they could be injected at successive intervals of time $\Delta t$.
	Nevertheless, in both treatments, 
fluctuations are implemented only in the coordinate space, i.e. they are projected on the spacial density.
	
	The difference between Eq.~(\ref{eq:BLOB}) and usual stochastic mean-field approaches is that those latter build fluctuations from introducing a well adapted external force or a distribution of initial conditions which should be accurately prepared in advance.
	On the contrary, Eq.~(\ref{eq:BLOB}) introduces fluctuations in full phase space intermittently, at any time, letting them develop spontaneously and continuously during the whole dynamical process.

	Table~\ref{tab:codes} collects the main numerical implementations of Boltzmann-Langevin approaches which led, so far, to a transport model; the list is not exhaustive because it does not consider approaches which derived from those given in the list, and some details are described thereafter.

\begin{table}[h!]
\begin{center}
   \caption{\label{tab:codes} Boltzmann-Langevin numerical implementations handling large amplitudes (other implementations exist but, to our knowledge, they are derived from those reported in this table).
	The table indicates, for each model, the formalism, the presence of a collision term, whether Pauli blocking is fully satisfied, the type of fluctuation seeds, the time when fluctuation seeds are injected (either intermittently, at time intervals $\Delta t$ or at the beginning only), and whether fluctuations act on full phase space or if they are projected on a suitable subspace. 
}
	\vspace{2.ex}
	\begin{tabular}{|l|c|c|c|c|c|c|}
		\hline
		\textit{Model} &\textit{formal.} &\textit{coll.} &\textit{Pauli} &\textit{seeds from}: &\textit{time} &\textit{phase sp.}\\
		\hline
		STDHF\hspace{0.5ex}\cite{Reinhard1992}   & quant.    & no      & yes     & m.f. subens.$^4$  & $\Delta t$ & full      \\
		Bauer\hspace{1.8ex}\cite{Bauer1987}      & semicl.   & yes$^1$ & no$^3$  & collisions        & $\Delta t$ & full      \\
		Suraud\hspace{1.0ex}\cite{Suraud1992}    & semicl.   & yes$^2$ & yes$^3$ & q/o momenta & $\Delta t$ & projected \\
		BOB\hspace{  2.6ex}\cite{Guarnera1996}   & semicl.   & yes$^2$ & yes     & $\rho$ ondulation & $t\!=\!0$  & projected \\
		SMF\hspace{  3.4ex}\cite{Colonna1998}    & semicl.   & yes$^2$ & yes     & external field    & $\Delta t$ & projected \\
		BLOB\hspace{ 0.5ex}\cite{Napolitani2013} & semicl.   & yes$^1$ & yes     & collisions        & $\Delta t$ & full      \\
		\hline
	\end{tabular}
	\vspace{-1.ex}\\
\end{center}
	$^1$ extended phase-space portions moved at once in one binary-collision event.\\
	$^2$ two test particles moved in one binary-collision event (standard UU).\\
	$^3$ see remarks in ref.~\cite{Chapelle1992} and \textsection~\ref{sec_exploitingcorrel}.\\
	$^4$ exploiting mean-field subensembles.
\end{table}

\section[Exploiting correlations in BLOB and handling metrics]{Exploiting correlations in BLOB and\\ handling metrics
\label{sec_exploitingcorrel}}

In the BLOB framework, eq.~(\ref{eq:BLOB}) is exploited to generate stochastic dynamical paths in phase space.
	The system is sampled through the usual test-particle method, often adopted for the numerical resolution of transport equations~\cite{Bertsch1988}, with the difference that, in the case of the BLOB implementation, the phase-space portions $A$ and $B$ involved in single two body collisions are not two individual test particles but rather agglomerates of $\Ntest$ test particles of equal isospin, where $\Ntest$ is the number of test particles per nucleon used in the simulations. 
	The initial states $A$ and $B$ are constructed by agglomeration around two phase-space sites, which are sorted at random, inside a phase space cell of volume h$^3$, according to the method proposed in ref.~\cite{Napolitani2013} and further improved in ref.~\cite{Napolitani2015}.
	At successive intervals of time, by scanning all phase space in search of collisions, all test-particle agglomerates are redefined accordingly in $h^3$ cells, so as to continuously restore nucleon-nucleon correlations.
	Since test particles could be sorted again in new agglomerates to attempt new collisions in the same interval of time, the nucleon-nucleon cross 
section $\sigma_{N\!N}$ contained in the transition rate $W$ should be divided by $\Ntest$ to give the scaled cross section $\sigma$ used in eq.~(\ref{eq:transition_rate}): 
\begin{equation}
	\sigma = \sigma_{N\!N} / \Ntest \;.
\end{equation}

	The metrics of the test particle agglomerates is defined in such a way that the packet width in coordinate space is the closest to $\sqrt(\sigma_{\textrm{NN}}^{\textrm{medium}}/\pi)$, where  $\sigma_{\textrm{NN}}^{\textrm{medium}}$ corresponds to the screened cross section prescription proposed by Danielewicz~\cite{Danielewicz2002,Coupland2011}, which was found to describe recent experimental data ~\cite{Lopez2014}. In this way, the spatial extension of the packet decreases as the nucleon density increases.

	Boltzmann-Langevin solutions where an ensemble of $\Ntest$ test particles are moved in one bunch and the nucleon-nucleon cross section is scaled by $\Ntest$ where already followed in the early approach by Bauer and Bertsch~\cite{Bauer1987}, or in more recent implementations~\cite{Mallik2015}.
	There is however a very fundamental difference: in the Bauer-and-Bertsch approach the Pauli-blocking term is not applied to the involved portions of phase space which are actually interested by the scattering at a given time $t$, as imposed by Eq.~(\ref{eq:Pauli_scaled}), but it is applied 
only to the centroids of
the two colliding packets.
	Such approximation makes the Pauli blocking satisfied only approximately,
with the drawback of loosing the Fermi statistics~\cite{Chapelle1992}.
	In the direction of BLOB, to prevent the above problem, a first practical solution 
was proposed in ref.~\cite{Rizzo2008}.
	These augmentations also complete the survey of table~\ref{tab:codes}.

%
%
\begin{figure}[b!]
\begin{center}
	\includegraphics[angle=0, width=.7\textwidth]{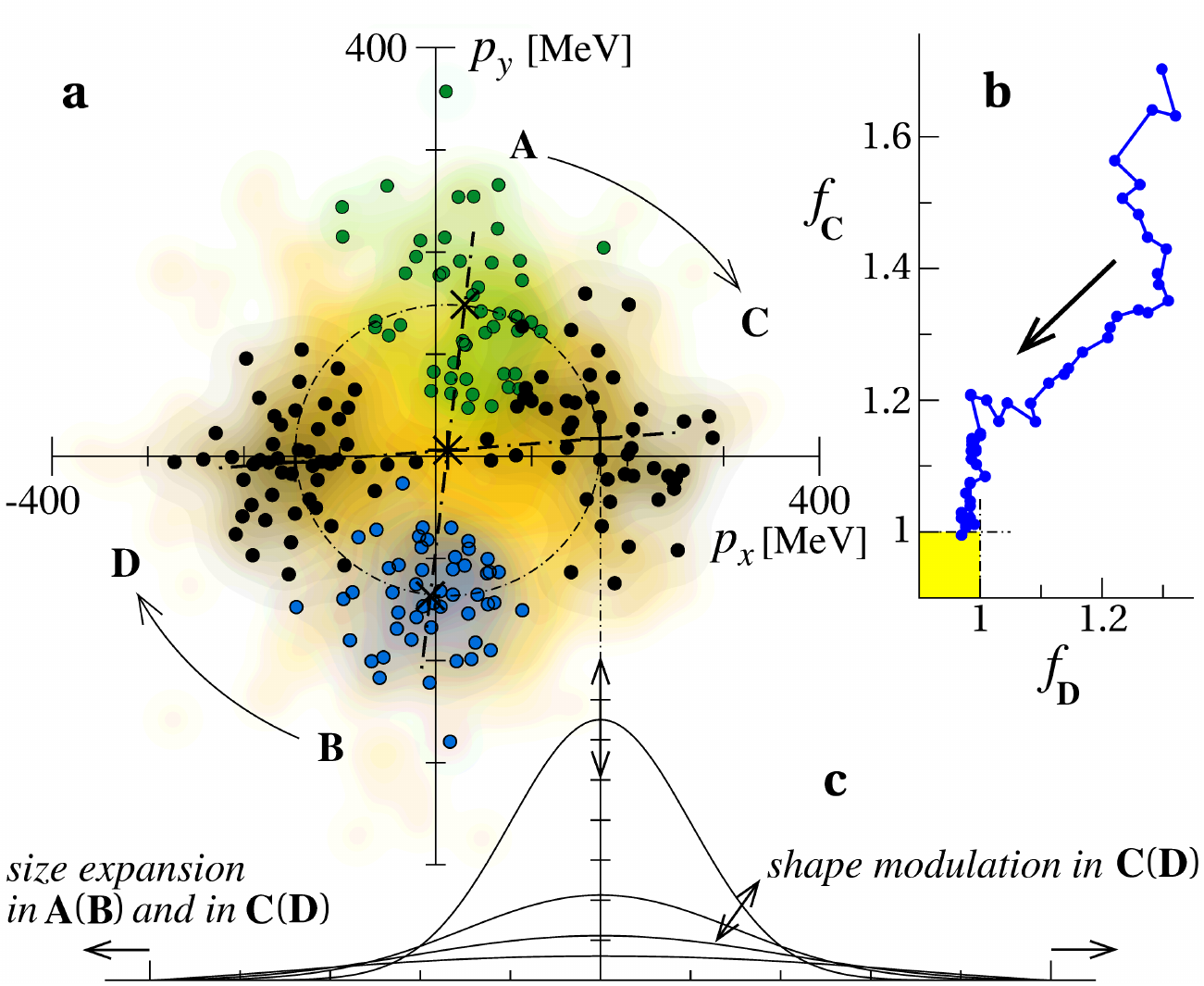}
\end{center}\caption{
(a) Definition of test-particle agglomerates in their initial ($A$,$B$) and final ($C$,$D$) states in momentum space in a $h^3$ volume.
(b) Convergence of a binary nucleon-nucleon collision configuration towards a situation where Pauli blocking is strictly satisfied. The path collects the sequence of modulations in phase-space of the test-particle clouds where the occupancy of the destination regions is iteratively optimised.
(c) Examples of possible shape modulations for a final state.
\label{fig_modulation} }
\end{figure}
	In BLOB, special attention is paid to the metrics when defining the test-particle agglomeration: the agglomerates are searched requiring that they are the most compact configuration in the phase space metrics which does neither violate Pauli blocking in the initial and in the final states, nor energy conservation in the scattering.
	For this purpose, when a collision is successful, its configuration is further optimised by modifying the shape and the width of the initial and final states~\cite{Napolitani2012}.
	Fig.~\ref{fig_modulation} illustrates the paths of a collision configuration which by a procedure of successive modulations is brought to a situation which respects Pauli blocking strictly. 
	The resulting occupancy functions of the modulated final-state density profiles should possibly approach unity.
	If such modulation procedure results unsuccessful, the collision is rejected.
	The rate of rejections due to unsuccessful modulation of the collision configuration is close to zero in open systems (collisions) so that the correlation between attempted and effective collision number is identical if a UU or a BLOB collision term is applied to the same mean-field, provided that the same nucleon-nucleon cross section is used.
	On the other hand, in uniform nuclear matter 
at equiliubrium, where only nucleons close to the Fermi surface can be involved in two-body collisions, the occurrence of such rejections becomes not negligible when the temperature $T$ considered is very low compared to the Fermi momentum.  
In this case, the perfect correspondence between attempted and effective collision rates in BUU (or SMF) and BLOB is lost (see \textsection~\ref{sub_noise} for further insight).

	A remarkable advantage of the renormalised form of the residual contribution in Eq.~(\ref{eq:BLOB}) is to connect directly the fluctuation variance to the local properties of the system, regardless the test-particle number.
	Such aspect has a general relevance because it makes their phenomenology independent from many aspects of the numerical implementation.
	The dependence on $\Ntest$ persists on the other hand in the mean-field representation, therefore when the fluctuation amplitude is small the global fluctuation phenomenology may suffer from noise effects produced by the use of a finite number of test particles in the numerical implementation of the transport equation.
	This remark should be taken in mind for the study of fluctuations of relatively small amplitude, like isovector fluctuations.
	The interplay between physical and numerical fluctuations will be
carefully investigated in the following.

\chapter{Inhomogeneity growth in two-component fermionic systems
\label{ch_inhomogeneities}}

	The advantage of one-body approaches with collisional correlations is to sample aspects of the behaviour of Fermi liquids~\cite{Lifshitz1958,Pines1966} and to allow including them in the description of heavy-ion collisions~\cite{Pethick1988}.
	In this chapter, we focus on the fluctuation phenomenology in nuclear matter at moderate temperature and in several density conditions.

	In particular we check how Eq.~(\ref{eq:BLOB}) handles the fluctuation variance of isoscalar and isovector one-body densities in equilibrated nuclear matter, in comparison with analytical expectations for fermionic systems interacting through effective forces.
	We aim to demonstrate that the implementation of Eq.~(\ref{eq:BLOB}) is better suited than approximate methods, like SMF,  to sample physical observables related to inhomogeneity development (which is equivalent to fragment formation in finite open systems).

	We focus on dynamical fluctuations, skipping discussions on the possible forms of the nuclear interaction and its isospin dependence, for which schematic descriptions are employed.
	We refer readers interested in the details of the nuclear interaction and its isospin dependence to the widespread literature on the topic, se e.g.~\cite{Baran2005,Li2008,Danielewicz2009,Danielewicz2014,EPJAtopicalNSE2014}.
	The purpose of this chapter is demonstrating that by solving the BLE in full phase space it is possible to describe the dispersion relation successfully and to enhance the isovector fluctuations with respect to the standard BUU.
	Such approach can therefore be applied to nuclear processes like heavy-ion collisions ensuring that observables related to the form of the nuclear potential and to the associated instabilities can be described efficiently.

\textit{Main sources for this chapter}: 
this chapter exploits a publication in preparation~\cite{Napolitani2017} and some recent talks~\cite{Napolitani_IWM2014}.

\section{Application to nuclear matter}

	Thereafter, both BLOB and SMF models are prepared as relying on a strictly identical implementation of the mean field, so that they differ only for the residual contribution.
	This study has a more general relevance, being intended to compare a BL approach where fluctuations are included as projected on the density landscape and a BL approach where fluctuations are introduced in full phase space. 
	We employ SMF and BLOB to represent the first and the second strategy, respectively.

	A simplified Skyrme-like (\textit{SKM}$^*$) effective interaction~\cite{Guarnera1996,Baran2005}, where momentum dependent terms are omitted, is employed in the propagation of the one-body distribution function, corresponding to the following definition of the potential energy per nucleon:
\begin{equation}
	\frac{E_{\textrm{pot}}}{A}(\rho) = 
		\frac{A}{2}u
		+\frac{B}{\sigma+1}u^\sigma
		+\frac{C_{\textrm{surf}}}{2\rho}(\nabla\rho)^2
		+\frac{1}{2}C_{\textrm{sym}}(\rho)u\beta^2 ,
\label{eq:pot}
\end{equation}
with $u=\rho/\rho_{\textrm{sat}}$, being $\rho_{\textrm{sat}}$ the saturation density 
and $\beta=(\rho_n-\rho_p)/\rho$.
	This parameterization, with $A\!=\!-356$ MeV, $B\!=\!303$ MeV and $\sigma\!\!=\!7/6$, corresponds to a soft isoscalar equation of state with a compressibility modulus $K\!=\!200$~MeV.
	An additional term as a function of the density-gradient introduces a finite range of the nuclear interaction and accounts for some contribution from the zero-point motion of nucleons~\cite{Guarnera1996}.
	$C_{\textrm{surf}}$ is related to various properties of the interaction range, like the surface energy of ground-state nuclei (the best fit imposing a value of $-6/\rho_{\textrm{sat}}$ MeV fm$^5$), the surface tension (light-fragment emission, in comparison to available data is better described for a smaller range given by $-7/\rho_{\textrm{sat}}$ MeV fm$^5$ in BLOB), and the ultraviolet cutoff in the dispersion relation for wavelengths in the spinodal instability (larger spectrum for a smaller range).
%
%
%
%
	Either a linear (asy-stiff) or a quadratic (asy-soft) density dependence~\cite{Colonna2014} of the potential part of the symmetry energy coefficient, $E_{\textrm{sym}}^{\textrm{pot}}$, is obtained by setting, respectively:
\begin{eqnarray}
\textrm{asy-stiff}&:& \qquad C_{\textrm{sym}}(\rho)\!=\!{\textrm{constant}}\!=\! 32\, \textrm{MeV} \;,\\
\textrm{asy-soft}&:& \qquad C_{\textrm{sym}}(\rho)\!=\! \rho_{\textrm{sat}}(482-1638\rho)\, \textrm{MeV} \;.
\label{eq:asy}
\end{eqnarray}
	If not otherwise specified, in this work the collision term involves an isospin- and energy-dependent free nucleon-nucleon cross section with an upper cutoff at $\sigma_{\textrm{NN}}=50$~mb.

	To simulate nuclear-matter properties, we fix the system density $\rho^0$ and the temperature $T$. 
	We prepare the system in a cubic periodic box of edge size $L=39$~fm, and we subdivide it in a lattice of cubic cells of edge size $l$ where we calculate density variances.
	For the sake of simplicity, we consider symmetric nuclear matter, i.e. with
equal number of neutrons and protons.
	We initially define the system imposing a uniform-matter effective field $U^0(\rho^0)$ depending only on the density considered, and a corresponding effective Hamiltonian $\epsilon(p) = h^0(p) = p^2/(2m) + U^0(\rho^0)$.
	Accordingly, the phase-space distribution function $f^0(\vecp)= \{1+\exp[\epsilon(p)-\mu]/T\}^{-1}$, not depending on configuration space (because the system is homogeneous), is the Fermi-Dirac equilibrium distribution at the given temperature $T$, for a chemical potential $\mu$.



\section[Fluctuations in nuclear matter and the BL equation]{Fluctuations in nuclear matter and \\the Boltzmann-Langevin equation}

%
%
	Either from the stochastic fluctuating residual term of the BLOB treatment or from an external stochastic force of the SMF approach we introduce a small disturbance in uniform matter 
\begin{equation}
	\delta f(\vecr,\vecp,t) = f(\vecr,\vecp,t) - f^0(\vecp,t)
\label{eq:pot}
\end{equation}
which lets a fluctuation develop in time around the mean trajectory $f^0$.

	By considering neutron and proton distributions functions, we can further decompose fluctuations in isoscalar modes $\delta f^\s$ and isovector modes $\delta f^\v$
\begin{eqnarray}
	\delta f^\s = (f_\n-f_\n^0)+(f_\p-f_\p^0) \;,\\
	\delta f^\v = (f_\n-f_\n^0)-(f_\p-f_\p^0) \;.
\end{eqnarray}
	The time evolution of both those modes is obtained by applying the BL equation (\ref{eq:SMF}) to the phase-space fluctuations.
For symmetric matter, and retaining only first order terms in $\delta f^q$, one obtains:
\begin{equation}
	\frac{\partial\delta f^q}{\partial t} + \frac{\vecp}{m}\cdot\nabla_\vecr\delta f^q - \frac{\partial f^0}{\partial \epsilon}\frac{\partial\delta U^q}{\partial \rho^q}\frac{\vecp}{m}\cdot\nabla_\vecr\delta\rho^q =   \frac{\partial f^0}{\partial \epsilon}\frac{\vecp}{m}\cdot\nabla_\vecr U'
	\;,
\label{eq:BLE_fluctuations}
\end{equation}
where the index $\q$ stands either for isoscalar ($\q=\s$) or isovector ($\q=\v$) modes, $f^0=f_\n^0+f_\p^0$ and
$U'$ is an external stochastic force (SMF) or a fluctuating stochastic field (BLOB).
	We dropped the average collision term $I_{\textrm{UU}}$ because we consider small temperatures. 

	To build our stochastic descriptions we assumed that, at least locally, fluctuation have small amplitude around their mean trajectory so that $\delta f^\q\!\ll\!f^\q$.  
%
	When the system is described as a periodic box, 
collective modes are associated to plane waves of wave number $\veck$.
	In this case, by expanding on plane waves expressed in Fourier components, we can study the evolution in time of phase space density fluctuations
\begin{equation}
	\delta f^\q(\vecr,\vecp,t)
	=\sum_k e^{(i\veck \cdot \vecr)} f_k^\q(\vecp,t) 
	= \sum_k e^{(i\veck \cdot \vecr)} e^{(i\omega_k t)} f_k^\q(\vecp)\;,
\label{eq:phasespacefluctuations}
\end{equation}
and undulations in the density landscape 
\begin{equation}
	\delta\rho^\q(\vecr,t)=\sum_k e^{(i\veck \cdot \vecr)} \rho_k^\q(\vecp,t)\;.
\label{eq:densityundulations}
\end{equation}
	Rewritten in Fourier components, 
and substituting $\partial_t\delta f_k^\q(\vecp,t) = i\omega_k f_k^\q(\vecp,t)$,
Eq.~(\ref{eq:BLE_fluctuations}) takes the form
\begin{equation}
	i\omega_k f_k^\q
+ i \veck\cdot\frac{\vecp}{m}f_k^\q - i \frac{\partial f^0}{\partial \epsilon}\frac{\partial U_k^\q}{\partial \rho^\q} \veck\cdot\frac{\vecp}{m}\rho_k^\q = i \frac{\partial f^0}{\partial \epsilon}\veck\cdot\frac{\vecp}{m}\F_k^\q
	\;,
\label{eq:BLE_Fourier}
\end{equation}
where $U_k^\q$ and $\F_k^\q$ are Fourier components of the potential $U$ and of the stochastic fluctuating field $U'$, respectively.


	When only stable modes can propagate, the response of the system to the action of the stochastic fluctuating field $\F_k^\q$ determines the equilibrium variance $(\sigma_k^\q)^2$ associated with the fluctuation $\rho_\veck^\q$.
	The inverse Fourier transform of $(\sigma_k^\q)^2$ 
gives the equilibrium variance of spacial density correlations 
\begin{equation}
	(\sigma_{\rho^\q})^2 \equiv \langle (\delta\rho^\q(\vecr))^2 \rangle = (2\pi)^{-3}\sum_\veck(\sigma_k^\q)^2 d\veck
\label{eq:eqvariancedensitycorr}
\end{equation}
in a cell of volume $\Delta V$ at temperature $T$. 
	At equilibrium, when the level density $\N\equiv(g/h^3)\int \partial_\epsilon f^0 \diff\vecp$ for a degeneracy $g$ can be defined, these variances are related to the curvature of the free energy density 
$F^\q(k)$ 
through the fluctuation--dissipation theorem so that
\begin{equation}
	(\sigma_k^\q)^2 = \frac{T}{F^\q(k)}\;; \;\;\; (\sigma_{\rho^\q})^2 = \frac{T}{\Delta V}\Big\langle\frac{1}{F^\q(k)}\Big\rangle_\veck
	\;,
\label{eq:fluctuation_dissipation}
\end{equation}
where $F^\q(k)= \partial_{\rho^\q} U_k^\q + 1/\N$, and for an average $\langle\cdot \rangle_\veck$ extending over all $\veck$ modes.

On the other hand, for unstable modes, 
the diffusion coefficient $\mathcal{D}$, or rather its 
projection
on a given unstable mode $k$,  $D_k$,  
determines the following evolution for the intensity of response $(\sigma^q_k)^2$ for the wave number $k$~\cite{Colonna1993,Colonna1994_a}:
\begin{equation}
	(\sigma^q_k)^2(t) \approx  D_k\tau_k(e^{2t/\tau_k}-1)+(\sigma^q_k)^2(t\!=\!0)e^{2t/\tau_k},
\label{eq:diffusion}
\end{equation}
where both the initial fluctuation seeds $(\sigma^q_k)^2(t\!=\!0)$ and the fluctuation continuously introduced by the collisional correlations contribute to an exponential amplification of the disturbance, characterised by the growth time $\tau_k$.

In the following, starting from Eq.~(\ref{eq:BLE_Fourier}), we select two very instructive situations which are isovector modes in uniform matter and, successively, isoscalar fluctuations in unstable matter.
	Translated into a violent nuclear-collision scenario, the first situation defines how isospin distributes among different phases and portions of the system, and the second situation coincides with the process of separation of those portions of the system into fragments.

\section{Isovector fluctuations in nuclear matter}


	Isovector effects in nuclear processes may arise from different mechanisms~\cite{Baran2012,DiToro2003}, like the interplay of isospin and density gradients in the reaction dynamics, 
or nuclear cluster formation, or the decay scheme of a compound nucleus.
	Alternately, in systems undergoing a nuclear liquid-gas phase transition, a role is played also by isospin distillation~\cite{Chomaz2004,Baran2005}, a mechanism which consists in producing a less symmetric nucleon fraction in the more volatile phase of the system along the direction of phase separation in a $\rho_n$--$\rho_p$ space, as an effect of the potential term in the symmetry energy~\cite{Ducoin2007,Colonna2008}.

Thus, it is particularly interesting to analyse the developing of isovector fluctuations in two-component nuclear matter.
	Those latter correspond to phase-space density modes where neutrons and protons oscillate out of phase.
	In processes where fragments arise rapidly, like in first-order phase transitions, isovector fluctuations contribute in determining the isotopic properties of the low- and high-density fractions which compose the mixed phase.

	Selecting isovector modes ($\q\rightarrow \v$) in Eq.~(\ref{eq:BLE_Fourier}), 
the phase-space density corresponds to $f^\v=f_\n-f_\p$.
	In order to isolate the isovector behaviour, we prepare nuclear matter in stable conditions.
	To keep nuclear matter uniform (no inhomogeneities will arise in configuration space), we keep only the isovector contribution in the potential, in absence of isoscalar terms, and we rely on the quantity
\begin{equation}
	U^\q\rightarrow U^\v = 2[(\rho_n-\rho_p)/\rho^0]E_{\textrm{sym}}^{\textrm{pot}}\;, 
\label{eq:uiso}
\end{equation}
where $\rho^0$ is the uniform-matter density and $E_{\textrm{sym}}^{\textrm{pot}}$ is the potential term in the symmetry energy.
	Following the procedure of ref.~\cite{Colonna2013}, $U_k^\v$ is obtained from the above quantity by introducing an interaction range through a Gaussian smearing $g_\sigma$ of width $\sigma$, and by taking the Fourier transform; its derivative with respect to $(\rho_n\!-\!\rho_p)$ yields 
\begin{equation}
	F^\v(k) = \frac{2}{\rho^0}E_{\textrm{sym}}^{\textrm{pot}}(\rho^0)g_\sigma(k) + \frac{1}{\N}\;.
\label{eq:DerivFtr}
\end{equation}
Substituting in Eq.~(\ref{eq:fluctuation_dissipation}) we obtain a relation between the isovector variance and the symmetry free energy
\begin{equation}
F_{\textrm{eff}}^\v = 
	\frac{T}{2\Delta V} \frac{\rho^0}{(\sigma_{\rho^\v})^2} = \frac{T}{2\Delta V} \frac{\rho^0}{\langle [\delta\rho_n(\vecr)-\delta\rho_p(\vecr)]^2 \rangle} 
	\;,
\label{eq:isovector_variance}
\end{equation}
where $F_{\textrm{eff}}^\v$ can be assimilated to an effective symmetry free energy which, at zero temperature and neglecting surface effects, coincides with the symmetry energy,  $E_{\textrm{sym}}(\rho^0)$.
In conventional BUU calculations however, the smearing effect of the test particles introduces a corresponding scaling factor~\cite{Colonna1994_a}, so that 
\begin{equation}
	F_{\textrm{eff}}^\v \approx \Ntest E_{\textrm{sym}}\;.
\label{eq:scalingfac}
\end{equation}
	Such scaling actually reduces drastically the isovector fluctuation variance produced by the UU collision term.
	In this paragraph we investigate how the collision term used in the BLOB approach differs from employing a UU treatment.
	Since the former is not an average contribution and it acts independently of the number of test particles, we expect a larger isovector fluctuation variance.

%
%
\begin{figure}[t!]\begin{center}
	\includegraphics[angle=0, width=.6\columnwidth]{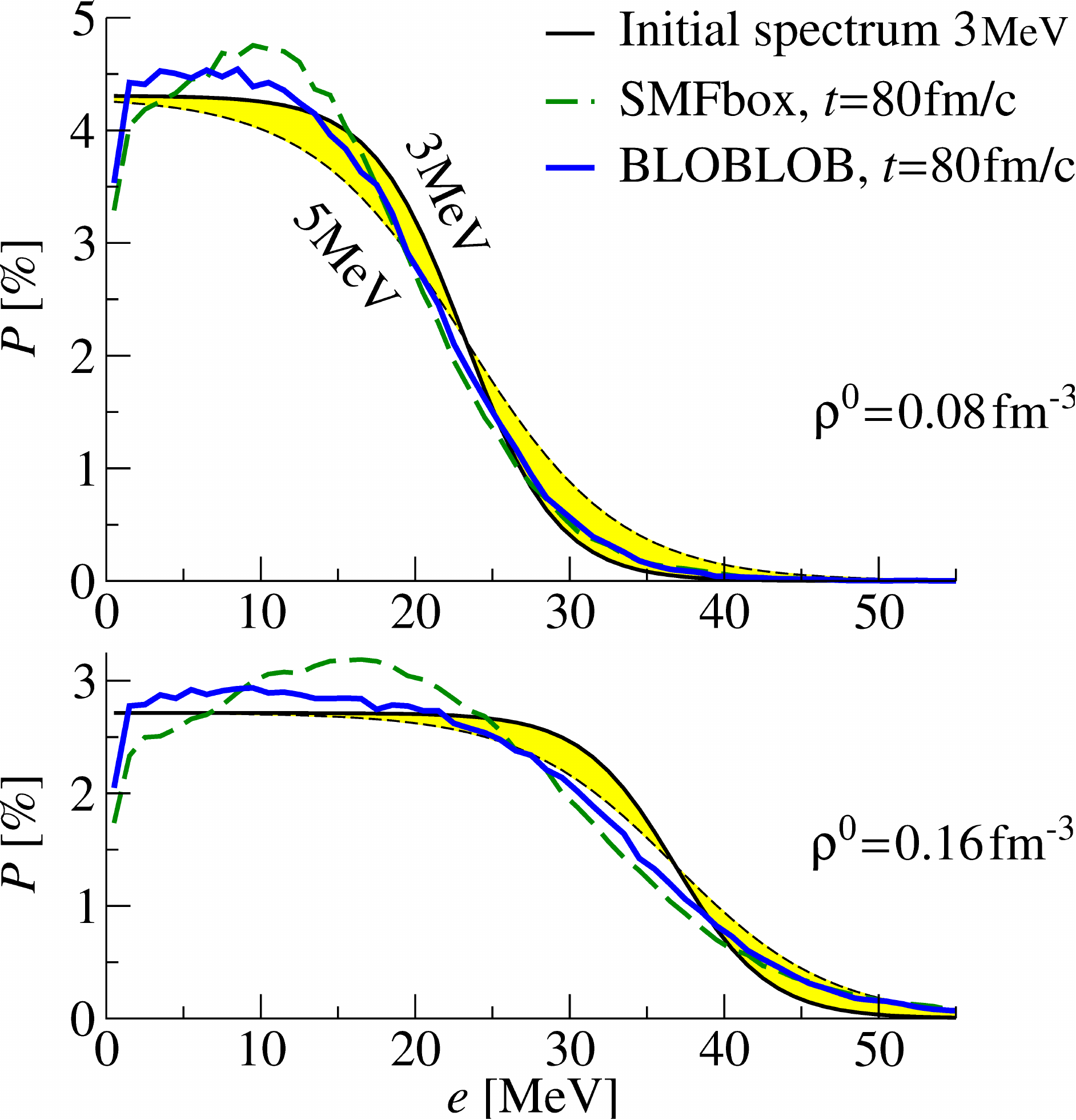}
\end{center}\caption
{
	Energy spectra for $\rho^0=0.8$ and $1.6$~fm$^{-3}$.
	The coloured bands indicate the variation of a Fermi-Dirac distribution from $T=3$ to $T=5$~MeV, the first value corresponding to the initial conditions.
	Spectra evaluated at $t=80$~fm/c, through the SMF and BLOB approaches are overlapped.
	An asy-stiff parameterisation is used.
}
\label{fig_Fermi_Dirac}
\end{figure}
	To prepare a transport calculation, the system is sampled for several values of $\rho^0$ and the potential, restricted to the only isovector contribution, is tested for a stiff and a soft density dependence of the symmetry energy.

	The system is initialised with a Fermi-Dirac distribution at a temperature $T_0=3$~MeV. 
	As shown in Fig.~\ref{fig_Fermi_Dirac}, both SMF and BLOB transport dynamics succeed to preserve the initial distribution quite efficiently as a function of time, even though a flattening of the spectrum, 
due to the fact that the Fermi statistics is not perfetly preserved, around an effective $T_{\textrm{eq}}$, should be accounted for.
	This temperature modification depends on the parameters of the calculation and is larger for larger densities.

%
%
\begin{figure}[b!]\begin{center}
	\includegraphics[angle=0, width=.64\columnwidth]{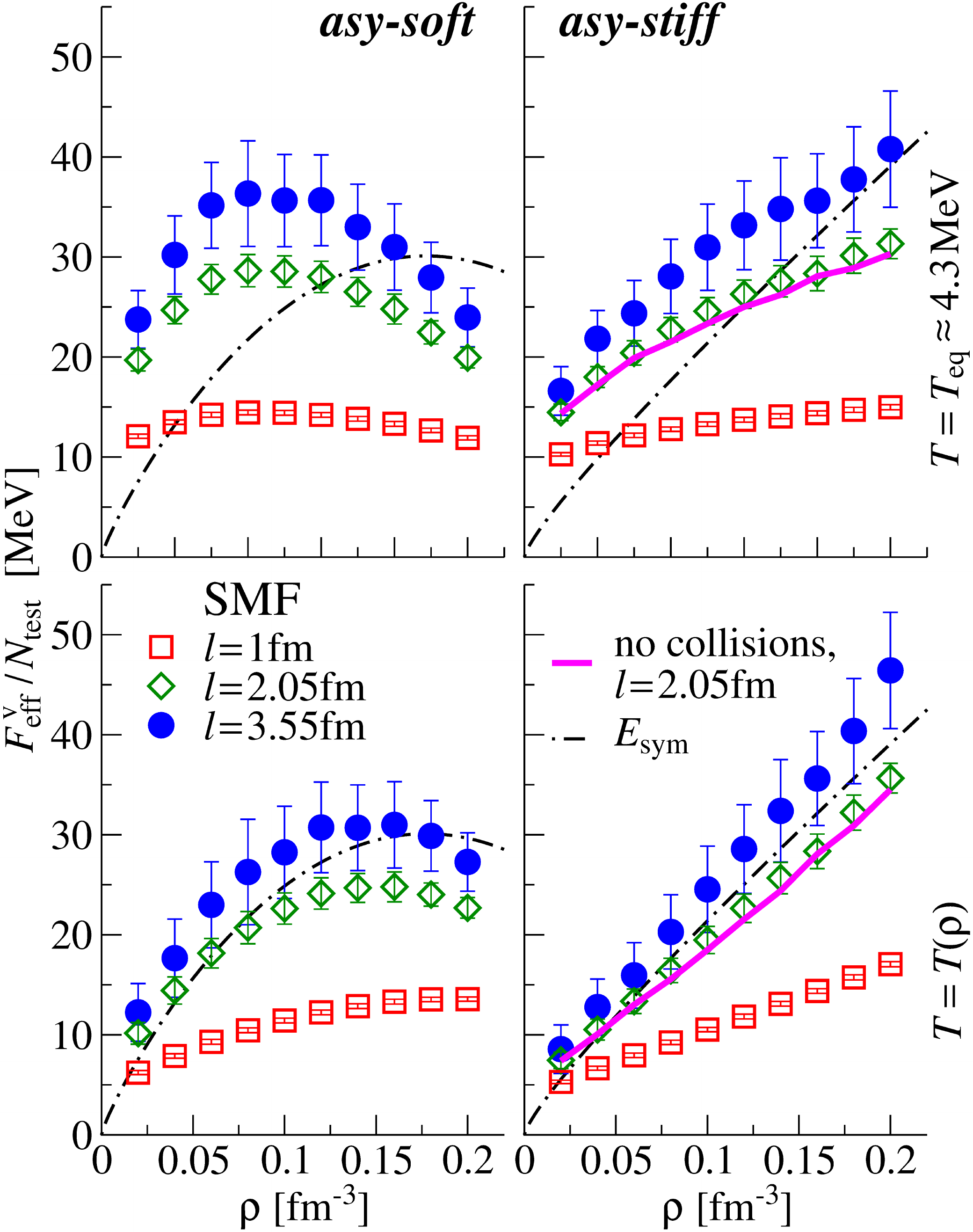}
\end{center}\caption
{
	Numerical solution of Eq.~(\ref{eq:isovector_variance}) for SMF, including the normalisation by $\Ntest$ for asy-stiff and asy-soft forms of the symmetry energy and for different sizes of the cells where the isovector variance is evaluated.
	The temperature correction is applied as equal to $T_{\textrm{eq}}$ calculated at saturation density, or extracted for each density bin.
	For the asy-stiff case a calculation relying only on a collisionless dynamics is added.
}
\label{fig_Fivrho}
\end{figure}
From a set of calculations for different densities ranging from $\rho^0=0.02$ to $\rho^0=0.2$~fm$^{-3}$ we obtain a numerical solution of Eq.~(\ref{eq:isovector_variance}) for SMF.
	We use the average equilibrium temperature $T_{\textrm{eq}}\approx4.3$~MeV, calculated at saturation density, for all other densities or, alternatively an equilibrium temperature extracted for each density bin from the slope of the Fermi-Dirac distribution evolved in time.
	The isovector variance $(\sigma_{\rho^\v})^2$ has been calculated in cells of edge size $l=1$, $l=2.05$ and $l=3.55$~fm and multiplied by $\Ntest$, in order to extract $F_{\textrm{eff}}^\v$ and to compare it with 
the symmetry energy $E_{\textrm{sym}}$.
	The comparison, shown in Fig.~\ref{fig_Fivrho}, is satisfactory and it is the closest in shape to $E_{\textrm{sym}}$ for larger cells than $l=1$~fm but, however, the large scaling factor $\Ntest$ had to be taken into account.
	The better agreement in larger cells reflects the decreasing importance of surface effects, which should allow recovering the (volume) symmetry energy.
	We notice that an equivalent calculation where the collision term is suppressed yields identical distributions; such collisionless calculation corresponds to switching off the collision term either in SMF or, equivalently, in BLOB, since the mean-field is implemented identically.
	The need of scaling by $\Ntest$, to recover the expected fluctuation value, reflects the fact that isovector fluctuations
are not correctly implemented in SMF, and the fluctuations which arise in the system are related to the use of a finite number of
test particles.  Indeed, in SMF, much attention is paid to a good reproduction of isoscalar fluctuations and amplification of mean-field unstable modes, by introducing an appropriate external field~\cite{Colonna1998}. 
On the other hand, explicit fluctuation terms are not injected in the isovector channel.  In this case one just obtains the fluctuations related to the use
of a finite number of test particles, which, as far as the Fermi statistics 
is preserved, amount to the physical ones divided by  $\Ntest$.


%
%
\begin{figure}[t!]\begin{center}
	\includegraphics[angle=0, width=.65\columnwidth]{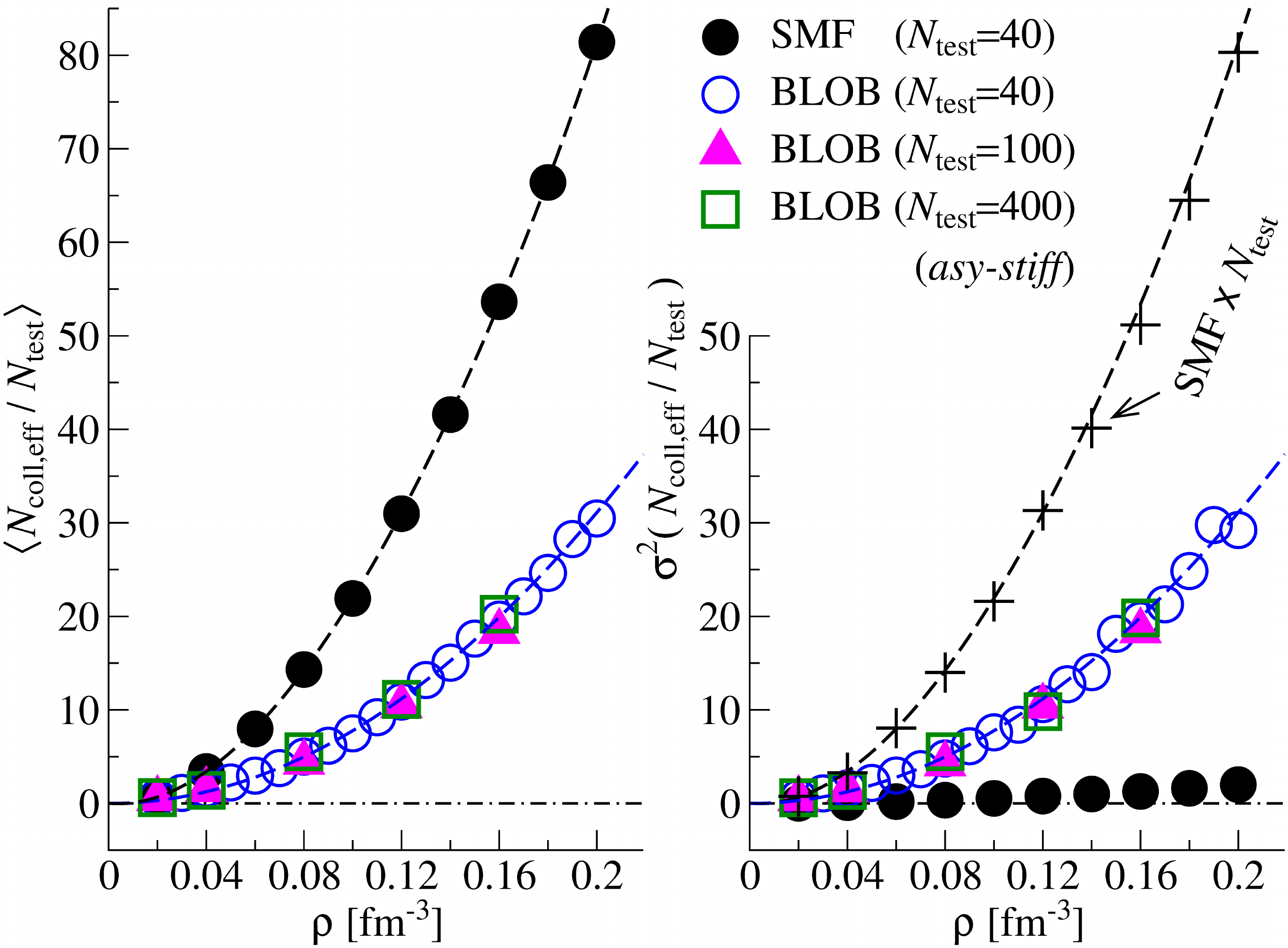}
\end{center}\caption
{
	Mean-value and variance of the number of effective collisions per nucleon for SMF and BLOB calculated at equilibrium (the average collision rate is constant in time) as a function of density.
	Dashed lines are quadratic fits to the mean-value specra; the fit to mean-values is repeated over the variance spectrum to show for the BLOB calculation the perfect correspondence, and for the SMF calculation the correspondence with the variance scaled by Ntest.
	Calculation with an increased number $\Ntest$ are presented for BLOB to show the independence on $\Ntest$.
	An asy-stiff parameterisation is used.
}
\label{fig_collisionvariance}
\end{figure}
	This fluctuation reduction in SMF is illustrated in Fig.~\ref{fig_collisionvariance}.
When studying the mean and variance of the number of effective collisions per nucleon in SMF, a negligible variance is found even if the mean grows with density.
	In practice, within the UU description, larger densities provide a larger number of collision candidates, so that, even if also the difficulty in relocating collision partners increases due to the Pauli blocking, the resulting number of effective collisions per nucleon grows significantly with density. 
	Though, the corresponding collision variance does not follow such trend, keeping 
a dependence on density reduced by a factor $\Ntest$ with respect to the mean.
This completes the study of ref.\cite{Colonna2013} concerning SMF.
Now we move to discuss also BLOB calculations.  
	Fig.~\ref{fig_collisionvariance} indicates that, despite the use of the same nucleon-nucleon cross section (which produces equal rates of attempted collisions per nucleon for all the employed approaches, not shown), the number of effective collisions per nucleon differs in the two models due to the different treatment of the Pauli blocking, which 
is more severe in BLOB, owing to the nucleon wave packet extension
(for instance, 98\% is the Pauli rejection rate in BLOB at $\rho^0=0.16$~fm$^3$).
	However, the study presented in Fig.~\ref{fig_collisionvariance} confirms that the variance of the number of effective collisions per nucleon in BLOB is large and exactly equals the mean value, according to the Poisson statistics~\cite{Burgio1991}.

%
%
\begin{figure}[b!]\begin{center}
	\includegraphics[angle=0, width=.65\columnwidth]{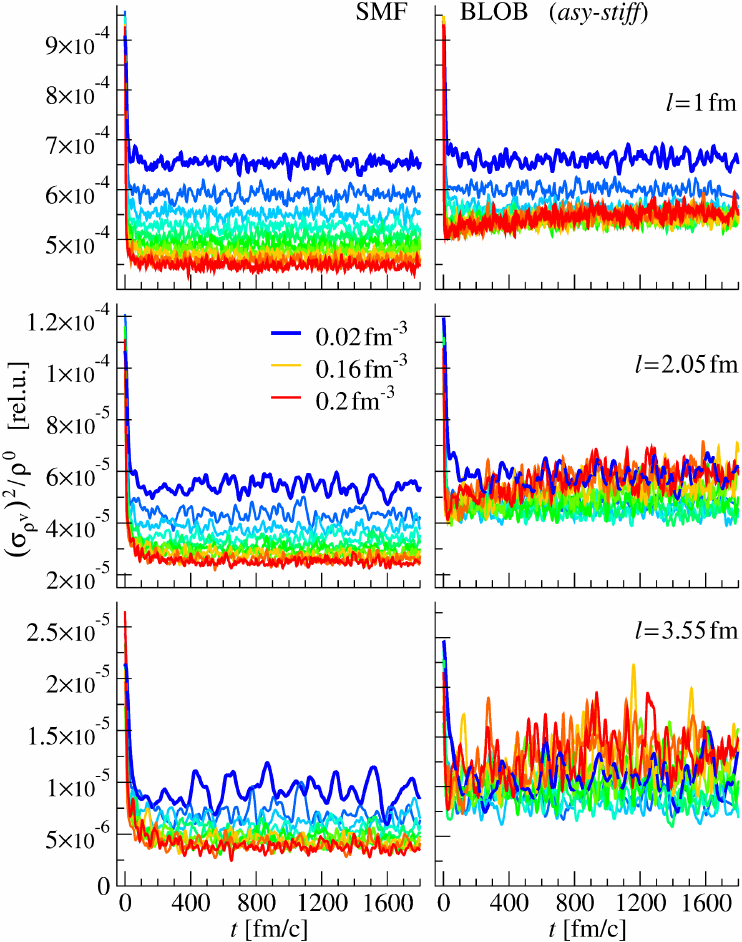}
\end{center}\caption
{
	Isovector variance as a function of time for different values of the system density for an asy-stiff form of the symmetry energy, for SMF and for BLOB, evaluated in cells of different size $l$.
	BLOB systematically shows larger values and a longer convergence time for large densities.
	Both calculation converge to the same variance at the small-density limit.
}
\label{fig_densityvariance_versus_t}
\end{figure}
%
%

%
%
\begin{figure}[t!]\begin{center}
	\includegraphics[angle=0, width=.65\columnwidth]{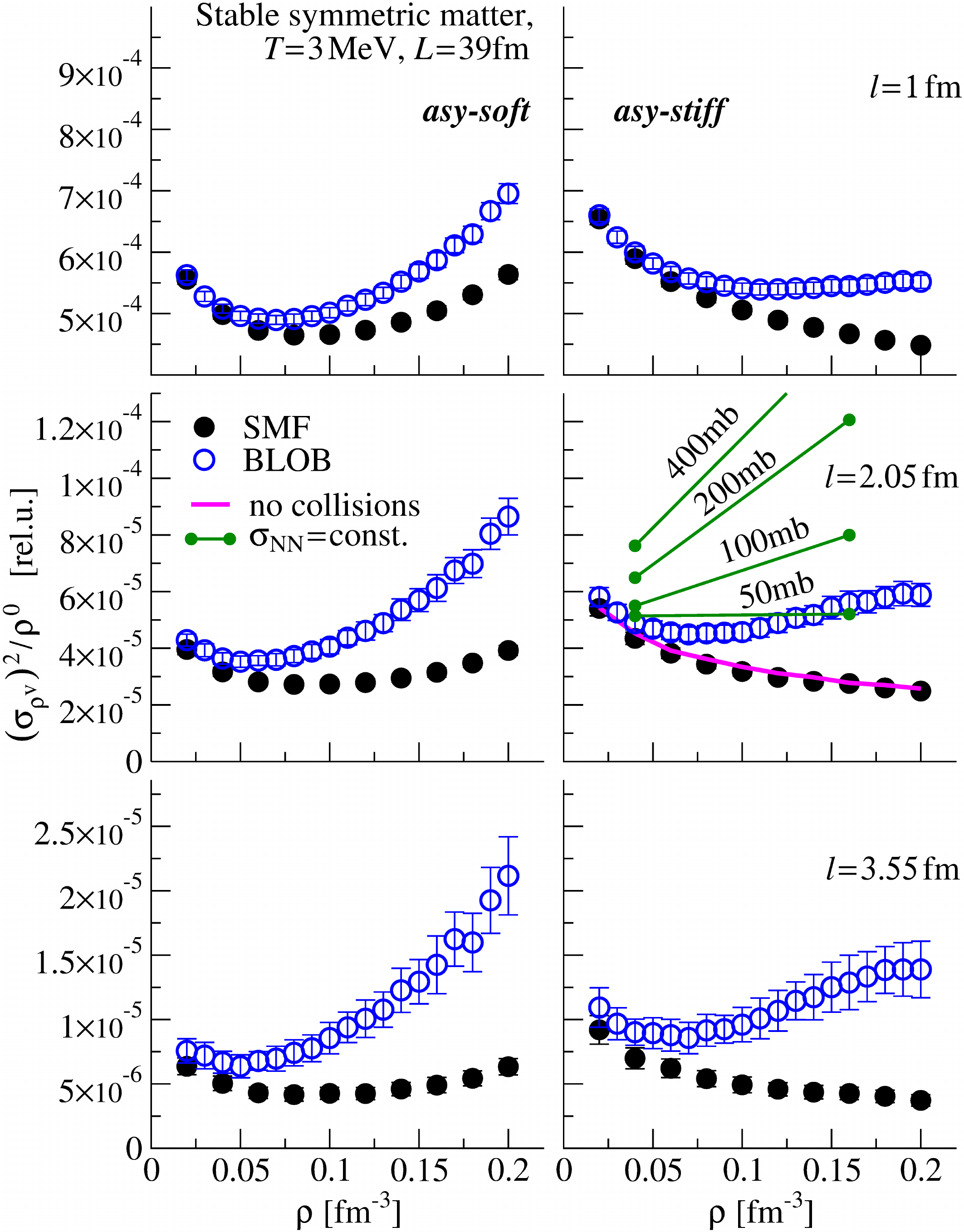}
\end{center}\caption
{
	Convergence value of isovector variance as a function of system density for asy-stiff and asy-soft forms of the symmetry energy evaluated in cells of different size $l$, for BLOB and SMF.
	For one case (asy-stiff, $l=2.05$~fm) other calculations are added: a collisionless calculation is identical to SMF; various BLOB calculations employing a constant $\sigma_{\textrm{NN}}$ with different values exhibits a dependence on the nucleon-nucleon cross section.
}
\label{fig_densityvariance_versus_rho}
\end{figure}

	Fig.~\ref{fig_densityvariance_versus_t} shows that the isovector variance in BLOB results larger than in SMF and, in general, larger than in a corresponding collisionless calculation. 
	Such difference is therefore the effect of the treatment of collisional correlations in BLOB, which displays a dependence with the system density.
	In particular, the low-density limit of the spectrum corresponds to a situation where the collision rate is vanishing.
	In this case, 
the BLOB procedure is practically ineffective (see \textsection~\ref{sub_noise})
and all approaches converge to the same isovector variance, just related to the finite number of test particles employed.
	At larger density than saturation ($\rho\ge 0.18$~fm$^{-3}$) BLOB displays a longer paths to convergence which is due to the difficulty of relocating large portions of phase space in binary collisions without violating Pauli blocking.
	Fig.~\ref{fig_densityvariance_versus_rho} condenses and extends the information of Fig.~\ref{fig_densityvariance_versus_t} by displaying the density evolution of the isovector variance attained at equilibrium as evaluated in cells of different size $l$, for asy-stiff and asy-soft forms of the symmetry energy.
	The SMF data correspond to those analysed in Fig.~\ref{fig_Fivrho}.
	The BLOB spectra progressively deviate from SMF data for increasing density.
	Such deviation 
increases for larger cell sizes indicating that the isovector fluctuations are better built in large volumes~\cite{Rizzo2008}.
	This is related to the variety of configurations, concerning 
shape and extension of the nucleon wave
packet, which occur in the implementation of the fluctuating collision
integral. This introduces a smearing of fluctuations on a scale comparable
to the wave packet extension in phase space.  
	However, the gain in isovector variance exhibited by the BLOB approach, indicates that the dependence on $\Ntest$ is partially reduced with respect to the SMF scheme.

\subsection{Interference between mean field, collisional correlations and numerical noise
\label{sub_noise}}

	According to Eq.~(\ref{eq:BLOB}), the BLOB approach should introduce and revive isovector fluctuations continuously.
	However, the procedure has chances to work only if there are no other antagonist sources which destroy isovector correlations.
	The agglomeration procedure employed in BLOB is actually able to construct agglomerates of test particles of the same isospin species and which are located around local density maxima in random selected phase-space cells: this technique should preserve at least partially the isovector correlations in the system, contrarily to the usual BUU technique which smears them out.
	This advance with respect to BUU is however far from being sufficient because the greatest smearing effect comes from the 
mean field itself which, even in absence of any explicit fluctuation seed, is actually affected by its own numerical noise, due to the use of a finite number of test particles in the numerical resolution of the transport equations
; such spurious contribution imposes the dependence of $(\sigma_{\rho^\q})^2$ on $\Ntest$~\cite{Colonna1993}.
	If this latter may be negligible with respect to the large isoscalar fluctuations introduced by the BLOB stochastic collision term (in presence of instabilities), it becomes a highly interfering contribution for the weaker isovector modes.
	As discussed in ref.~\cite{Reinhard1992}, the use of a finite number of test particles, i.e. the approximate
mapping of the one-body distribution function, induces a numerical noise that may even cause deviation
from the fermionic statistics, towards a classical behaviour of the system.  This effect is more pronounced
when the collision integral is neglected.  Indeed the latter contains explicit Pauli-blocking factors and helps
restoring the fermionic behaviour. 
The numerical noise leads, on a short time scale,  to fluctuations corresponding
to the expected value, but reduced by $\Ntest$ (as far as the Fermi statistics is still preserved).  In other words, the numerical noise induces an effective 
diffusion coefficient $D_k' = D_k + D_k^{\textrm{noise}}$ and an effective relaxation time
$1/\tau_k' = 1/\tau_k^{\textrm{coll}} + 1/\tau_k^{\textrm{m.f.}}$.  $\tau_{\textrm{m.f.}}$ depends on the test-particle number (becoming infinite when the test-particle number goes to infinity). 
If a small number of test particles is considered,  and two-body collisions
are not so frequent, then $D_k^{\textrm{noise}}$ will prevail over $D_k$ and   $1/\tau_{\textrm{m.f.}}$ over $1/\tau_k^{\textrm{coll}}$.   For this reason, though in principle the fluctuation
equilibrium value should not depend on the nucleon-nucleon cross section, our results are sensitive to the cross section amplitude and to the number of test particles employed.   

%
%
\begin{figure}[b!]\begin{center}
	\includegraphics[angle=0, width=.65\columnwidth]{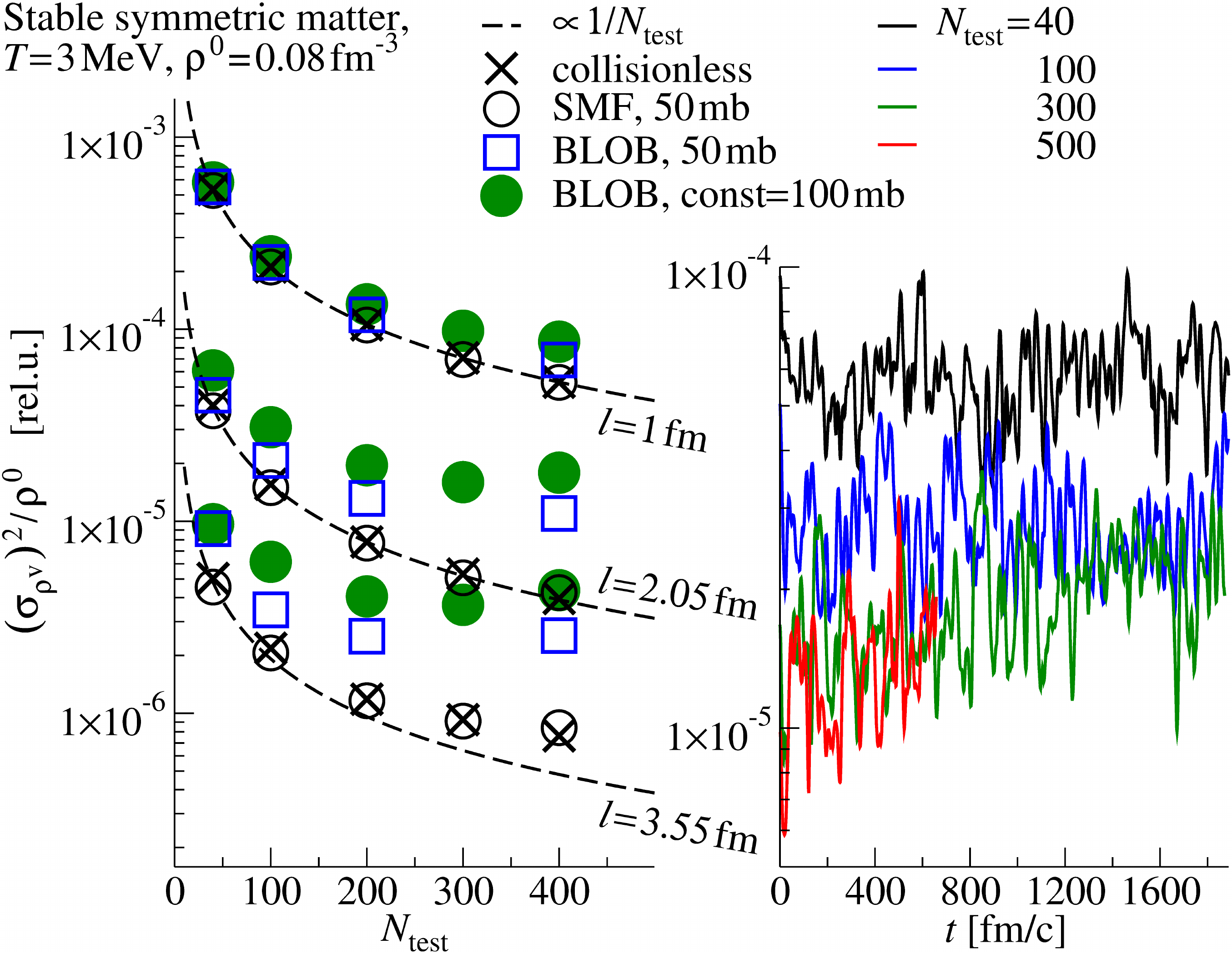}
\end{center}\caption
{
	Study of $\Ntest$ dependence of isovector variance for asy-stiff forms of the symmetry energy 
$\rho^0=0.08$~fm$^{-3}$, evaluated in cells of different size $l$, for a collisionless approach, SMF with constant $\sigma_{\textrm{NN}}$=50~mb, BLOB with constant $\sigma_{\textrm{NN}}$=50 and 100~mb.
	Left. dependence of the mean value at equilibrium as a function of $\Ntest$ for various models compared to a $1/\Ntest$ low.
	Right. Time dependence showing the tendency to lose $\Ntest$ dependence for BLOB with constant $\sigma_{\textrm{NN}}$=100~mb.
}
\label{fig_Ntest}
\end{figure}
	Two ways can be tested to overcome this problem: either the collision term should be considerably enhanced, or fluctuations generated by the action of test particles should be controlled.

	The first solution can be achieved by simply multiplying the nucleon-nucleon cross section by a large factor, with the drawback of then handling incorrect collision rates which would be a severe concern when describing out-of-equilibrium processes, like heavy-ion collisions.
	Some tests in the first direction are proposed in Fig.~\ref{fig_densityvariance_versus_rho}, by employing a constant $\sigma_{\textrm{NN}}$ with progressively larger values, showing that the isovector variance grows with the collision rate as expected.
	As shown in Fig.~\ref{fig_densityvariance_versus_rho}, we observe that the fluctuation variance built by BLOB may deviate significantly, up to a factor 10, from the SMF results, especially when considering larger cells ($l\approx 2-3 fm$) to evaluate the one-body density.   

	The second solution would consist in employing the largest possible number of test particles per nucleon.
	In this case, the collisionless transport model would ideally correspond to the Vlasov approach and, when collisional correlations are introduced, interferences with spurious stochastic sources can be highly reduced.
	As far as numerical complexity can be handled, Fig.~\ref{fig_Ntest}, left, illustrates such situation:
SMF and collisionless calculations show the same behaviour $\propto 1/\Ntest$ and no dependence on $\sigma_{\textrm{NN}}$.
	On the other hand, BLOB calculations show 
a $\sigma_{\textrm{NN}}$ dependence, which deviates more and more, for large test particle numbers, 
from the SMF results, especially in the
largest cells, where the BLOB fluctautions are better entertained.
	Fig.~\ref{fig_Ntest}, right, illustrates that small, progressively increasing values of $\Ntest$, are related to a systematically decreasing isovector variance, which is still completely dominated by the noise.
	Only when the number of test particles per nucleon becomes very large, the isovector variance loses its dependence on $\Ntest$ and exhibits a clear tendency to grow in time towards a larger value, signing that isovector correlations are not only preserved, but they are also revived.
	However, since we are interested in physical conditions at low temperature, 
the number of nucleon-nucleon collisions is extremely low and insufficient to rapidly introduce a pattern of isovector correlations: the isovector variance shows in fact a very gentle growth.

	In conclusion, the BLOB fluctuation source term works well in conditions where the collision rate is large enough, as compared to the spurious dissipative terms associated with the finite number of test particles and to the mean-field propagation. 
	These conditions are likely reached in the first, non equilibrated stages of heavy ion collisions at Fermi and intermediate energies, but not necessarily for equilibrated nuclear matter at low temperature. In the latter case, the variance associated with the fluctuating collision integral can be recovered by artificially increasing the n-n cross section employed in the calculations.

\section[Isoscalar fluctuations in stable and unstable nuclear matter]{Isoscalar fluctuations in mechanically stable and unstable nuclear matter}

	If fluctuation seeds are introduced in homogeneous neutral nuclear matter at low temperature, Landau zero-sound~\cite{Landau1957} collective modes should stand out and propagate in the system. 
	We analyse in the present section whether, as aimed, the BLOB approach is able to develop isoscalar fluctuations of correct amplitude spontaneously, and not from an external contribution, in nuclear matter when the system is placed in a dynamically unstable region of the equation of state~\cite{Belkacem1994}, like the spinodal zone, where a density rise is related to a pressure fall.
	In this circumstance, as soon as fluctuation seeds are generated, unstable zero-sound waves should be amplified in time.
	In the opposite situation, in conditions of mechanical stability, undamped stable zero-sound waves propagate.
	When unstable modes succeed to get amplified, inhomogeneities develop and eventually lead to mottling patterns at later times.
	This mechanism has been intensively investigated~\cite{Chomaz2004} foremost because in open dissipative systems, like heavy-ion collisions, it corresponds to a catastrophic process which can lead to the formation of nuclear fragments~\cite{Tabacaru2003,Borderie2008}.	
	When the temperature is significant, two-body collision rates become prominent and these mean-field dominated zero-sound waves are absorbed and taken over by hydrodynamical first-sound collective modes.
	Since our approach exploits two-body collisions to introduce fluctuations in a self-consistent mean field, we expect the possible occurrence of a zero-to-first-sound transition which, at variance with other Fermi liquids~\cite{Abel1966}, should be even smeared out due to the small values taken by the Landau parameter $F_0$ in nuclear matter.
	It was found that, depending on how the system is prepared and on the type of collective motion, such transition should arise in a range of temperature from 4 to 5 MeV and occur as late as 200~fm/c~\cite{Larionov2000, Kolomietz1996}.
	In practice, zero-sound modes associated to wave vectors $\veck$ characterise the system as long as the corresponding phase velocity exceeds the velocity of a particle on the Fermi surface $\vF$ or, equivalently, as long as the corresponding frequency $\omega_k$ is much higher then the two-body collision frequency $\nu$.
	These premises imply that after defining a homogeneous initial configuration at a suited finite and not so large temperature, we should study early intervals of time to extract properties of the response function which can be compared with zero-sound conditions.

%
%
\begin{figure}[b!]\begin{center}
	\includegraphics[angle=0, width=.7\textwidth]{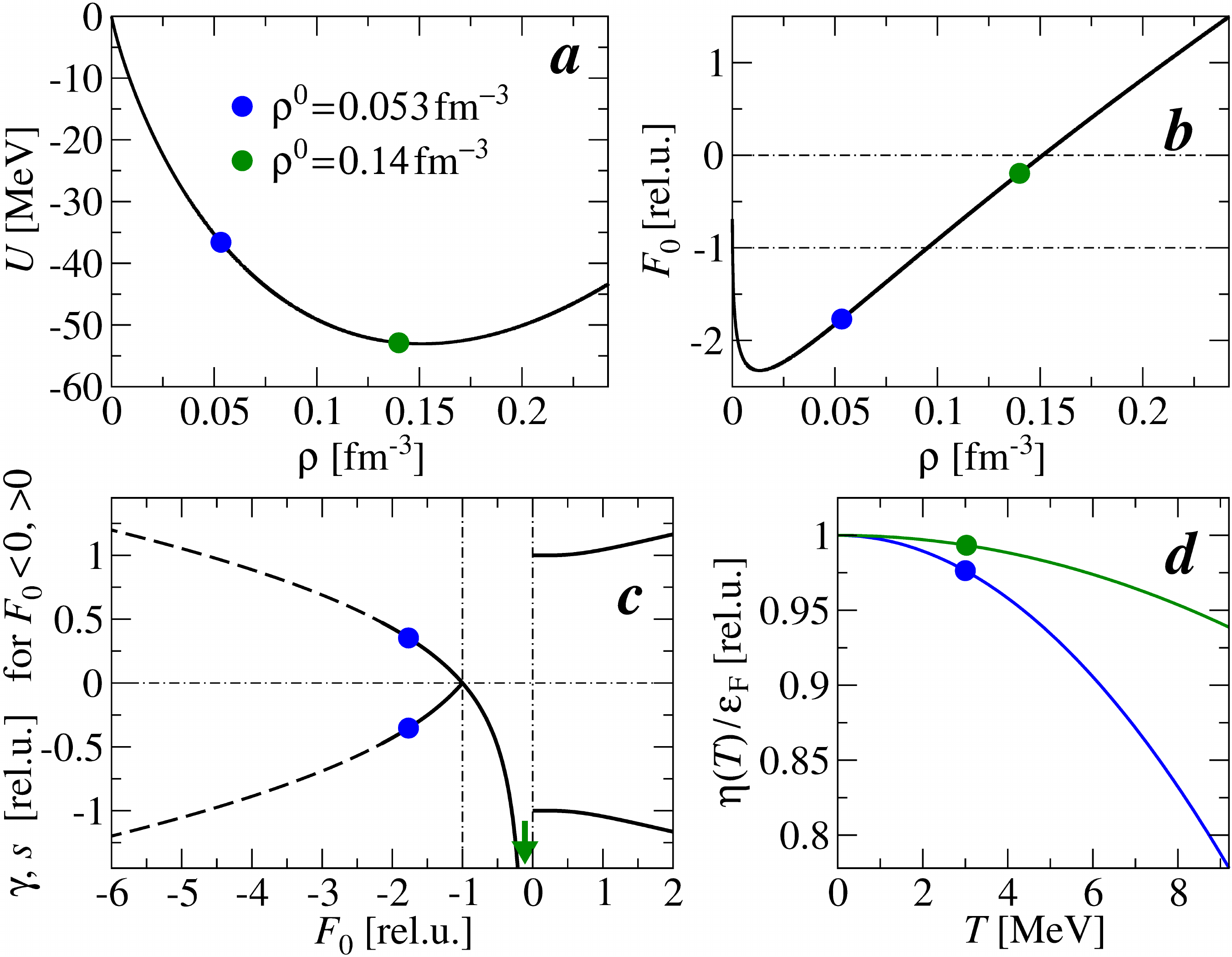}
\end{center}\caption
{
	Two choices for the system density are illustrated in relation with corresponding quantities. 
	(a) Nuclear potential. (b) Landau parameter. (c) real ($F_0>0$) and imaginary ($F_0<0$) roots of the dispersion relation; the spinodal region corresponds to $F_0<-1$, the region for $-1<F_0<0$ corresponds to Landau damping and positive values of $F_0$ define stable modes. The arrow indicates that the point for $0.14$~fm$^{-3}$ is situated at a very negative value of the ordinate. (d) Temperature effects on the dispersion relation for the two system densities.
}
\label{fig_disp_rel_analytical}
\end{figure}
	For the numerical approach we keep the previous scheme for the definition of the box metrics; the isoscalar density variance is calculated over cells of edge size $l=1$fm.
	We now use the full parametrisation of the energy potential per nucleon Eq.~(\ref{eq:pot}), where we use a stiff density dependence of $E_{\textrm{pot}}$ (the same parametrisation was analysed in ref.~\cite{Colonna1997}).
	A value of $C_{\textrm{surf}}\!=\!-7/\rho_0$ MeV fm$^5$ is chosen for the surface term.
	Nuclear matter is isospin symmetric and is initially uniform and prepared at a temperature $T=3$MeV and a densities equal to $\rho^0\!=\!0.053$ and $0.14$~fm$^{-3}$. 
	Fig.~\ref{fig_disp_rel_analytical}a illustrates the potential values related to these choices.
	$\Ntest=40$ test particles per nucleon are employed.
	The collision term involves the usual isospin- and energy-dependent free nucleon-nucleon cross section with an upper cutoff at $\sigma_{\textrm{NN}}=50$mb.

\subsection{Sampling zero-sound propagation}

	The early growth of fluctuations in nuclear matter can be described in a linear-response approximation~\cite{Colonna1994_a} as far as deviations from the average dynamical path are small.
	In Eq.~(\ref{eq:BLE_Fourier}), by selecting isoscalar modes ($\q\rightarrow \s$, we drop the $\s$ index in the following), and setting residual contributions equal to zero, we obtain a linearised Vlasov equation in terms of frequencies $\omega_k$ to describe 
nuclear matter with isoscalar contributions:
\begin{equation}
	\omega_k f_k + \veck\cdot\frac{\vecp}{m}f_k - \frac{\partial f^0}{\partial \epsilon}\frac{\partial U_k}{\partial \rho} \veck\cdot\frac{\vecp}{m} \rho_k = 0
	\;,
\label{eq:linearised_Vlasov}
\end{equation}
	Different wave numbers $\veck$ are decoupled, each linked to a collective solution $f_k$ given by the Fourier-transformed equation of motion.
	By applying the self-consistency condition
\begin{equation}
	\rho_k(t) = (g/h^3)\int f_k(\vecp,t) \diff\vecp
\label{eq:selfcon}
\end{equation}
we obtain the dispersion relation for the propagation of density waves in Fermi liquids at $T=0$ :
\begin{equation}
	1 = \frac{g}{h^3}  \frac{\partial U_k}{\partial\rho} \int \frac{\partial f^0}{\partial\epsilon}  \frac{\veck\cdot\vecp /m}{\omega_k+\veck\cdot\vecp /m} \diff\vecp
	\;.
\label{eq:dispersion_relation_primitive}
\end{equation}
where $\omega_k$ and $-\omega_k$ are pair solutions due to the invariance $\vecp \leftrightarrow -\vecp$.
	As well documented in the literature, at $T=0$, eigenmodes $f_k$ depend on states near the Fermi level.
	The momentum integral should therefore be restricted to the Fermi surface so that $\partial_\epsilon f^0 \approx -\delta(\epsilon-\eF)$, being $\eF$ the Fermi energy, and angular and energy dependencies can be decoupled so that 
the dispersion relation reduces to an expression where solutions correspond to sound velocities 
\begin{equation}
	s=\omega_k / (k v_\textrm{F})
\label{eq:sound}
\end{equation}
in units of Fermi velocity  $v_\textrm{F}=p_\textrm{F}/m$.
	In this case, introducing the Landau parameter 
\begin{equation}
	F_0(\veck)=\N_0\partial_\rho U_k= (3/2)(\rho^0/\eF)\partial_\rho U_k
\label{eq:Landauparameter}
\end{equation}
linked to the number of levels at Fermi energy $\eF$, the dispersion relation takes the form of the Lindhard function~\cite{Kalatnikov1958}
\begin{equation}
	- \frac{1}{F_0} = L(s) = 1  - \frac{s}{2} \textrm{ln} \left(\frac{s+1}{s-1} \right)
	\;,
\label{eq:disprel_T0}
\end{equation}
where the dependence on $k$ has been removed by the introduction of the sound speed $s v_\textrm{F} = \omega_k/k$, equal for all $k$ waves.
	For the two selected system densities,
	Fig.~\ref{fig_disp_rel_analytical}b illustrates the Landau parameter and
	Fig.~\ref{fig_disp_rel_analytical}c presents the roots of the dispersion relation.

	Such expression, derived for zero temperature, is not consistent with the incorporation of temperature effects through a short-mean-free-path two-body dissipative mechanism.
	Nevertheless, it was proposed~\cite{Yannouleas1992} that the inclusion of a temperature dependence is still possible in a semiclassical picture when supposing that a moving boundary of the system is involved in the dissipation of energy from collective to microscopic degrees of freedom.
	Such so-called wall-dissipation model~\cite{Blocki1978} can be applied to a Fermi gas at finite but small temperature $T$ by identifying the moving boundary with the Fermi surface of the system.
	This argumentation results in including the ratio between the chemical potential $\eta(T)$ at a temperature $T$ and the Fermi energy $\eF$, which carries the temperature dependence
\begin{equation}
	\frac{\eta(T)}{\eF} \approx 1-\frac{\pi^2}{12} \left(\frac{T}{\eF}\right)^2 
	\;,
\label{eq:wall_dissipation}
\end{equation}
as illustrated in Fig.~\ref{fig_disp_rel_analytical}d for the two selected densities. 

	As a further modification, we consider that zero-sound conditions also present a strong dependence on the interaction range.
	This latter can be included in the dispersion relation by applying a Gaussian smearing factor of the mean-field potential $\sigma$, 
which is related to the nuclear interaction range in configuration space~\cite{Colonna1994,Kolomirtz1999}.

\begin{equation}
	U \rightarrow U \otimes g(k), \;\;\textrm{with} \;\; g(k)= \textrm{e}^{-\frac{1}{2} (k\sigma)^2}
	\;,
\label{eq:Gaussian_smearing}
\end{equation}

From Eq.~(\ref{eq:wall_dissipation}) and Eq.~(\ref{eq:Gaussian_smearing}), 
the dispersion relation, Eq.~(\ref{eq:disprel_T0}), will involve  
an effective Landau parameter, 
\begin{equation}
	\widetilde{F}_0(k,T) = \frac{\mu(T)}{\eF}F_0g(k)
	\;.
\label{eq:effectiveF}
\end{equation}

	Mechanically unstable conditions are experienced when the evolution of local density $\rho$ and pressure $P$ implies that the incompressibility 
is negative, so that 
this situation is reflected by an effective Landau parameter $\widetilde{F}_0(k=0,T)$ 
smaller than $-1$:   
\begin{equation}
	\rho \frac{\partial P}{\partial\rho} = \frac{2}{3}\rho\eF
[1+\widetilde{F}_0(k=0,T)] \;\; <\,0
	\;,
\label{eq:chi_Landau}
\end{equation}
and it corresponds to imaginary solutions of the dispersion relation~\cite{Pomeranchuk1959}.
	By replacing $s\rightarrow i\gamma$, the relation yielding imaginary solutions can be put in the form:
\begin{equation}
	1 + \frac{1}{\widetilde{F}_0(k,T)}
= \gamma\,\textrm{arctan}\frac{1}{\gamma}
	\;,
\label{eq:disprel_gamma}
\end{equation}

where $\gamma = i~s$. 
The growth rate $\Gamma_k=1/\tau_k$ is obtained from the solutions of the dispersion relation 
\begin{equation}
|\gamma| = \frac{|\omega_k|}{k\vF} = 
\frac{1}{\tau_k k \vF}
	\;,
\label{eq:gamma}
\end{equation}

	As far as the Fermi statistics is kept in a sufficiently large periodic portion of mechanically unstable nuclear matter, and a fluctuation source term
is acting,  
the expectation is that the intensity of the response should be amplified with the growth rate 
$\Gamma_k$ 
imposed by the mean-field potential $U$ as a function of the unstable mode $\veck$.

%
%
\begin{figure}[t!]\begin{center}
	\includegraphics[angle=0, width=.7\textwidth]{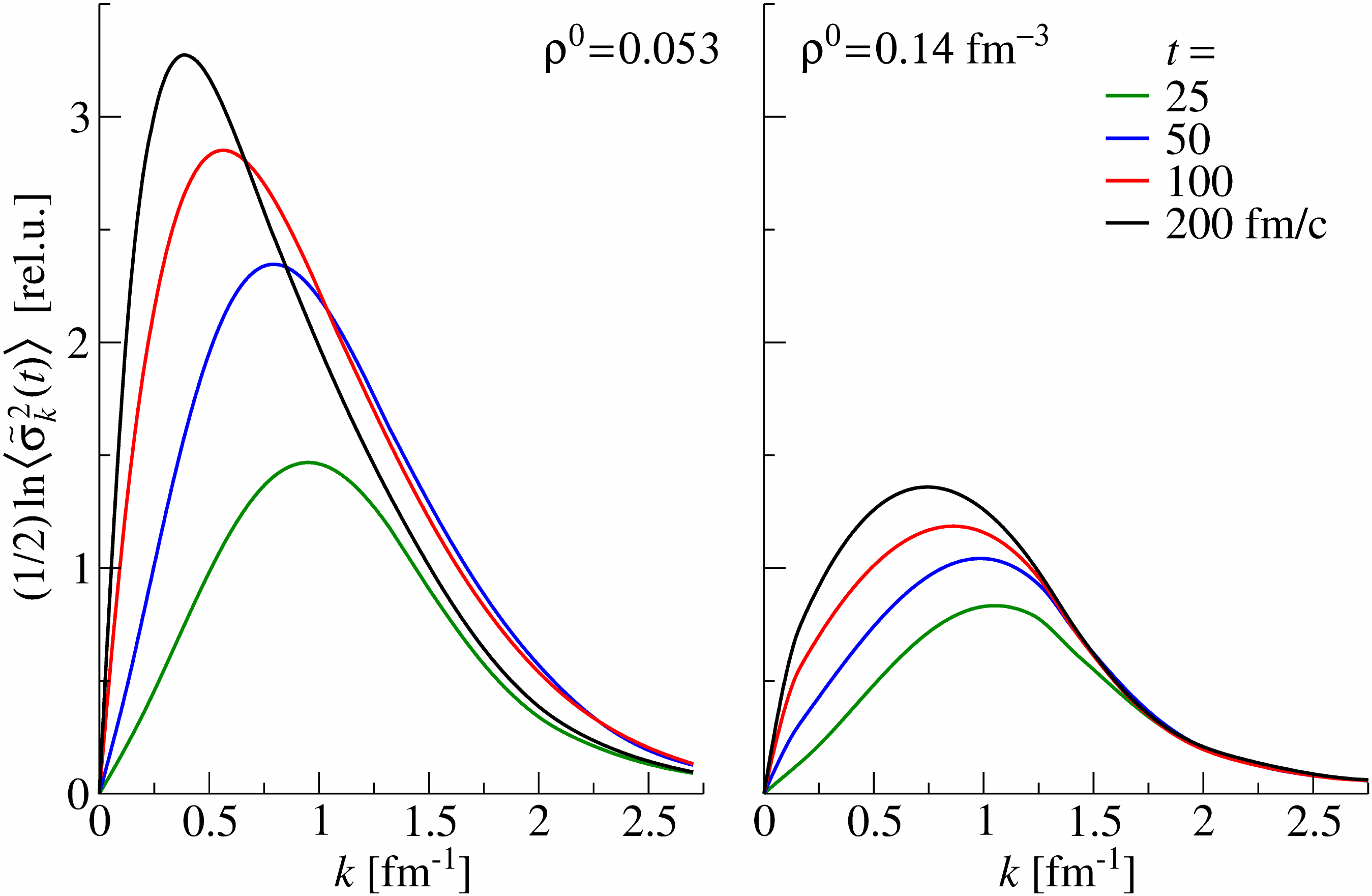}
\end{center}\caption
{
	BLOB calculation. Response intensity $\tilde{\sigma}^2_k(t) = \sigma^2_k(t)/ \sigma^2_k(t=0)$ (fitted curves) at different time intervals for $\rho^0\!=\!0.053$~fm$^{-3}$ (spinodal) and $0.14$~fm$^{-3}$ (Landau damping), averaged over several dynamical paths.
}
\label{fig_isoscalar_variance}
\end{figure}
	To check such expectation numerically through a BL transport approach we should register at each interval of time $t$ the density in all cells of edge size $l$ of the lattice which constitutes the periodic system of edge size $L$.
	A specific cell can be identified by the vector $\vecn'$. 
Having introduced such a lattice, the perturbation wave number $k$ 
can be expressed as $k=2\pi n /L$, where $n$ is the modulus of 
a vector ranging from 1 to $n_{\textrm{max}} = L/l$ along each of the three spacial directions.  
	Then the amplitude of the isoscalar fluctuation of a mode $k$ is obtained from the Fourier transform of the space density

\begin{eqnarray}
	\sigma^2_k(t) &=& \langle F_\veck^2(t)\rangle = 
	\frac{1}{l^3}\Big\langle\Big[\sum_{\vecn'}\rho_{\vecn'}(t)\,\exp\big(a\vecn\!\cdot\!\vecn'\big)\Big]^2\Big\rangle \\\notag
	&\propto& \Big\langle\Big[\sum_{\vecn'}\rho_{\vecn'}(t)\textrm{cos}(a\vecn\!\cdot\!\vecn')\Big]^2 \!+\! \Big[\sum_{\vecn'}\rho_{\vecn'}(t)\textrm{sin}(a\vecn\!\cdot\!\vecn')\Big]^2\Big\rangle 
	\;,
\label{eq:isoscalar_amplitude}
\end{eqnarray}
where $a=2\pi i l/L$ and
the average is extended over all orientations of $\veck$. 
	The distribution of ratios $\tilde{\sigma}^2_k(t) = \sigma^2_k(t)/ \sigma^2_k(t=0)$, averaged over several dynamical paths, are shown at different time intervals in Fig.~\ref{fig_isoscalar_variance} for the two density choices. 
	The initial fluctuation amplitude is an effect of the finite number of test particles employed in the calculations while, as soon as the BLOB term starts to act, fluctuations of larger amplitude are built up (see \textsection~\ref{sub_noise}), and further amplified by the unstable mean-field. 

%
%
\begin{figure}[b!]\begin{center}
	\includegraphics[angle=0, width=.65\textwidth]{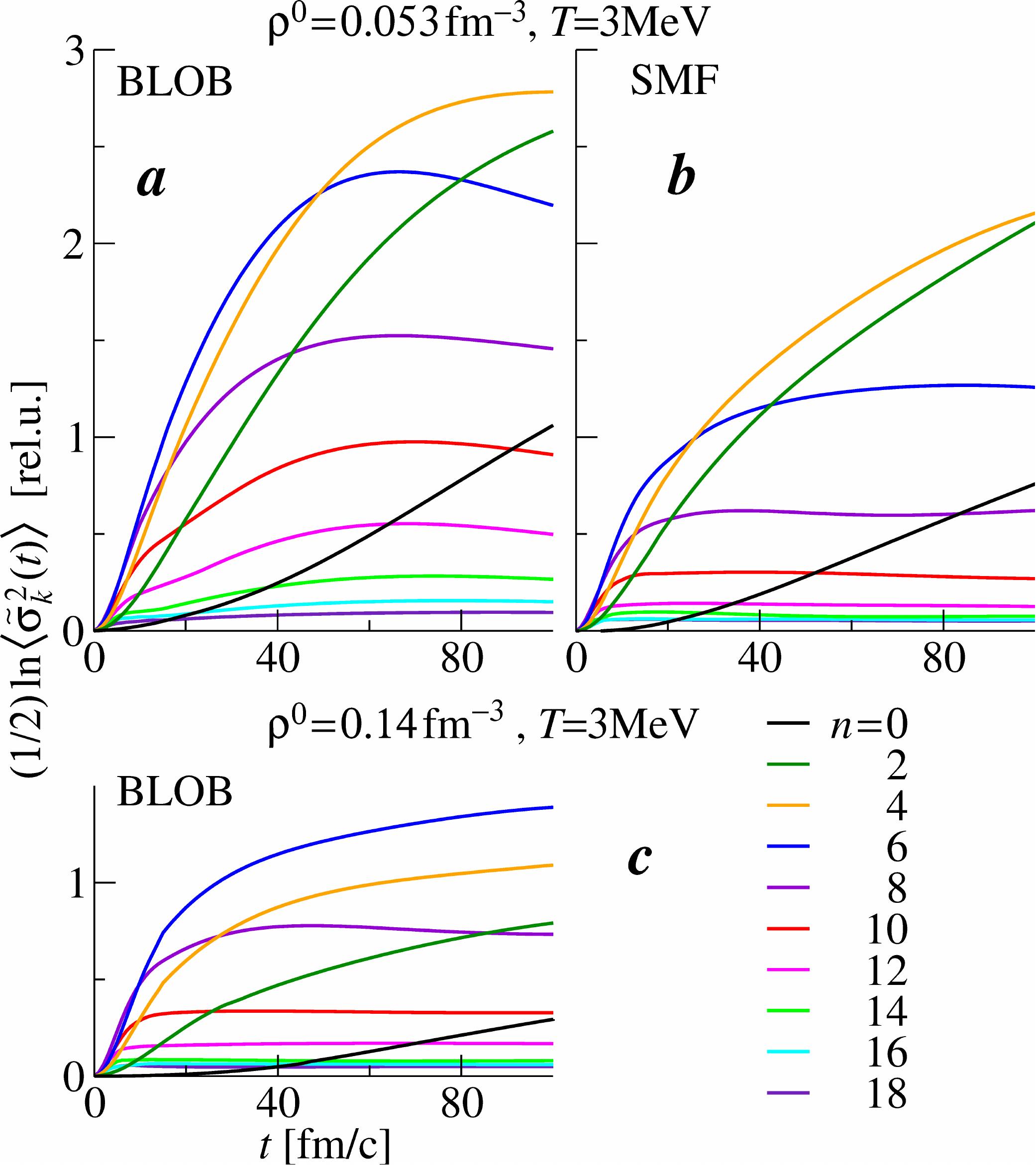}
\end{center}\caption
{
	Early evolution in time of the response intensity $\tilde{\sigma}^2_k(t)$ for several modes ($n=j$ stands for all $k$ modes within $j-1\le n< j$).
	(a) BLOB approach for $\rho^0\!=\!0.053$~fm$^{-3}$ (spinodal). The leading modes are compared to linear fits.
	(b) SMF approach for $\rho^0\!=\!0.053$~fm$^{-3}$.
	(c) BLOB approach for $\rho^0\!=\!0.14$~fm$^{-3}$ (Landau damping).
}
\label{fig_is_variance_early}
\end{figure}
%
%
%
%
	The system prepared at $\rho^0\!=\!0.053$~fm$^{-3}$, inside of the spinodal region, exhibits a clear growth of instabilities as a function of time for some $k$ waves, while the system prepared at $\rho^0\!=\!0.14$~fm$^{-3}$, outside of the spinodal region, presents an evolution of the response intensity which does not lead to the developing of disturbances for any $k$ wave, as we would expect for a phenomenon of Landau damping. However, the fluctuations reach a 
significant amplitude, owing to the small compressibility value in the density
region considered. 
	Correspondingly, the early evolution in time of $\tilde{\sigma}^2_k(t)$ is analysed in Fig.~\ref{fig_is_variance_early} for the two density choices. In Fig.~\ref{fig_is_variance_early}a the leading modes can be compared to linear fits applied to intervals ranging from around 20fm/c to time instants close to saturation. 
	The very initial path is excluded because the fluctuation mechanism sets in spontaneously after that a sufficient number of collisions has occurred, and does not emerge from suited initial conditions.
	Differences from the ideal linear response in the growing side of single modes indicate a more complex
behaviour, resulting from the coupling of different wavelengths and the tendency toward a chaotic evolution~\cite{Baldo1995}; 
	A SMF calculation is also presented for the unstable system, where the linear growth of the leading modes is initially comparable to the BLOB approach and deviates at later times.
	This behaviour is due to the different efficiency of the collision term 
in building fluctuations in the two models.  
	Indeed, the saturation regime
is reached earlier in BLOB, because of the more efficient fluctuation source. 

%
%
\begin{figure}[t!]\begin{center}
	\includegraphics[angle=0, width=.7\textwidth]{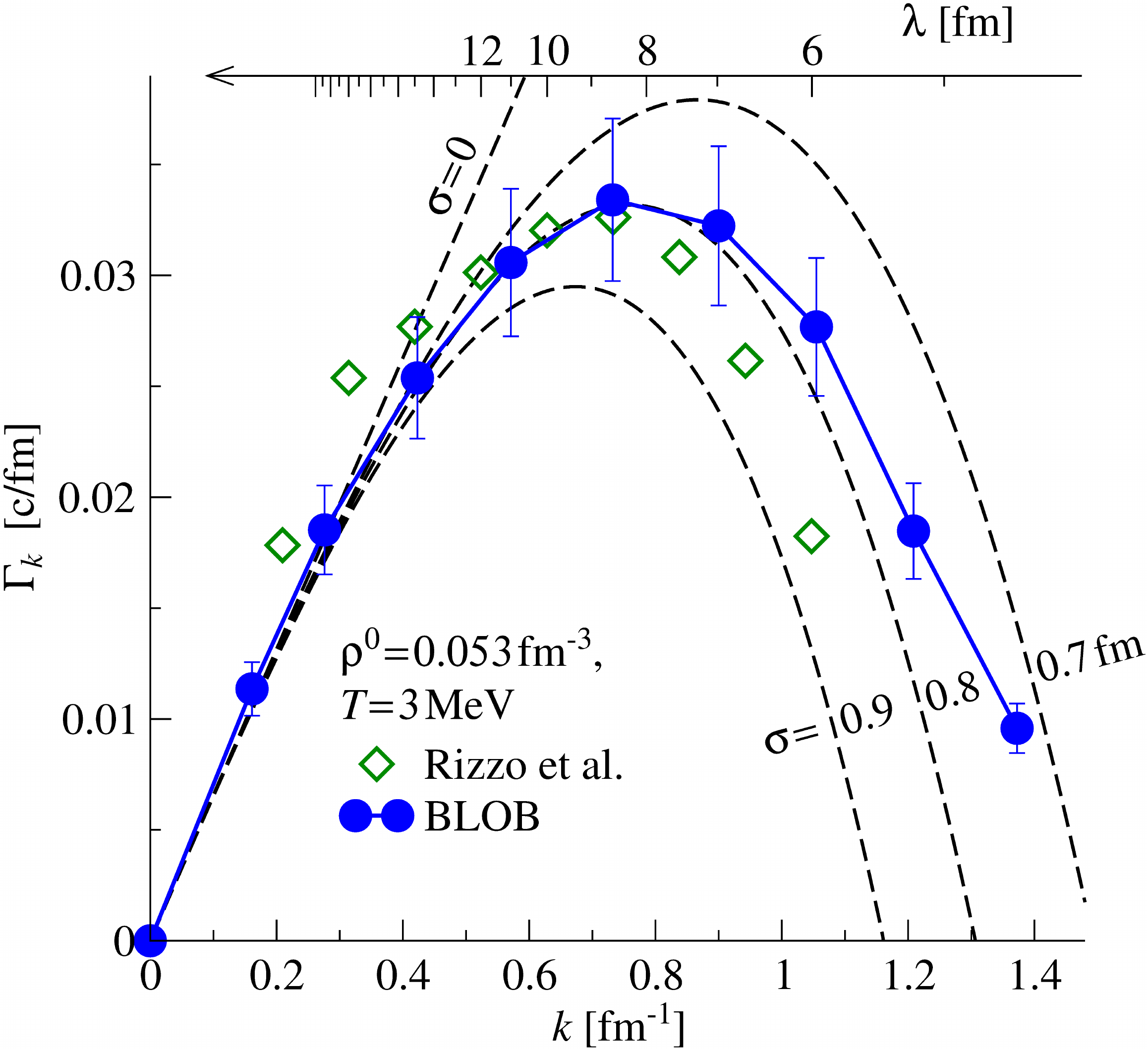}
\end{center}\caption
{
	Dispersion relation: BLOB calculation compared to the analytic relation of Eq.~(\ref{eq:gamma}).
	$k$ and $\Gamma$ values are averaged over modes belonging to discrete intervals of $n$.
	Uncertainties are evaluated from variances around the linear fits to the unstable modes.
	For comparison, a calculation in one-dimension within the approach of ref.~\cite{Rizzo2008} is added.
}
\label{fig_growth_rate}
\end{figure}
	The numerical extraction of the growth rate $\Gamma_k$, equivalent to the analytic relation of Eq.~(\ref{eq:gamma}) is obtained from the time derivative of the amplitude of the isoscalar fluctuation for a given mode $k$ as
\begin{equation}
\Gamma_k = \frac{1}{2}\frac{\partial}{\partial t} {\textrm ln}\prec \tilde{\sigma}^2_k(t) \succ 
	\;,
\label{eq:tau_numerical}
\end{equation}
where the average $\prec \cdot \succ$ is done on several stochastic dynamical trajectories.
	Such analysis is presented in Fig.~\ref{fig_growth_rate}, where the numerical calculation, averaged over 100 events, is compared to the analytic result of Eq.~(\ref{eq:gamma}).
	The range of the interaction, as an effect of the implemented surface term, would correspond to a Gaussian smearing of around $\sigma=0.8$~fm to $\sigma=0.9$~fm.
	We infer that BLOB reproduces consistently the expected dispersion relation within the uncertainties of the linear regression.
	Another calculation, also based on the same mean-field, but which employs the earlier approach of ref.~\cite{Rizzo2008}, also solved in three dimensions but with fluctuations developing along one axis of configuration space, produces a similar result.
	While BLOB keeps the different unstable modes decoupled for a more extended interval of time during their early growth, resulting into a larger ultraviolet cutoff, the other approach presents some alterations due to the combining of unstable modes, where small-wavelength ($n>7$) are gradually absorbed by large-wavelength ($n=0,n=1$).
	The effect in this case is an increase of the grow rate for small $k$ values and it signs the entrance of the chaotic behaviour which characterises larger times~\cite{Jacquot1996}. 
	The largest $k$ modes, corresponding to wavelengths which drop below the Gaussian smearing width $\sigma$ are meaningless.
	The leading modes are found in a wavelength range from 8 to 9fm, and for a growth time $\tau_k$ of around 30fm$/$c. 


\section{From nuclear matter to open systems}
%
%
\begin{figure}[b!]\begin{center}
\includegraphics[angle=0, width=1\textwidth]{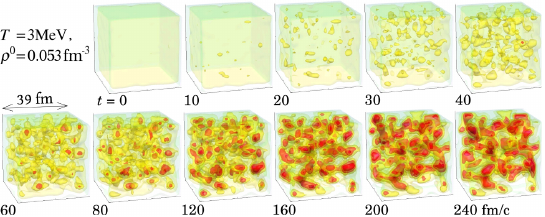}
\end{center}\caption
{
	Evolution of density landscape in configuration space in nuclear matter in a periodic box at $\rho^0\!=\!0.053$~fm$^{-3}$. The simulation employs the BLOB approach as defined in the text.
The arising of spinodal fragmentation occurs in the interval of time ranging from about $t=80$ to about $160$ fm/c.
}
\label{fig_mottling_NM}
\end{figure}
%

%
%
	The subject of this section is studying the effect of isovector and isoscalar fluctuations separately.
	The ultimate purpose of the transport approaches discussed therein is describing the formation of nuclear fragments in a fermionic system and their properties through the combination of these two fluctuating modes, as will be detailed more diffusely in forthcoming works.
	In particular, isovector fluctuations, on top of other isospin transport effects, impose that the isospin content is distributed through a density-dependent process of distillation.
	The onset of isoscalar modes are then responsible for breaking the uniformity of the density landscape and eventually partitioning it into nuclear fragments, where the isospin properties of the initial nesting sites are preserved.
	The isoscalar and isovector mechanisms should therefore be intimately connected in order to describe fragment formation.

%
%
	Qualitatively, we may underline some connection between the wavelengths involved in the dispersion relation analysed in Fig.~\ref{fig_growth_rate}, and fragment formation~\cite{Matera2000}, considering that at the system density $\rho^0$ the leading modes correspond to fragments of mass 
\begin{equation}
	A\approx \rho^0 \lambda^3\;; 
\label{eq.Afromlambda}
\end{equation}
for the leading wavelengths, this corresponds to a distribution of sizes peaked around Neon. 
	These results are also in agreement with other previous studies where quantum effects were taken into considerations explicitly~\cite{Ayik1995,Jacquot1997,Colonna1998bis,Norenberg2000,Ayik2008} despite a more schematic treatment of fluctuations, or of the whole dynamics (2-dimension treatments, fluctuations propagated from an initial state, spherical geometries).
	In this respect, BLOB extends these previous attempts to a model that can be applied at the same time to nuclear matter and, rather successfully, to heavy-ion collisions in three dimensions and without any preliminary initialisation of fluctuation seeds \cite{Napolitani2013,Napolitani2015}.

%
%
	From the growth time of the leading modes in Fig.~\ref{fig_growth_rate}, we infer that the corresponding process of fragment formation would be rather short, progressing from when the system has been largely diluted.
	This suggest that the scenario studied in nuclear matter can be quite directly translated to the phenomenology of open systems~\cite{Chomaz2004}.
	As an example, Fig.~\ref{fig_mottling_NM} illustrates the evolution of a periodic portion of unstable nuclear matter (simulated for $T=3$~MeV and $\rho^0\!=\!0.053$~fm$^{-3}$ for an interaction defined as in Eq.~(\ref{eq:pot}).
	We remark that inhomogeneities emerge in nuclear matter rather early: the process starts at around 20fm/c by exhibiting a spinodal signal (equal-size fragments), which is then smeared out by fragment recombination when exceeding about 150fm/c.
	In the box calculation clusters continue interacting with each other while in the open system they split apart.
	Open systems are the subject of the following chapters.


\chapter{Fragmentation scenarios in spallation reactions
\label{ch_spallation}}

The transport description of mean-field dynamics and phase-space fluctuations in nuclear matter undertaken in the previous chapters can be applied to open nuclear systems.
	A natural transition towards the modelling of heavy-ion collisions is suggested by spallation reactions induced by light high-energy projectiles on heavy targets.
	At variance with dissipative ion-ion collisions, where the excitation of the system is mostly determined by mechanical perturbations, in the spallation process the excitation originates from an almost isotropic propagation of the energy deposited by the light projectile. 

	Like in simulation of periodic portions of nuclear matter, excited spallation systems are also expected to undergo phase-space fluctuations of large amplitude and activate the spinodal behaviour and the amplification of mechanically unstable modes.
	Also, like in nuclear-matter simulations (see \textsection~\ref{ch_inhomogeneities}), the mean field may then have the effect of reverting the whole system, or part of it, into a compact shape, smearing, modifying, or completely erasing the fragment configuration.
	Besides many important effects related to the finite size, to the surface and to the Coulomb field, a major difference in passing from nuclear matter to open systems is that the time evolution should be accurately described: equilibrium descriptions do not necessarily apply to the initial instants of the process and, later on, the system cools down along a decay path.
	It is therefore crucial to correctly describe the time duration and sequence of the various interplaying processes.

	Perhaps the most suggestive finding reported in this chapter is the possibility that binary decays, which are frequent in spallation reactions, are not exclusively assimilated to ordinary asymmetric fission but, when related to fast splits, they often match spinodal-instability scenarios.

\textit{Main sources for this chapter}: 
this chapter puts into context the refs.~\cite{Napolitani2015,Napolitani_IWM2016} and profits from the experimental results of refs.~\cite{Napolitani2011,Napolitani07,Napolitani04,Napolitani2003}.

\section[Need of a dynamical description; experimental digression]{Need of a dynamical description: an experimental digression
\label{sec_expdigression}}

	This section advocates the applicability of a microscopic dynamical model to spallation in the 1$A$GeV range.
	Such application is in fact less intuitive than for ion-ion collisions at Fermi energies and it may require some justification.
Nevertheless, as this section is mostly an historical-experimental review, it may be skipped without discontinuity.

\subsection{Spallation pioneers}

	The traditional Serber's scheme~\cite{Serber1947} for spallation depicts a process of excitation followed by a sequence of compound-nucleus decay steps which can be fully described through a statistical model.
	The underlying assumption of equilibrium relies in this case on the expectation that light high-energy projectiles induce negligible mechanical perturbations when traversing intermediate-mass or heavy elements.
	In this respect, spallation would result in an essentially different process than nucleus-nucleus collisions at Fermi energies, where dissipative effects dominate.
	Yet, in spallation, especially at bombarding energies of around 1$A$GeV, where the whole IMF production has a minor contribution to the total reaction cross section, statistical approaches give an incomplete description of the intermediate-mass fragments (IMFs).
	The distribution of the IMF production attracted much attention since the earliest observations by  Nervik, Serber and Robb Grover~\cite{Nervik1954,RobbGrover1962}: the corresponding yields can in fact fill the gap between the lightest fission-evaporation residues and the light charged particles (indicatively, from lithium to argon), just as in dissipative heavy-ion collisions at Fermi energies.

	On the phenomenological side, as discussed in the following sections, 1$A$GeV proton or deuteron projectiles are sufficient to access density and temperature conditions for collective unstable modes to be amplified: in this case, large-amplitude fluctuations arise, determining the bulk behaviour of the target nucleus, heated up by the interaction with the light projectile.
	The corresponding exit channel should therefore be at the threshold between compound-nucleus decay and multifragmentation, manifesting a rich variety of dynamical trajectories.
	The microscopic stochastic dynamical approach developed in the previous chapters should therefore be suited to such situation, where we also need to drop any a-priori assumption on the degree of equilibration.

	On the experimental side, the possibility that the the multifragmentation mechanism, i.e. the simultaneous disassembly of the system in several IMF, could occur in spallation in the 1$A$GeV-energy regime has been advocated since early works (e.g. ref.~\cite{Hirsch84,Andronenko86,Barz86,Korteling90,Kotov95,Hsi97,Avdeyev98}.
	However, such conclusion was deduced quite indirectly, requiring to import some theoretical concepts in the experimental analysis itself, and opened therefore the way for a variety of alternative physical interpretations, from attributing all IMFs to sequential fission processes, to the opposite extreme that all IMFs signal multifragmentation events.

	Further research focused on the study of thermodynamic observables from spallation reactions in the relativistic domain with light projectiles including antiprotons~\cite{Botvina85b,Lynch87,Botvina90,Karnaukhov99,Karnaukhov03,ISIS}, in connection with the liquid-gas phase transition in nuclear matter~\cite{Binder1984,Chomaz2000,Chomaz2003}, and in parallel with the research on the multifragmentation process observed in ion-ion collisions in the Fermi-energy domain~\cite{Bowman1991,Bondorf95,Borderie2008,Moretto2011,Dagostino2000}.
	At large energies, the multifragmentation scenario and its consistency with the nuclear liquid-gas phase transition picture was well established since early works; on the other hand, the mechanisms in the 1$A$GeV region and their compatibility with a threshold towards multifragmentation has been a source of debate for years.

\subsection{First direct experimental evidence of multifragmentation in spallation in the 1$A$GeV energy range}

%
%
\begin{figure}[t!]\begin{center}
	\includegraphics[angle=0, width=1\textwidth]{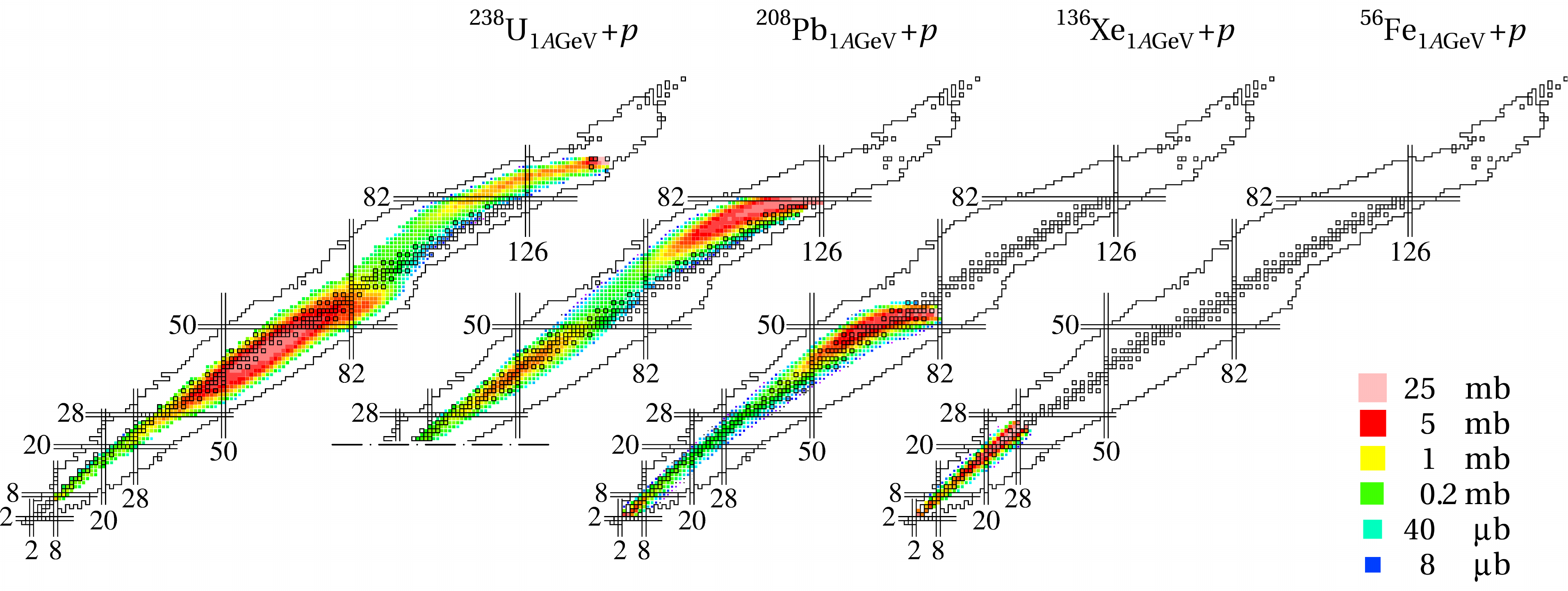}
\end{center}
\caption{
	Survey of isotopic production cross-sections for proton-induced spallation on 
$^{238}$U, $^{208}$Pb, $^{136}$Xe and $^{56}$Fe at 1$A$GeV incident energy, from the experiments of refs.~\cite{Wlazlo2000,Benlliure01,Rejmund01,Enqvist01,Enqvist02,Taieb03,Bernas03,Armbruster04,Bernas06,Napolitani04,Napolitani07}.
}
\label{fig_charts}
\end{figure}

	A deeper insight in the spallation mechanism in the 1$A$GeV-energy regime was achieved in a vast experimental campaign at GSI (Darmstadt) in inverse-kinematics with proton, deuteron or beryllium targets: the FRagment Separator~\cite{Geissel92} (FRS) was employed as an achromatic magnetic spectrometer~\cite{Schmidt87} to measure the nuclide production and the associated kinematic properties with unprecedented high-resolution~\cite{Wlazlo2000,Benlliure01,Rejmund01,Enqvist01,Enqvist02,Taieb03,Bernas03,Armbruster04,Bernas06,Casarejos06,Pereira07,Napolitani04,Napolitani07,Villagrasa07,Benlliure08}.
	Fig.~\ref{fig_charts} presents some production-cross-section surveys obtained in these experiments.
	Particularly interesting is the measurement of the IMF production in systems where binary fission is more or less predominant and presents different mass-asymmetry distributions depending on the distance from the Businaro-Gallone point~\cite{Businaro55a,Businaro55b}.

\begin{figure}[t!]\begin{center}
	\floatbox[{\capbeside\thisfloatsetup{capbesideposition={left,top},capbesidewidth=.35\textwidth}}]{figure}[\FBwidth]
	{\caption{
	Experimental raw zero-angle invariant-velocity distributions for a set of isotopes measured in the beam frame for $^{136}$Xe$+p$ (left) and $^{136}$Xe$+$Ti (right) at 1$A$GeV.
		A combination of two modes, with either a concave or a convex shape, characterise each single isotope.
		The convex shape seems to characterise a more violent mechanism and is in fact predominant in the ion-ion system.
	}
	\label{fig_v_distr_raw}}
	{\includegraphics[angle=0, width=.53\textwidth]{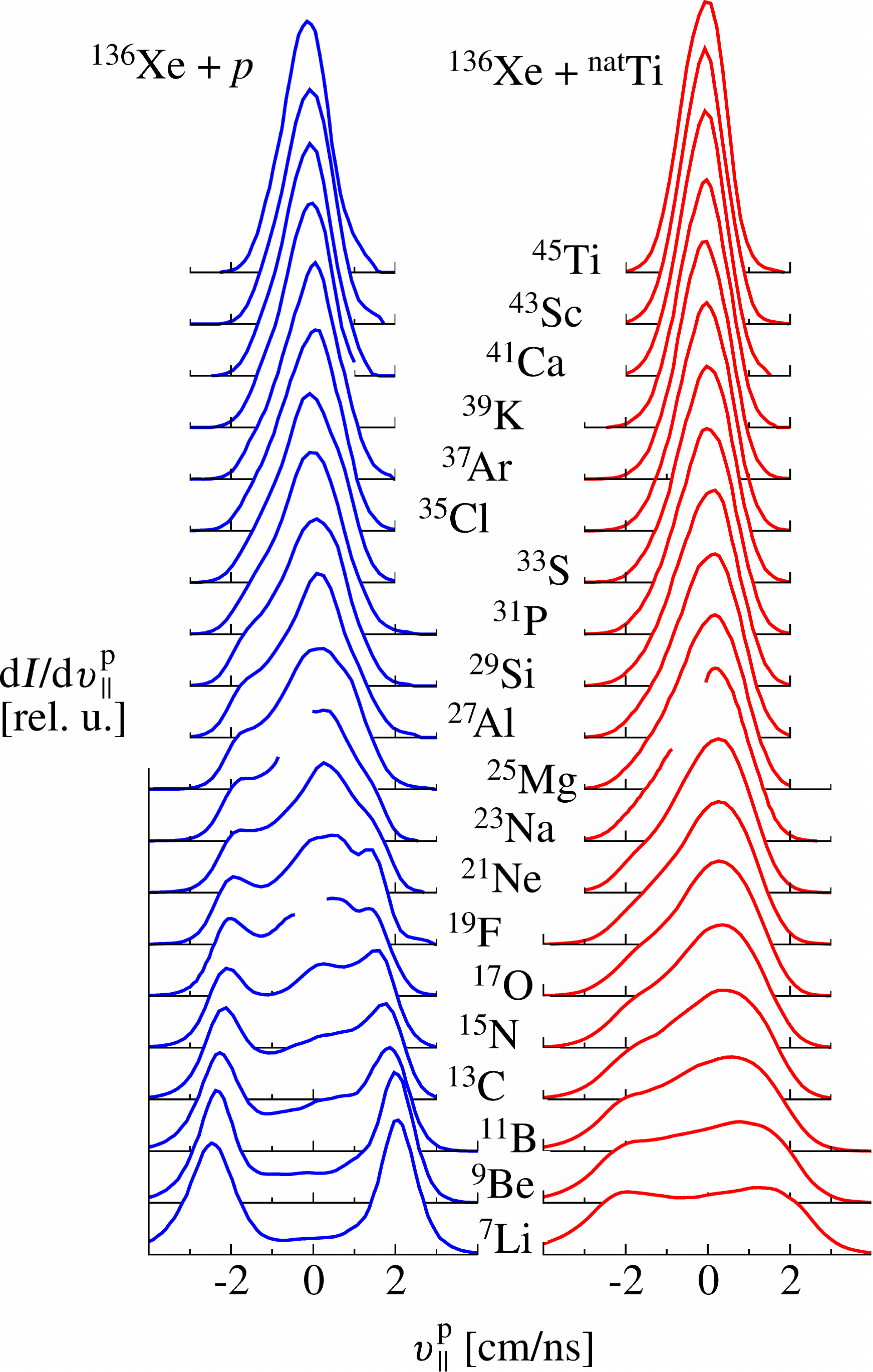}}
\end{center}\vspace{-5ex}
\end{figure}

	From these experiments also emerged the first direct experimental evidence of multifragmentation in spallation in the 1$A$GeV-energy regime.
	Thanks to an extremely high degree of selectivity in the data~\cite{NapolitaniLabtalk2011}, velocity spectra for both multifragmentation and binary splits could be measured simultaneously and distinctly for one given nuclide within the same analysis~\cite{Napolitani04,Napolitani07,Napolitani2011}.
	Fig.~\ref{fig_v_distr_raw} presents some of those experimental results for the spallation systems $^{136}$Xe$+p$ at 1$A$GeV, compared to the relativistic ion-ion collision system $^{136}$Xe$+$Ti at 1$A$GeV.
	The observables are raw velocity spectra for a set of isotopes measured in the beam frame (which is $^{136}$Xe, in inverse kinematics) and mostly along the beam axis, i.e. with very small transverse velocity component due to the geometric acceptance of the spectrometer: they correspond closely to zero-angle invariant-velocity distributions.
	The distributions are given by the combination of two modes, with either a convex or a concave (showing two peaks) shape, with different relative weights for each single isotope.
	Already these raw observables suggest that the convex contribution in the spallation system recalls the much more violent ion-ion system, while the concave contribution seems to be a less violent mechanism characterising predominantly the spallation system.

%
%
%
\begin{figure}[t!]\begin{center}
	\includegraphics[angle=0, width=.8\textwidth]{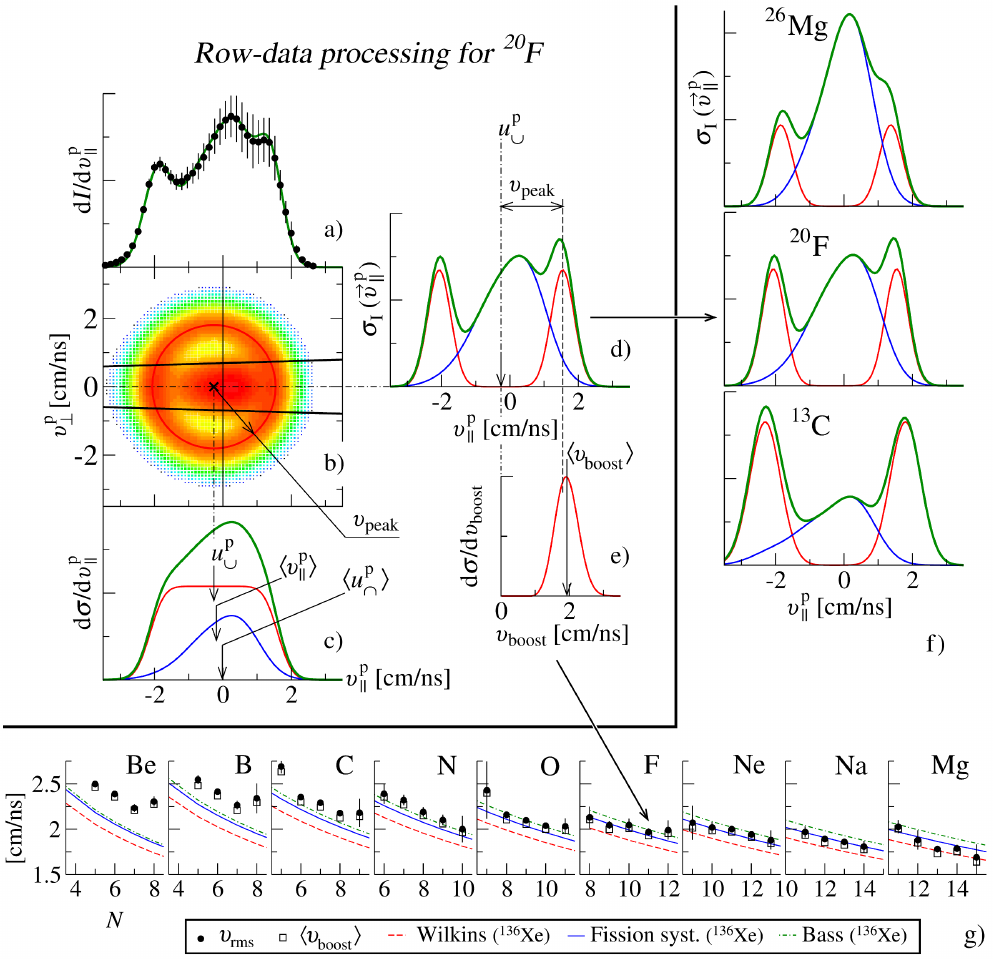}
\end{center}
\caption{
	Extraction of zero-angle invariant cross sections for $^{20}$F (a,b,c,d,e) and some results (f,g). 
(a)	Raw velocity spectrum (black dots, from Fig.~\ref{fig_v_distr_raw}) compared to the 
	reconstruction from the data-processing.
(b)	Planar cut along the beam axis of the reconstructed full distribution 
	$\diff\sigma/\diff\vvec^\beam$ in the projectile ($^{136}$Xe) frame. 
	Fragments were measured within the angular acceptance, indicated by two lines.
	A Coulomb-shell of radius $v_{\mathrm{peak}}$ is associated to the concave mode.
(c)	Reconstructed full distribution $\diff\sigma/\diff\vvec^\beam$
	projected on the beam axis. Arrows indicate the source positions
	of the concave mode $u_{\smallsmile}^\beam$, the average source position of the
	convex mode $\langle u_{\smallfrown}^\beam\rangle$ 
	and the average longitudinal velocity $\langle\vpar^\beam\rangle$.
(d)	Reduction from the two-dimensional cut (b) to a zero-angle invariant-cross-section
	distribution along the beam axis; convex/concave components are indicated.
(e)	Reconstructed cross-section distribution for the 
	concave component as a function of the boost
	velocity in the source frame $v_{\mathrm{boost}}$.
(f)	Same as (d), with additional spectra for $^{13}$C and $^{26}$Mg.
	Convex and concave components exhibit evolving relative shares and a recoil mismatch.
(g)	Mean boost of the concave mode as a function of the neutron number,
	compared to three prescriptions, see text.
}
\label{fig_v_distr}
\end{figure}

	Further analysis steps~\cite{Napolitani2011}, resumed in Fig.~\ref{fig_v_distr}, allowed to precisely select the velocity vectors aligned along the beam direction (their distribution is different from longitudinal projections of the whole velocity distribution only), separate precisely the concave and convex components in the zero-angle invariant-velocity distributions, and measure their relative shares as a function of the nuclide.	
	Though, the presence of two different contributions is evident only in the kinematics (possibly along the beam axis, where also the recoil is measured), while the nuclide production selected for either the concave or the convex mode contribute to the same region of the nuclide chart, in general situated in the neutron-rich side with respect to $\beta$ stability: this is the reason why the presence of these two modes was invisible in many experiments.
Moreover, the two contributions are associated with a shift in the mean value of the spectra, indicating incompatible values for the mean momentum transfer: this latter reveals in fact the violence of the entrance channel.

%
%
	The concave shape is reflected in a Coulomb-shell velocity distribution probed at zero angle.
	The mean value evolves coherently with empirical systematics for the mean recoil momentum as a function of the fissioning system (i.e. compatible with the systematics of Morrissay~\cite{Morrissey89}).
	The corresponding boost $\langle v_{\mathrm{boost}}\rangle$~\footnote{
The velocity $v_{\mathrm{rms}} = \sqrt{\langle v_{\mathrm{boost}}\rangle^2 + \sigma^2_v}$,
where $\sigma^2_v$ is the variance of the velocity distribution in the representation (e) of Fig.~\ref{fig_v_distr},
corresponds to the mean value of the total kinetic energy distribution and is slightly larger than the radius of the Coulomb shell, equal to the mean value of the velocity distribution $\langle v_{\mathrm{boost}}\rangle$.
}, is globally compatible with the empirical systematics of total kinetic energy for asymmetric binary splits condensed in the parametrisation of Beck and Szanto de Toledo~\cite{Beck96}, and further adapted to describe asymmetric splits by using the prescription of ref.~\cite{Napolitani04}.
	For the lightest IMF, $\langle v_{\mathrm{boost}}\rangle$ is compatible and may even exceed the boost which the light fragment and its heavy partner would explore in a fusion reaction, according to the empirical model of Bass~\cite{Bass79, Bass80}: within this description, the Coulomb repulsion acts on non-deformed fission fragments and results in the largest possible boost in absence of additional contributions. 
	$\langle v_{\mathrm{boost}}\rangle$ is on the other hand larger that expected from the scission-point model of Wilkins et al.~\cite{Wilkins76, Bockstiegel97}, which deduces the total kinetic energy from an empirical liquid-drop description of a deformed system in presence of a neck.

	The convex shape is one broad hump, often asymmetric, which signs 
the folding of many different contributions; it is associated with a mean recoil momentum which does not follow any empirical systematics.
	This indicates that the fissioning configuration is not achieved, either because the multiplicity of fragments is not equal to two, or because the kinematics is not consistent with a conventional fission configuration.
	In this respect, the convex shape is rather compatible with multifragmentation.

	Such observations led to the conclusion that the IMF production should combine asymmetric-fission (concave shapes) and multifragmentation (convex shapes).
	On the one hand, this description was rigorously established because these two velocity contributions were measured at the same time for each single nuclide~\cite{Napolitani2011}.
	On the other hand, the problem of assuring such interpretation was that particle-particle correlations and IMF multiplicities were not measured in the inclusive approach.
	Especially for the convex shape this information is important as multifragmentation is usually associated with a large number of IMF of similar size.


	To characterise the process, in addition to the high-resolution zero-angle velocity spectra, an ideal experimental approach should measure event-by-event particle-particle and kinematic correlation observables.
	A second experimental technique was adapted to add this information on exclusive (i.e. event-by-event) correlations.
	Some of the systems previously measured inclusively at the FRagment Separator were successively measured again with an exclusive apparatus in the Spalladin experimental campaign~\cite{LeGentil08,Gorbinet2011,Gorbinet2012} at GSI (Darmstadt).
	These experiments indicated that IMF are observed in events where fragment multiplicity is prevalently equal to two, and that events with larger multiplicity were more rare.
	On the one hand, this confirmed that there are actually two contributions to the IMF production, a binary channel and a higher multiplicity channel. 
	On the other hand, the events exceeding two IMF were not easily identifiable with ordinary multifragmentation because of the low IMF multiplicity and the size asymmetry.
	This encouraged interpretations fully relying on statistical models, where the IMF production is obtained either from a sequence of (asymmetric) fission contributions, or sampled from an ensemble of possible multifragmented configurations.
	In general, these approaches are both an efficient workaround because they define directly the outcome of the reaction on the basis of the involved excitation energy, and they can yield quite similar results for the IMF production despite implying different physical pictures.
%
	Then the question arises whether we can include spallation in the multifragmentation picture.

	The experimental situation discussed above is a specific example implying that the same considerations could follow from other experimental analyses.
	It is our intention to illustrate that the key to achieve a fully coherent understanding of this apparently self-contradictory experimental information is a microscopic dynamical description of the process, conciliating the possibility to emit only few IMFs with the observation of new kinematic properties.

%
\section[Boltzmann-Langevin One Body description of spallation]{Boltzmann-Langevin One Body description of a spallation system}
%

%

	We recalled that spallation is a favoured processes to produce a thermalised remnant and that therefore statistical models, proceeding from the usual assumption of a hot and fully equilibrated source, are well suited to reproduce large part of the experimental observables~\cite{Napolitani04}.
	On the other hand, it is not the purpose of these approaches to investigate the reaction mechanism which leads to the final configuration or to describe threshold effects or out-of-equilibrium mechanisms as those possibly related to the onset of mechanical instabilities and IMF production.

	To directly address this issue, stochastic one-body approaches are well suited to sample the variety of possible dynamical trajectories that unstable conditions may produce, and they avoid a priori assumptions on the degree of equilibration of the system at a given reaction time.
	With the purpose of investigating fragment formation in spallation, early attempts~\cite{Colonna1997bis} which did not treat fluctuations in full phase-space yielded incomplete success.
	We proceed therefore through the BLOB approach as described in \textsection~\ref{ch_inhomogeneities}, in order to handle fluctuations in full phase space, and to profit from an efficient description of the dispersion relation as already tested in nuclear matter.

	The initial-state configuration of the target heavy nucleus is defined by distributing the test-particles in configuration space, inside of a sphere $r_0$, and in momentum space inside of a corresponding local Fermi sphere, with density-dependent Fermi momentum.
	$r_0$ is obtained from a procedure of minimisation of the total energy.
	
	Both for handling the interaction with the projectile and for describing the dynamics, two-body collisions are introduced.
	Calculations presented in this chapter employ a nucleon-nucleon cross section $\sigma_{\textrm{NN}}$ equal to the free nucleon-nucleon cross section, with a cutoff at $100$~mb.	
	Moreover, the differential cross section depends on the scattering angle according to the prescription of ref.~\cite{Bertsch1988}.

\subsection{Definition of the heated system}

	After a suitable initialisation, only the dynamics of the heated heavy nucleus is followed.
	As usual, the system is initially defined by organising the test particles in a minimum-energy configuration in accordance with the form chosen for the nuclear interaction.
	In order to define the heated system, this configuration is redefined by processing a simplified cascade induced by the incoming light projectile: the amount of energy deposited by the projectile in traversing the nucleus is calculated as well as the corresponding distribution in phase space.
	A time-dependent calculation would require very small time steps and a relativistic formalisation of the dynamics.
	Due to the rapidity of the spallation process with respect to the dynamics of the heated system, it is convenient to reduce the cascade to an approximate description where only test-particles from the incoming relativistic projectile are followed along space trajectories and target test-particles are not displaced during the cascade.
	In practice, this simplification is made by reducing the cascade process to a calculation of the energy loss of the projectile, modifying the momentum landscape of the target test-particles without processing any time evolution of the system in coordinate space.
	For relativistic projectiles this choice is not incompatible with the observation that the projectile leaves the target nucleus before that the swarm of the first fastest ejectiles appears at the surface of the target nucleus~\cite{Cugnon1981,Cugnon87}.

%
%
\begin{figure}[b!]\begin{center}
	\includegraphics[angle=0, width=.5\textwidth]{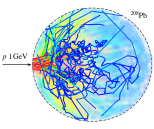}
\end{center}
\caption{
	Energy-deposition map with one bunch of spacial cascade trajectories in relief, calculated for a central impact parameters in the reaction $p+^{208}$Pb at 1$A$GeV.
	Each trajectory is associated to one test particle of the incoming projectile.
}
\label{fig_cascade}
\end{figure}
	Fig.~\ref{fig_cascade} shows a bunch of spacial cascade paths corresponding to a $^{208}$Pb target nucleus bombarded by 1 GeV proton projectiles with a central impact parameter;
the resulting excitation energy distribution corresponds therefore to the most violent events.
	The test particles composing the projectile hit the nucleus along the projectile direction within a cylinder of radius equal to the projectile radius.
	Each projectile test particle produces a cascade path inside of the target nucleus, which is redefined at each scattering occurrence:  after scattering, the projectile test-particle trajectory continues to be followed along the fastest scattered particle and the other particle, after being assigned a new momentum, is no more followed.
	Between two scattering points the path is a straight segment.
	All cascade paths traced by the projectile test particles are followed in coordinate space simultaneously. 
	For a couple of target and projectile test particles, the collision is searched according to the closest-approach criterion applied to the corresponding centre-of-mass energy $\sqrt{s}$~\cite{Bertsch1988} and by using the same nucleon-nucleon cross section used for the transport calculation.
	All collisions are considered as elastic scatterings;  
the model could be improved by including the $\Delta$ production-absorption mechanism, but we consider the present simplified treatment sufficient for the purpose of obtaining the excitation energy of the target nucleus.
	A strict Pauli-blocking condition here is imposed, so that only scattering events which create a hole and a particle outside of the Fermi sphere are accepted; otherwise, the target test particle could participate to a scattering with another projectile test-particle.
	When the cascade trajectories hit the inner potential boundary of the system, they can traverse the boundary according to the corresponding transmission probability, calculated with the relativistic formalisation proposed in ref.~\cite{Cugnon1997}; the potential depth used for calculating the transmission probability is 40MeV, which represents the average value characterising the bulk of the system.
	This transmission probability is used to calculate an additional portion of the total energy of the projectile, which is considered dissipated in the target system and which corresponds to the reflected wave.

%
%
\begin{figure}[b!]\begin{center}
	\includegraphics[angle=0, width=.85\textwidth]{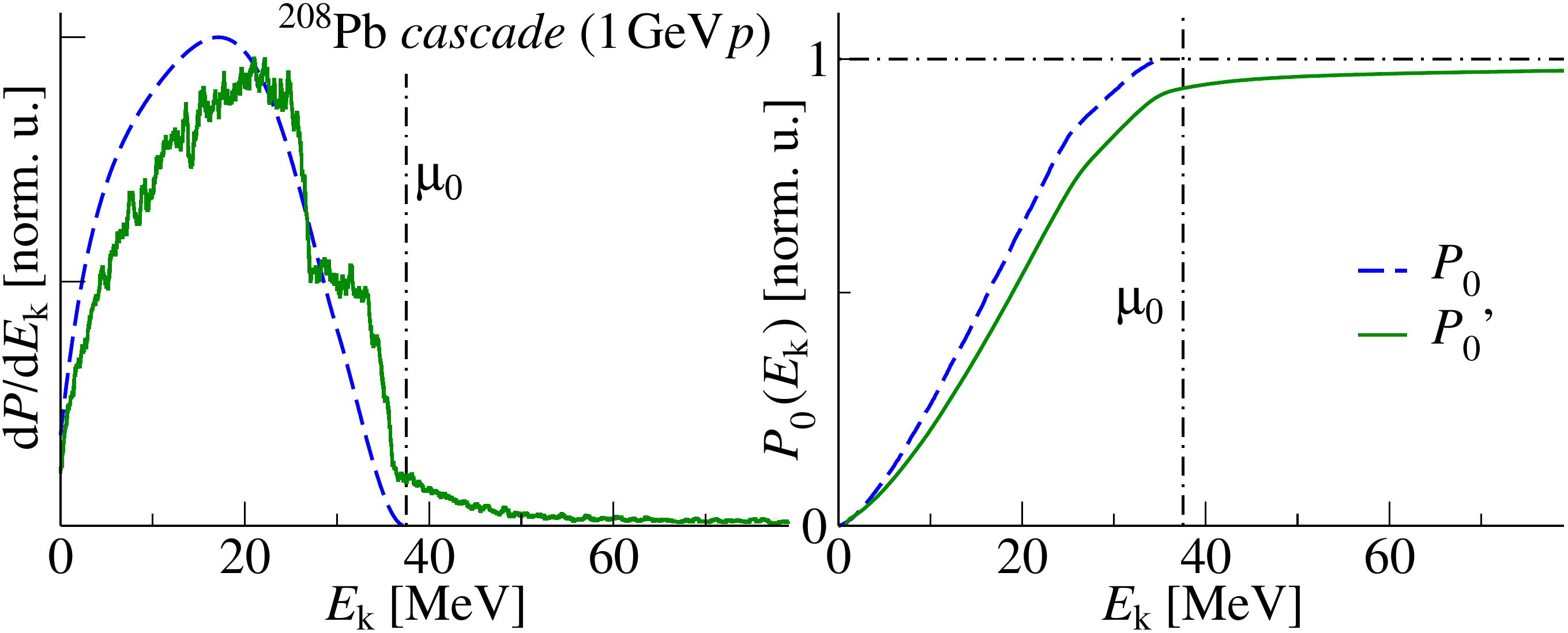}\\
\end{center}
\caption{
	Effect of the cascade in modifying the momentum space from the initial state configuration (dashed line)
to the excited configuration (full line), calculated for one event in the reaction $p+^{208}$Pb at 1 $A$GeV 
(see text).
	The panel on the left represents the nucleon energy distribution, whereas on the right the integrated energy distribution is presented.
}
\label{fig_finalstate}
\end{figure}
%

\subsection{Dynamical evolution
\label{BLOB_opensystems}}
%
%
\begin{figure}[t!]\begin{center}
	\floatbox[{\capbeside\thisfloatsetup{capbesideposition={left,top},capbesidewidth=.35\textwidth}}]{figure}[\FBwidth]
	{\caption{
	Upper panel. Evolution of the mean number of emitted nucleons per interval of time in the reactions $p+^{208}$Pb at 1 GeV.
	Lower panel. Evolution of the mean excitation energy calculated for the fraction of bound matter in the reaction $p+^{208}$Pb.  
	Central, intermediate and peripheral impact parameters are tested.
	The width of the bands give the standard deviation around the trajectories for the central collisions; other trajectories have a comparable standard deviation (not indicated).
	}
	\label{fig_collcasc}}
	{\includegraphics[angle=0, width=.6\textwidth]{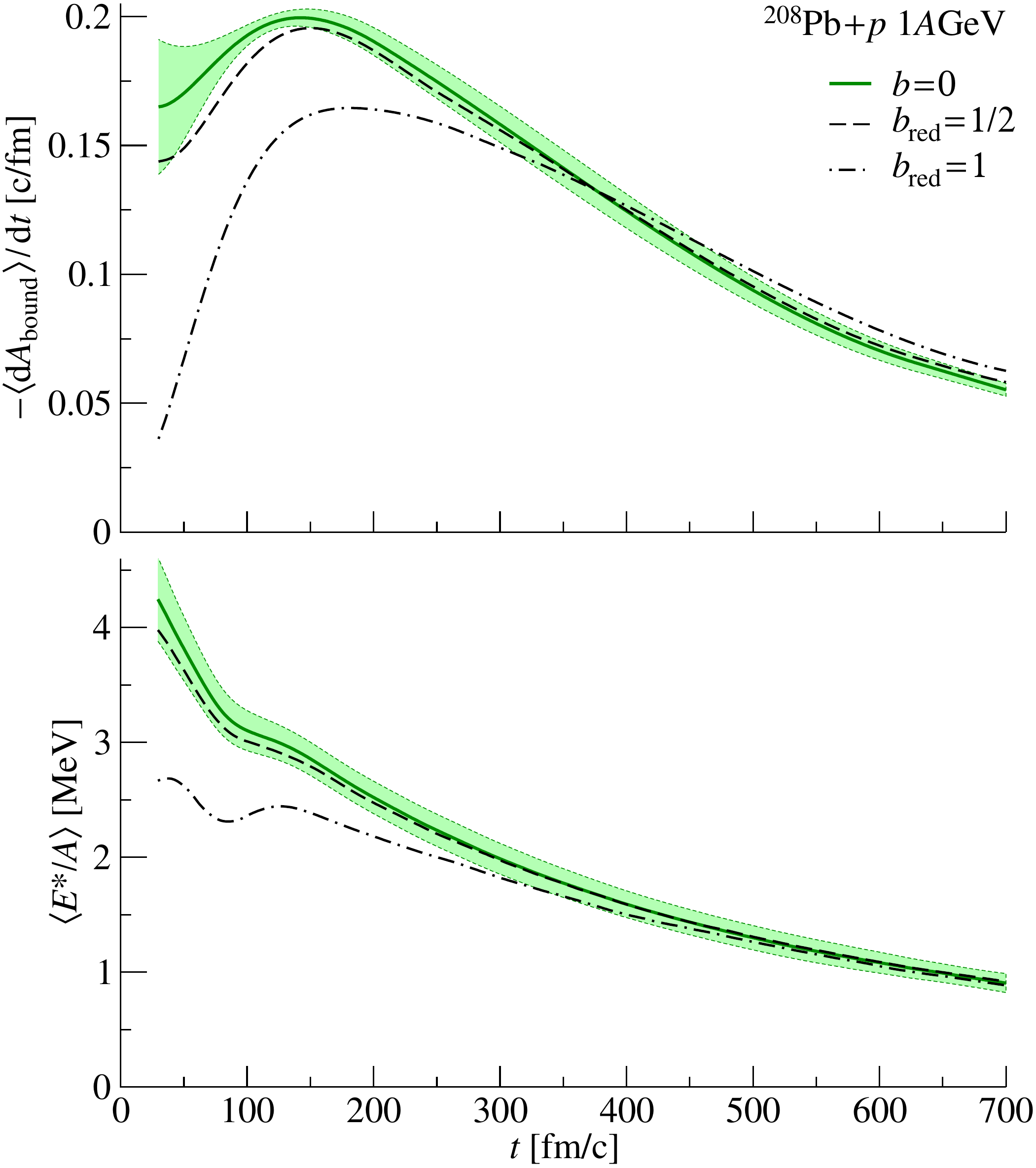}}
\end{center}
\end{figure}
	While the coordinate space is frozen to its initial configuration, the initial momentum distribution is updated according to the cascade scatterings.
	The energy deposited by the projectile in the system is then obtained by considering the momentum variation, supplemented by the reflection contribution at the potential boundary of the system.	
	Accordingly, as shown in Fig.~\ref{fig_finalstate} for the reaction $p+^{208}$Pb at 1 GeV, the initial integrated kinetic energy distribution $P_0(\Ek)$ is modified into a new distribution $P_0'(\Ek)$.

	The cooling process of the excited system is then followed in time with BLOB. 
The latter is shown in Fig.~\ref{fig_collcasc}, setting the reduced impact parameter $b_{\textrm{red}}$ (impact parameter divided by the target radius) equal to 0, 0.5 and 1 for the system $p+^{208}$Pb at 1 GeV.
	The evolution of the mean fraction of bound matter $\langle \diff A_{\textrm{bound}}\rangle / \diff t$ tracks the mean number of emitted nucleons per interval of time: central and intermediate impact parameters act almost equally in removing a large part of nucleons, while peripheral collisions favour the formation of heavier remnants.
	The corresponding information is carried by the evolution of the mean excitation energy per nucleon $\langle E^*/A\rangle$ averaged over all portions of bound matter in the system.

	Nuclear fragments are identified through a coalescence 
	algorithm in phase-space which defines the corresponding mass and charge 
	content. The fragment charge Z and mass A are approximated to integer numbers 
	under the constraint of mass, charge, momentum and energy conservation.


%
\section{Nuclide production and kinematics}
%
	The model described above was applied to six systems, chosen because close to some significant experimental data and because they constitute a series of successive variations of only one parameter among projectile, target and energy:
$^{208}$Pb+$p$ at 1 $A$ GeV,
$^{208}$Pb+$p$ at 750 $A$ MeV,
$^{208}$Pb+$d$ at 750 $A$ MeV,
$^{197}$Au+$d$ at 750 $A$ MeV,
$^{136}$Xe+$p$ at 1 $A$ GeV and
$^{124}$Xe+$p$ at 1 $A$ GeV.
%
%
\begin{figure}[b!]\begin{center}
	\includegraphics[angle=0, width=1\textwidth]{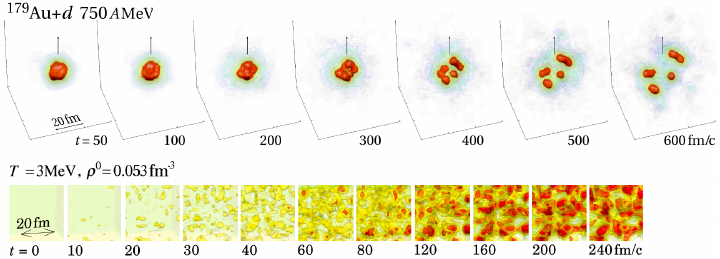}
\end{center}
\caption{
	Top. Study of the fragment configuration for one event of the system $^{197}$Au+$d$ at 750 $A$ MeV.
The event shown in the figure is selected among those giving the largest fragment multiplicity.
The arrow indicates the relative direction of the deuteron.
	After a interval of time corresponding to the expansion of the system and the developing of a mottling pattern in the density distribution, the fragmentation pattern stands out at around 300 fm/C.
	Bottom. For comparison, a study of isoscalar-fluctuation growth is shown (calculation discussed in \textsection~\ref{ch_inhomogeneities}), presenting a mottling pattern 
which can be compared to the spallation system in timing and size of the inhomogeneities.
}
\label{fig_animation_exit}
\end{figure}
	The dynamical calculations were performed reducing to central impact parameters in the range $0<\!b\!<0.75$fm, with the purpose of restricting to the small portion of geometric cross section where the contribution of heavy residues is not dominant, and where IMF formation is enhanced. 
	 The remaining fraction of cross section favouring compound-nucleus decays can be efficiently described through statistical approaches.
	Such a choice is however schematic because, due to fluctuations in the cascade trajectories, the impact parameter is not directly characterising the entrance channel, and violent collisions may arise also in less central configurations with smaller probability.
	Conversely, less excited configurations are also associated to central impact parameters with smaller proportion than in peripheral collisions.
	A statistics of about 1500 
stochastic events per system have been collected, using a 32 CPU parallel computing station.

\subsection{Dynamical description up to the formation of primary fragments}

%
%
\begin{figure}[t!]\begin{center}
	\floatbox[{\capbeside\thisfloatsetup{capbesideposition={left,top},capbesidewidth=.35\textwidth}}]{figure}[\FBwidth]
	{\caption{
	Upper panel. As in Fig.~\ref{fig_collcasc}: evolution of the mean number of emitted nucleons per interval of time for the spallation reactions described in the text.
Middle panel. Probability of split as a function of time for  $p+^{208}$Pb and $p+^{136}$Xe at 1GeV.
Lower panel. Saturation of the fragment multiplicity $M_{\textrm{frag}}$ for $p+^{208}$Pb at 1GeV; Mean and standard deviation are shown for the multiplicity of fragment with $Z\!>\!4$.
	}
	\label{fig_mass_loss}}
	{\includegraphics[angle=0, width=.6\textwidth]{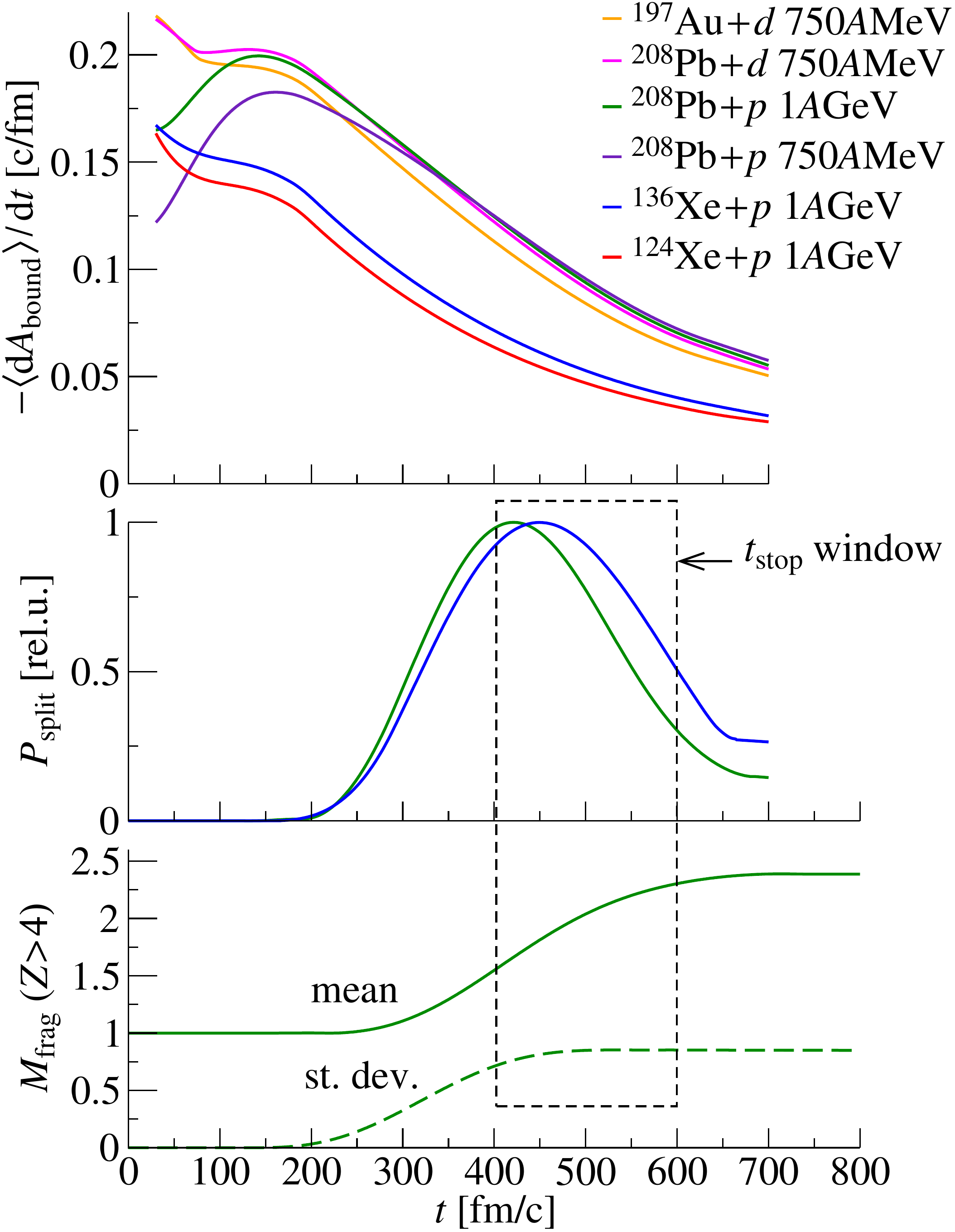}}
\end{center}
\end{figure}

	Within the model described above, in the upper sequence of Fig.~\ref{fig_animation_exit} we represent one possible evolution of the density profile of the systems $^{197}$Au+$p$ at 750 $A$ MeV for a central impact parameter; this is a rather rare event corresponding to the fragmentation of the target nucleus in more than three fragments.
	At early times, the system develops inhomogeneities; both in shape and chronology, they recall the mottling pattern obtained in unstable nuclear matter in the study of \textsection~\ref{ch_inhomogeneities}, reported in the lower sequence of Fig.~\ref{fig_animation_exit}.
	The system breaks into three asymmetric parts visible at 400~fm/c.
At later times, further splits may proceed from some individual largely deformed sources, as displayed in Fig.~\ref{fig_animation_exit} for the time 500~fm/c.
	In these spallation processes the fragment multiplicity saturates after 700~fm/c.
	This is shown in Fig.~\ref{fig_mass_loss} (bottom), in correlation with the particle emission and the corresponding reduction of bound mass as a function of time for all the simulated systems (top).
The middle panel of the figure shows the probability of observing a split in the system, as a function of time, for two
of the reactions considered. 
 
\begin{figure}[t!]\begin{center}
	\includegraphics[angle=0, width=.7\textwidth]{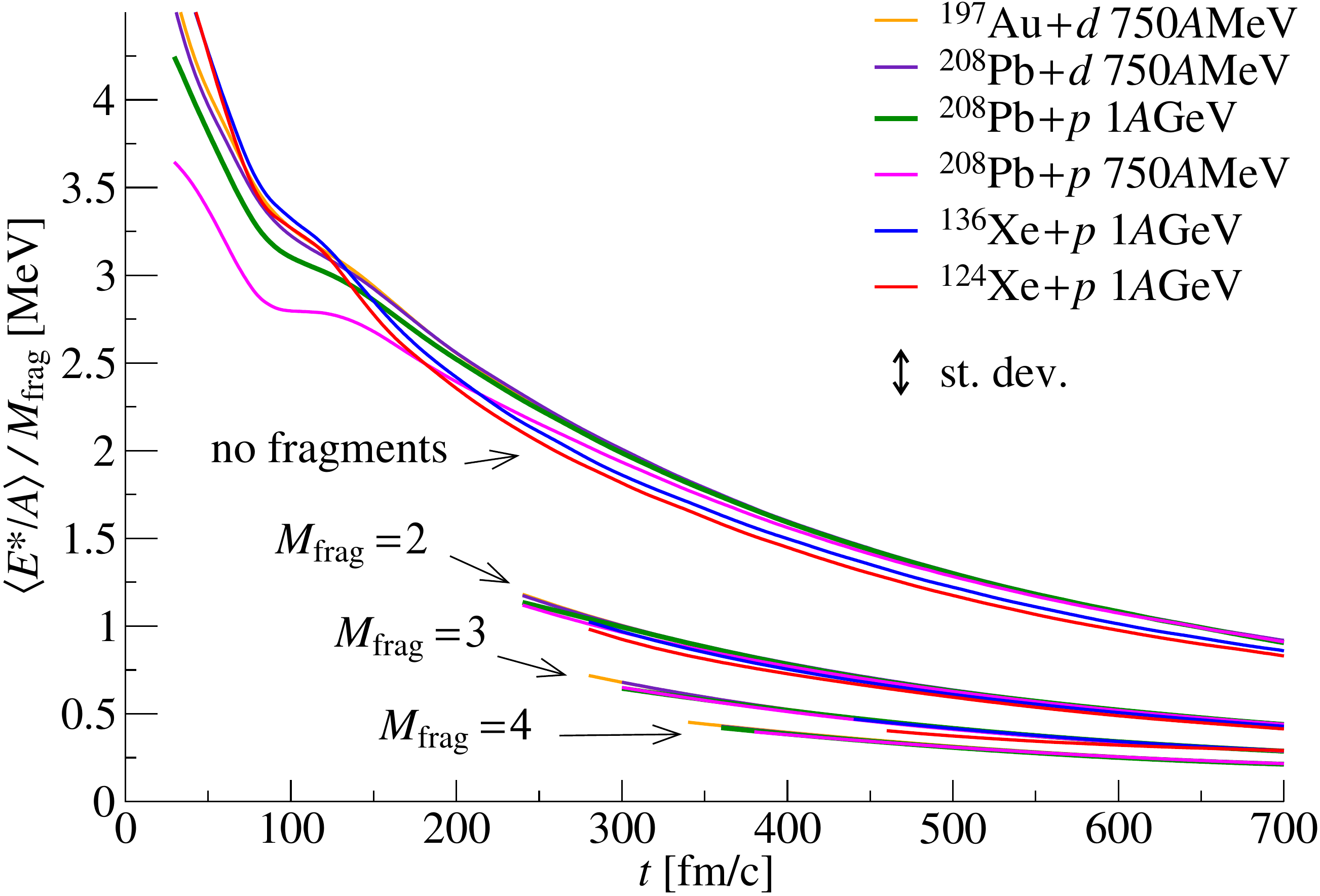}
\end{center}
\caption{
	Evolution of the average excitation energy $\langle E^*/A\rangle$ for the fraction of bound matter during the reaction. The double arrow gives the average uncertainty in terms of standard deviation.
	The bunch of lines extending over the whole time evolution describes events where only a heavy residue is present; bunches of lines for systems which split into $M_{\textrm{frag}}$ equal to two, three or four fragments are divided by $M_{\textrm{frag}}$ for better visibility as they would all collapse on the line for $M_{\textrm{frag}}$=1.
}
\label{fig_energy_time}
\end{figure}

	The cooling process is reflected in the decrease of the average thermal excitation energy per nucleon shown in Fig.~\ref{fig_energy_time}.
	A backbending appearing between around 50 and 100 fm/c indicates the attempt of the system to revert the initial pure expansion dynamics into the mechanism of fragment formation.  Indeed, in presence of instabilities, it is energetically convenient for the system to break up into fragments.   
	This also causes a slight increase of the temperature and thus of the thermal excitation energy.
	Event by event, we consider as freeze-out time, $t_{\textrm{stop}}$, the instant between $t\!=\!400$~fm/c and $t\!=\!600$~fm/c where the last split has occurred. Our choice is motivated by the fact that at $t\!\approx\!400$~fm/c the split 
probability is maximum, whereas at $t\!\approx\!600$~fm/c it reduces to a quite low constant value.
For events where a residue is observed, we adopt $t_{\textrm{stop}}\!=\!400$~fm/c. 
	Beyond the time $t_{\textrm{stop}}$ the decay process slows down and only sequential binary splits become possible, which can be efficiently described through a transition-state model. 
	The dynamical calculation is therefore completed with the model SIMON~\cite{Durand1992}, 
which incorporates in-flight Coulomb repulsion.
	
%
%
%

\subsection{On the way to the residue corridor and generalisation to peripheral relativistic ion-ion collisions}

	In correlation with the excitation energy, also the isospin content of fragments and residues evolves in time.
	In general, when a compound nucleus is formed, its excitation energy is extinguished in an attempt of balancing proton and neutron decay widths, so that the bound matter of the systems tends to accumulate along the residue corridor~\cite{Charity98}, which is located in the neutron-deficient side of the nuclide chart with respect to beta stability, and any further decay occurs only along this line in average.
	If however part of the excitation energy is spent in fragmenting the system, neutron rich fragments stop their decay path before reaching the residue corridor, in locations of the nuclide chart which are closer to beta stability, or that are even neutron rich~\cite{Napolitani2011}. 

%
%
\begin{figure}[b!]\begin{center}\vspace{-2ex}
	\includegraphics[angle=0, width=.71\textwidth]{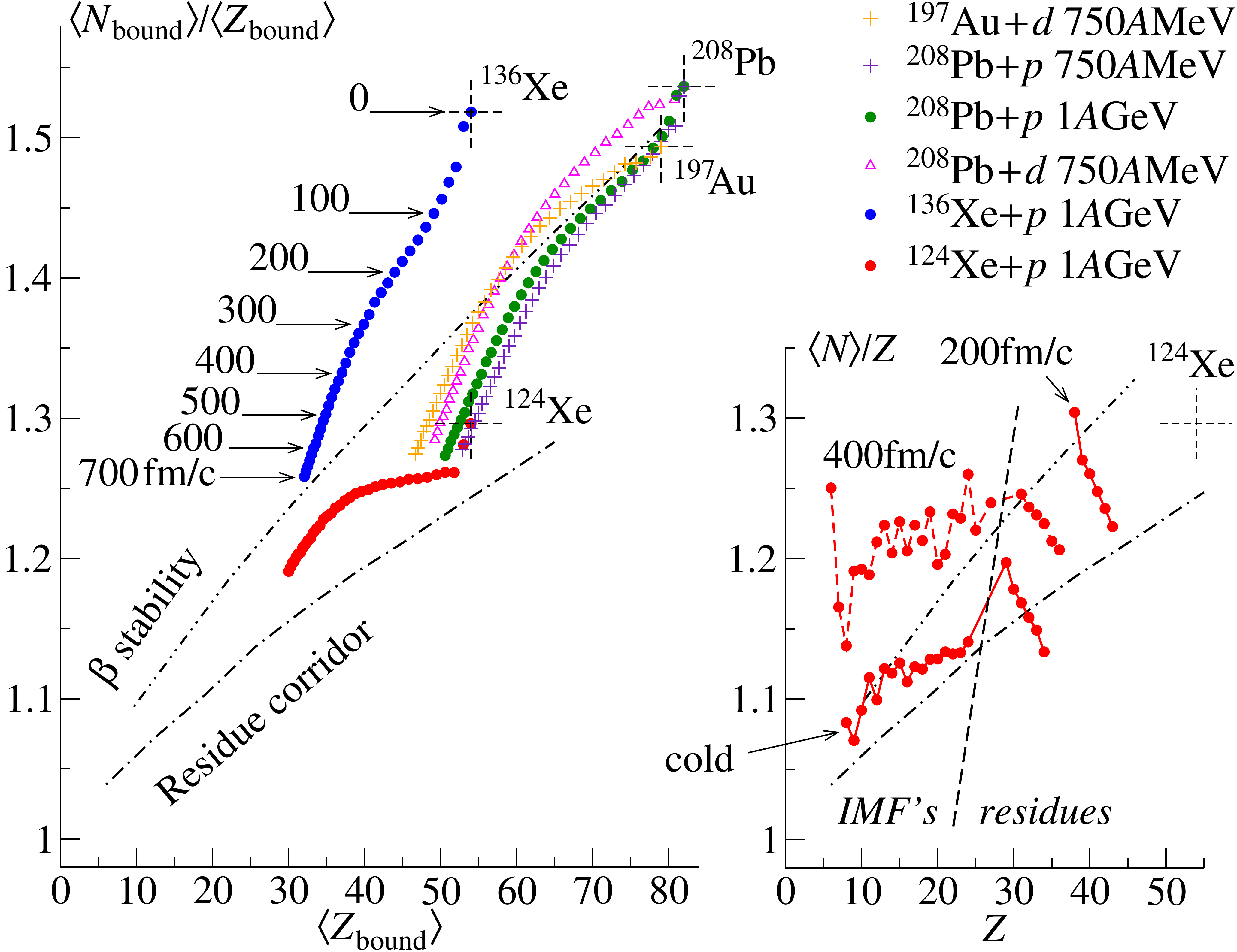}\\
	\includegraphics[angle=0, width=.71\textwidth]{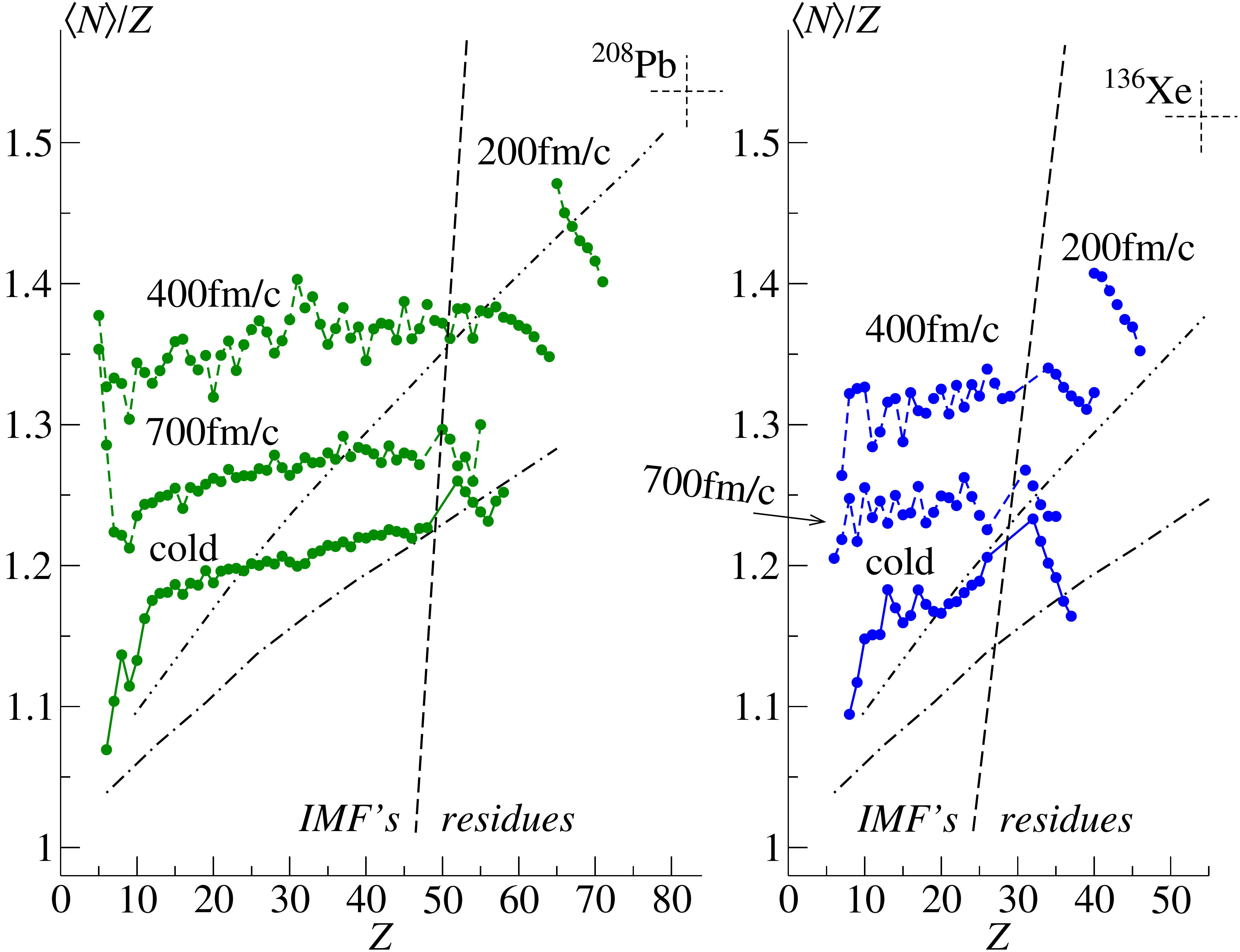}\\\vspace{-2ex}
\end{center}
\caption{
	Top left. Evolution of the average isotopic content of bound matter constituting different hot systems as a function of time (calculated for central impact parameters).
	Top right, bottom right, bottom left:
	Distributions of the average isotopic content of single elements produced in the systems 
$^{124}$Xe, $^{136}$Xe and $^{208}$Pb (moving clockwise) bombarded by 1 GeV protons as a function of the element number at 200fm/c (before fragmentation), at 400fm/c (latgest fragmentation probability), at 700fm/c (when the fragment multiplicity saturates in the dynamical calculations) and after secondary-decay progressing from $t_{\textrm{stop}}$.
	The $\beta$-stability and the residue corridor are indicated (see text). 
Residues and IMF regions are also indicated.
}
\label{fig_NoverZ}
\end{figure}
	This process inspired several experiments and simulations with statistical models where an assumption of thermal equilibrium of the system was imposed and a temperature was assigned~\cite{Schmidt2002} (the so-called `limiting temperature for fragmentation', corresponding to about 5~MeV).
	The dynamical approach handles this process without any hypothesis of equilibrium.
	Fig.~\ref{fig_NoverZ} (top left panel) examines the evolution of the isotopic content for the six different hot systems for central impact parameters: the average isotopic content of bound matter, obtained by dividing the average number of bound neutrons constituting the system $N_{\textrm{bound}}$ by the average bound charge $Z_{\textrm{bound}}$ is tracked as a function of time until 700~fm/c. 	
	In this interval of time the path moves in average in the direction of the residue corridor while removing mass.
	The whole distribution of the isospin content $\langle N\rangle/Z$ of hot fragments is given in Fig.~\ref{fig_NoverZ} for $^{208}$Pb+$p$, for the neutron-deficient system $^{124}$Xe+$p$ and for the neutron-rich systems $^{136}$Xe+$p$ 
(bottom left, top right and bottom right panels); the following times are analysed: $t\!=\!200$~fm/c, before fragmentation, $t\!=\!400$~fm/c, after fragmentation, and $t\!=\!700$~fm/c, when the fragment multiplicity saturates.
	In all the three systems, the distribution at 400~fm/c covers the region of neutron-rich nuclei as a flat function of the element number and its distance from the residue corridor depends on the isospin content of the target nucleus; it drops to smaller values of $\langle N\rangle/Z$ for later times.
	As a function of the available excitation (i.e. of the time), the corresponding distribution of cold fragments is found displaced in the direction of the residue corridor.
	The cold IMFs align along the residue corridor only for the neutron poor target $^{124}$Xe; they do not reach it completely for the neutron rich target $^{136}$Xe, ending their decay path in the vicinity of $\beta$-stability; in the case of the heavy neutron rich target $^{208}$Pb, only the largest fragments can reach the residue corridor at the end of their (shorter) decay path. 
	A complete study of this reaction, involving also less violent events for peripheral impact parameters would extend the distribution of residues to larger mass numbers which would accumulate along the residue corridor.
	The same behaviour characterises also the other heavy systems (not shown) and it recalls closely the experimental results for peripheral relativistic heavy-ion collisions~\cite{Schmidt2002,Henzlova2008}.

%
%
\begin{figure}[b!]
\begin{center}
	\includegraphics[angle=0, width=1\textwidth]{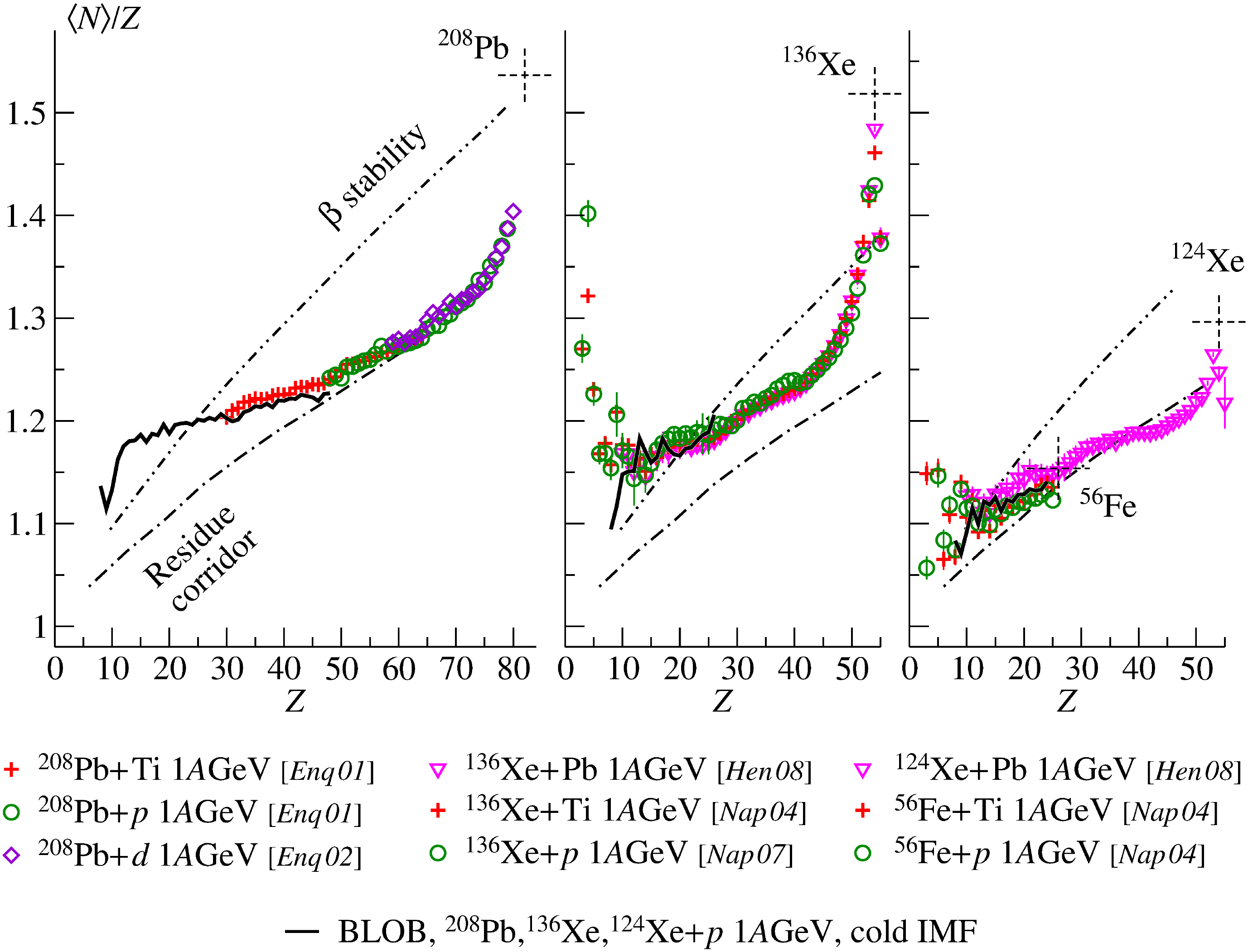}
\end{center}
\caption{
Comparison of the average isotopic content of IMF cold fragments, calculated with BLOB (already shown in Fig.~\ref{fig_NoverZ}) for the systems $^{208}$Pb$+p$, $^{136}$Xe$+p$ and $^{124}$Xe$+p$ at 1$A$GeV, with a set of experimental data.
Data were measured at FRS (Darmstadt) for fragments produced in proton/deuteron-induced spallation and for fragments emitted by spectator sectors in relativistic peripheral ion-ion collisions~\cite{Enqvist01,Enqvist02,Napolitani2003,Napolitani04,NapolitaniPhD04,Napolitani07,Henzlova2008}.
}
\label{fig_NoverZ_data}
\end{figure}
	We may also suggest that these results, in good agreement with previous studies based on statistical approaches, indicate that the transport description is well adapted to follow the reaching of equilibrium conditions, through a chaotic population of the available phase space,  
within the dynamical process~\cite{Raduta2006}. 

	This scenario of frustrated multifragmentation does not only recall heavy-ion collisions approaching Fermi energies, but it can also extend to the spectator region in peripheral heavy-ion collisions at relativistic energies.
	This similarity emerges from quantitative comparisons with experimental data.
	Fig.~\ref{fig_NoverZ_data} selects the average isotopic content of the cold IMF fragments calculated in Fig.~\ref{fig_NoverZ} for the spallation of the systems $p+^{208}$Pb $p+^{136}$Xe $p+^{124}$Xe at 1$A$GeV.
	This observable is then compared with several experimental data measured at FRS (Darmstadt), both from spallation reactions and from peripheral heavy-ion collisions~\cite{Enqvist01,Enqvist02,Napolitani2003,Napolitani04,NapolitaniPhD04,Napolitani07,Henzlova2008}.
	The agreement with data is quite successful.
	As expected~\cite{Schmidt2002}, the average isotopic content of cold IMF only shows a dependence on the properties of the remnant, independently on the excitation mechanism (spallation or peripheral relativistic heavy-ion collisions): as far as the system is too neutron rich to decay till reaching the residue corridor ($^{208}$Pb, $^{136}$Xe), data and calculations exhibit the same dependence on the isotopic content of the excited remnant.
	When the system is not neutron rich ($^{124}$Xe, $^{56}$Fe), data and calculations yield results compatible with reaching the residue corridor and losing the memory of the entrance channel.

	Up to this stage, this analysis agrees with inclusive data and statistical simulations, but it is not sufficient to characterise the mechanism: in principle, both fission and multifragmentation can populate the neutron-rich side of the nuclide chart, due to the curvature of the $\beta$-stability valley.

\subsection{Fragmentation in few IMF}

%
%
\begin{figure}[b!]\begin{center}
	\includegraphics[angle=0, width=.7\textwidth]{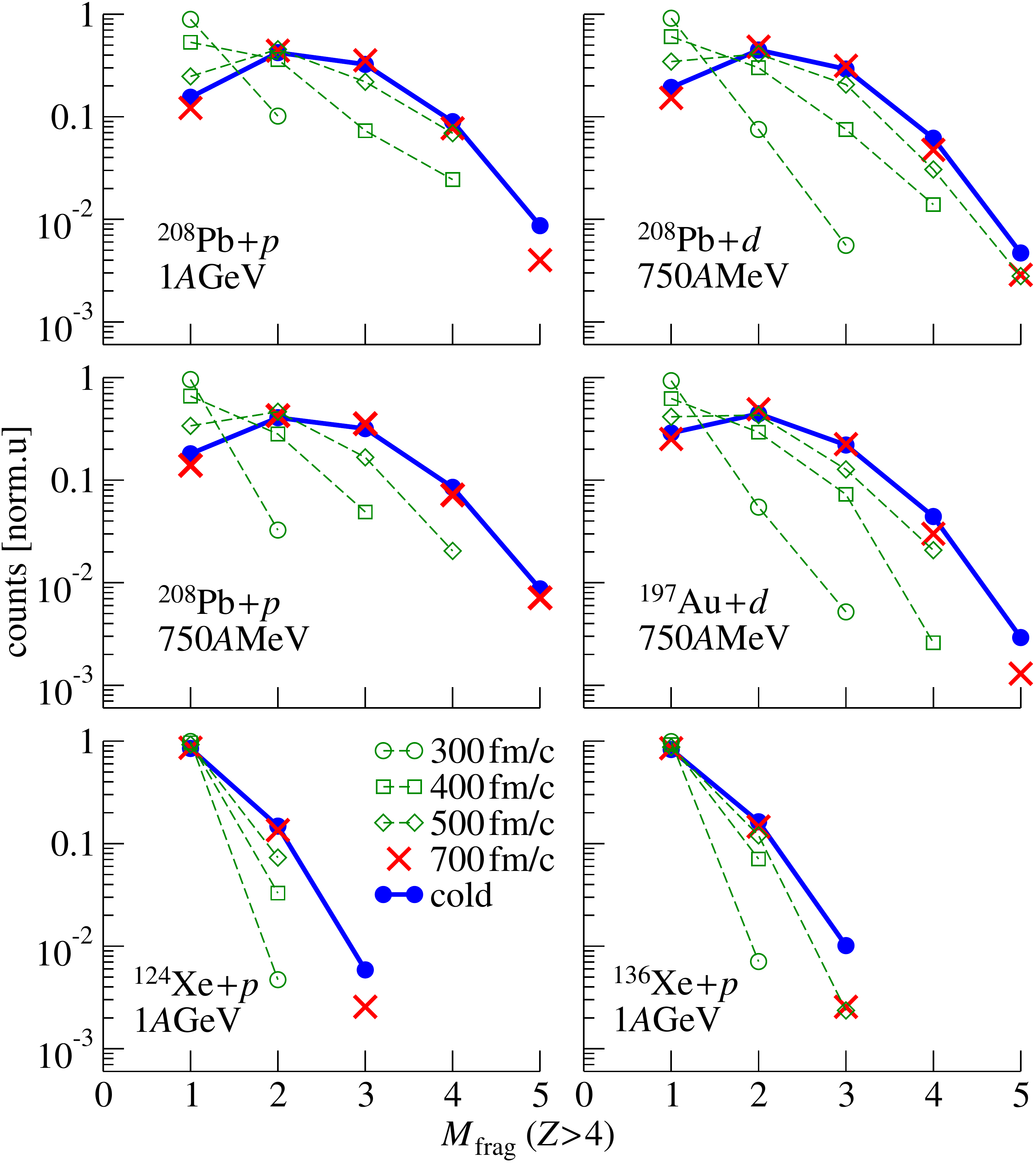}
\end{center}
\caption{
	Normalised yields as a function of the multiplicity of fragments with $Z>4$, for central impact parameters, at various time-steps of the dynamical process, and after secondary-decay.
}
\label{fig_multiplicity}
\end{figure}
	From the analysis of the multiplicity of fragments with $Z>4$ at 700 fm/c, studied for central impact parameters, we found that the lighter systems (Xe) prevalently recompact into one compound nucleus, or they undergo binary splits with about one order of magnitude smaller probability, and multiple splits are rare. 
	The heavier systems, despite also displaying some tendency to recompacting, are on the other hand dominated by binary splits, and ternary splits are also relevant.
	This analysis is presented in Fig.~\ref{fig_multiplicity}.
	The evolution of the fragment-multiplicity spectrum is also shown as a function of time: we observe that, even if density inhomogeneities arise at earlier times, the system starts separating into fragments rather late, at around 300~fm/c.

%
%
\begin{figure}[b!]\begin{center}
	\includegraphics[angle=0, width=.7\textwidth]{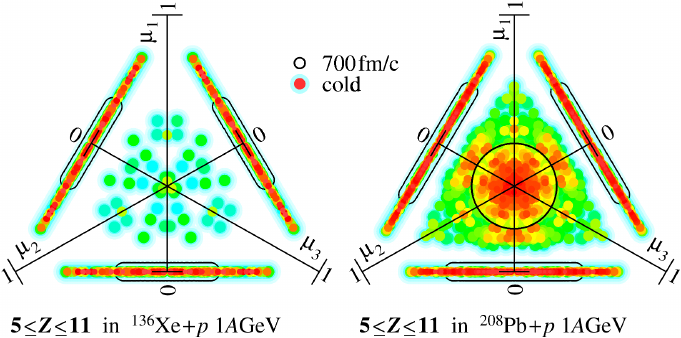}
\end{center}
\caption{
	Correlation among the three heaviest IMFs of mass number $A_1$, $A_2$ and $A_3$ produced in the same event, for all events where at least two fragments are found in the range $5\le Z\le 11$.
	Combinations of relative sizes $\mu_1$, $\mu_2$ and $\mu_3$ (where $\mu_i=A_i/(A_1+A_2+A_3)$) are studied in a Dalitz plot for the systems  $^{136}$Xe and $^{208}$Pb, for central impact parameters.
	Colour maps refer to cold systems after secondary-decay and the configurations at 700fm/c are indicated by black contour lines.
}
\label{fig_Dalitz}
\end{figure}

	An insight about the asymmetry of the splits is proposed in Fig.~\ref{fig_Dalitz} by analysing the size correlation among the three heaviest IMFs, of mass number $A_1$, $A_2$ and $A_3$, produced in the same event, for events where at least two fragments are found in the range $5\le Z\le 11$.
	All combinations of the relative sizes $\mu_1$, $\mu_2$ and $\mu_3$, where $\mu_i=A_i/(A_1+A_2+A_3)$, are used as coordinates in Dalitz plots.
	The size correlations are investigated for the systems $^{136}$Xe and $^{208}$Pb, for central impact parameters, both for the hot (at 700fm/c) and for the cold systems.
	From this analysis we infer that, even when the fragment multiplicity is larger than two, the splits exhibit a large asymmetry.
	In the $^{136}$Xe hot system, represented by black contours positioned on the sides of the plot, at maximum two IMFs are found in the range $5\le Z\le 11$, and the fragment multiplicity is completed by a heavier residue.
	The action of the secondary decay may turn some few events into three-IMF patterns which enter the selection and fill the centre of the plot.
	In the $^{208}$Pb system, hot and cold, symmetric splits are still rather rare with respect to events where one heavier fragment is present.
	The configuration of the splits has an obvious consequence on the kinematics.

%
%
%
%
\begin{figure}[b!]\begin{center}
	\includegraphics[angle=0, width=.7\textwidth]{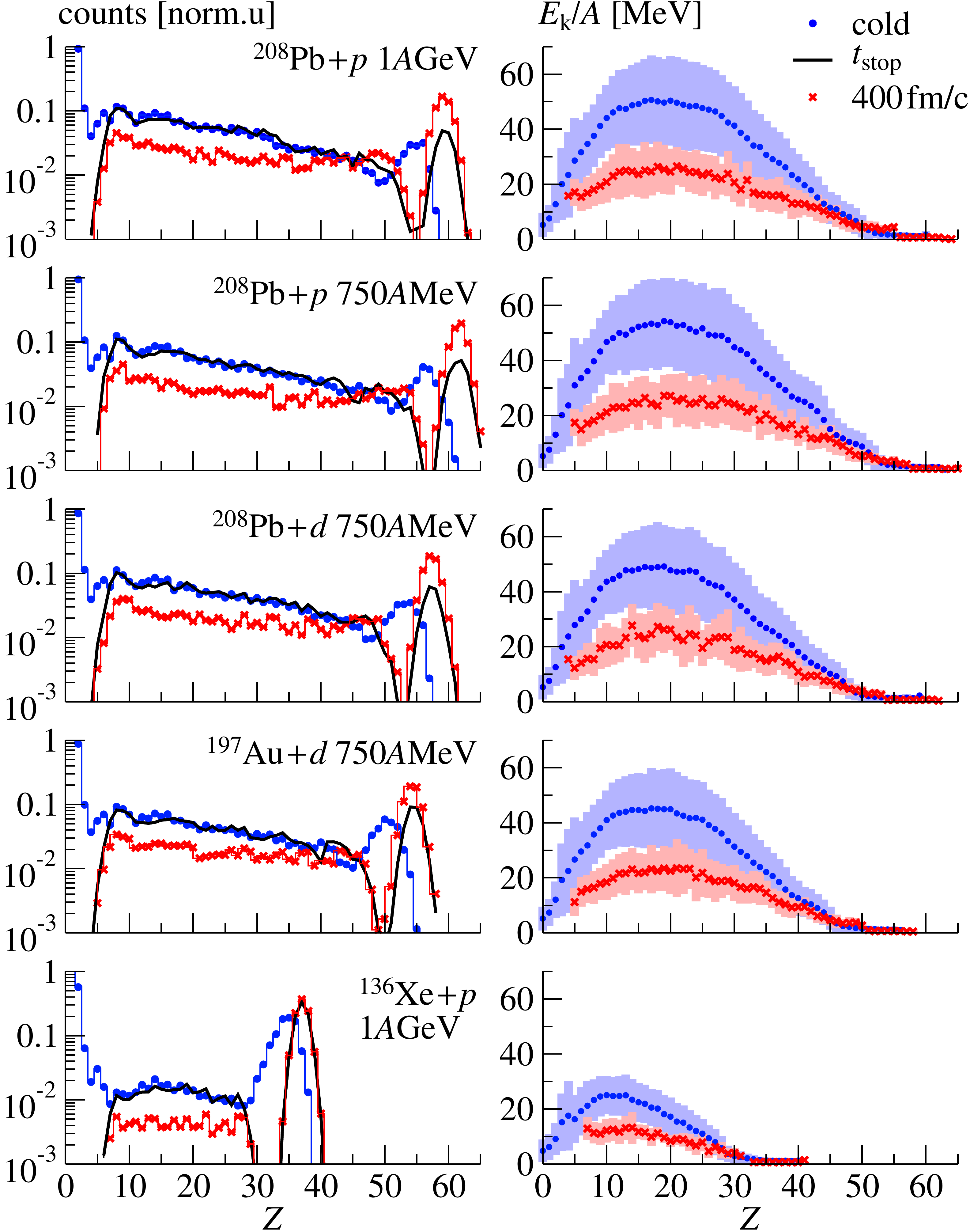}
\end{center}
\caption{
Production yields and kinetic energies for five spallation reactions, calculated as a function of the mass number, for central impact parameters, at 400fm/c, at the time $t_{\textrm{stop}}$, and after secondary-decay (cold).
}
\label{fig_Adistr}
\end{figure}
%

%
%
%
\begin{figure}[b!]
\begin{center}
	\includegraphics[angle=0, width=.8\textwidth]{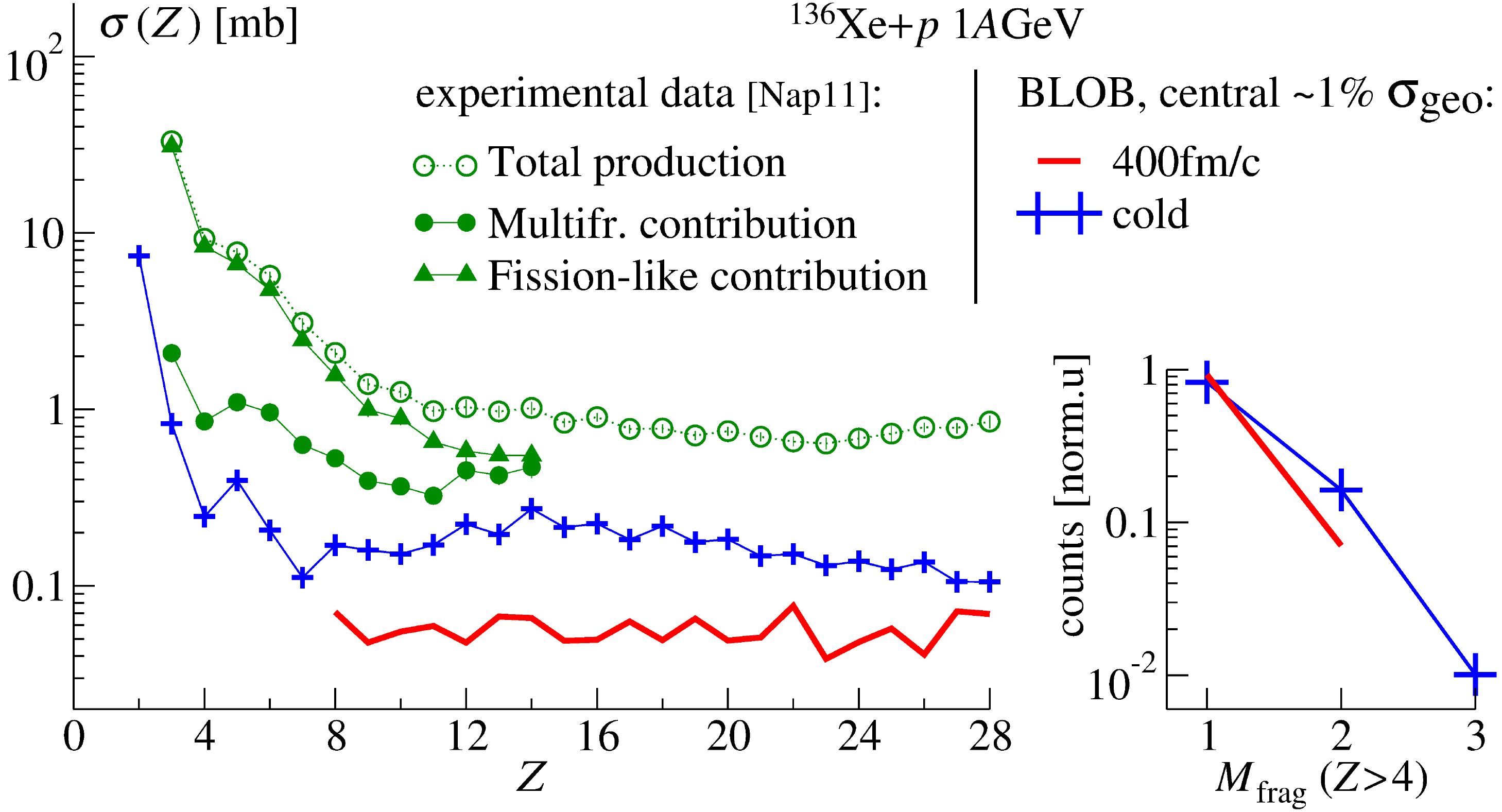}
\end{center}\caption{
Production cross section of light elements produced in $p+^{136}Xe$ at 1$A$GeV.
Data were measured at the FRS (Darmstadt)~\cite{Napolitani07}; contributions related to multifragmentation or to fission-like events could be separated~\cite{Napolitani2011}.
The full production of light elements (for all impact parameters) is compared to a BLOB simulation of IMF cross sections restricted to central impact parameters covering 1\% of the total geometric cross section at 400fm/c, and after secondary-decay (cold).
}
\label{fig_Zdistrdata}
\end{figure}
%
%
\subsection{Charge distribution and kinematics: \\two emission modes for IMFs}

	The fragment-mass yields are shown in Fig.~\ref{fig_Adistr}.
	The spectra at $t_{\textrm{stop}}$ and at the end of the sequential decay are similar except for the extremities, corresponding to the lightest and the heaviest masses, which have been modified by a prominent light-particle evaporation process and by asymmetric fission.
	Fig.~\ref{fig_Zdistrdata} shows the calculated spectrum for the system $p+^{136}Xe$ at 1$A$GeV, converted into cross section for the approximately 1\% of the total geometric cross section, corresponding to the choice of selecting central impact parameters for the simulation; the calculated spectrum is compared with the experimental total production cross section of the corresponding system~\cite{Napolitani2011}: a similar trend, corresponding to a flat cross section for IMF, ending up in a steep slope for the lightest IMF is shown. 
	Turning back to Fig.~\ref{fig_Adistr}, it is interesting to notice that the heavy-residue region is already filled at $t\!=\!400$~fm/c, whereas IMFs are also produced at later times. Moreover, their final yield, after de-excitation has been considered, is quite close to the yield given by the BLOB simulations at $t_{\textrm{stop}}$.    
	Therefore, within our calculation, the kinematics of the cold IMFs should mostly reflect the kinematics of the hot IMFs, when they are related to the most violent entrance channels.
	As shown in Fig.~\ref{fig_Adistr}, the kinematics reveals therefore the explosive character of the process and is then modified by the Coulomb propagation.

\begin{figure}[p!]\begin{center}
	\includegraphics[angle=0, width=.7\textwidth]{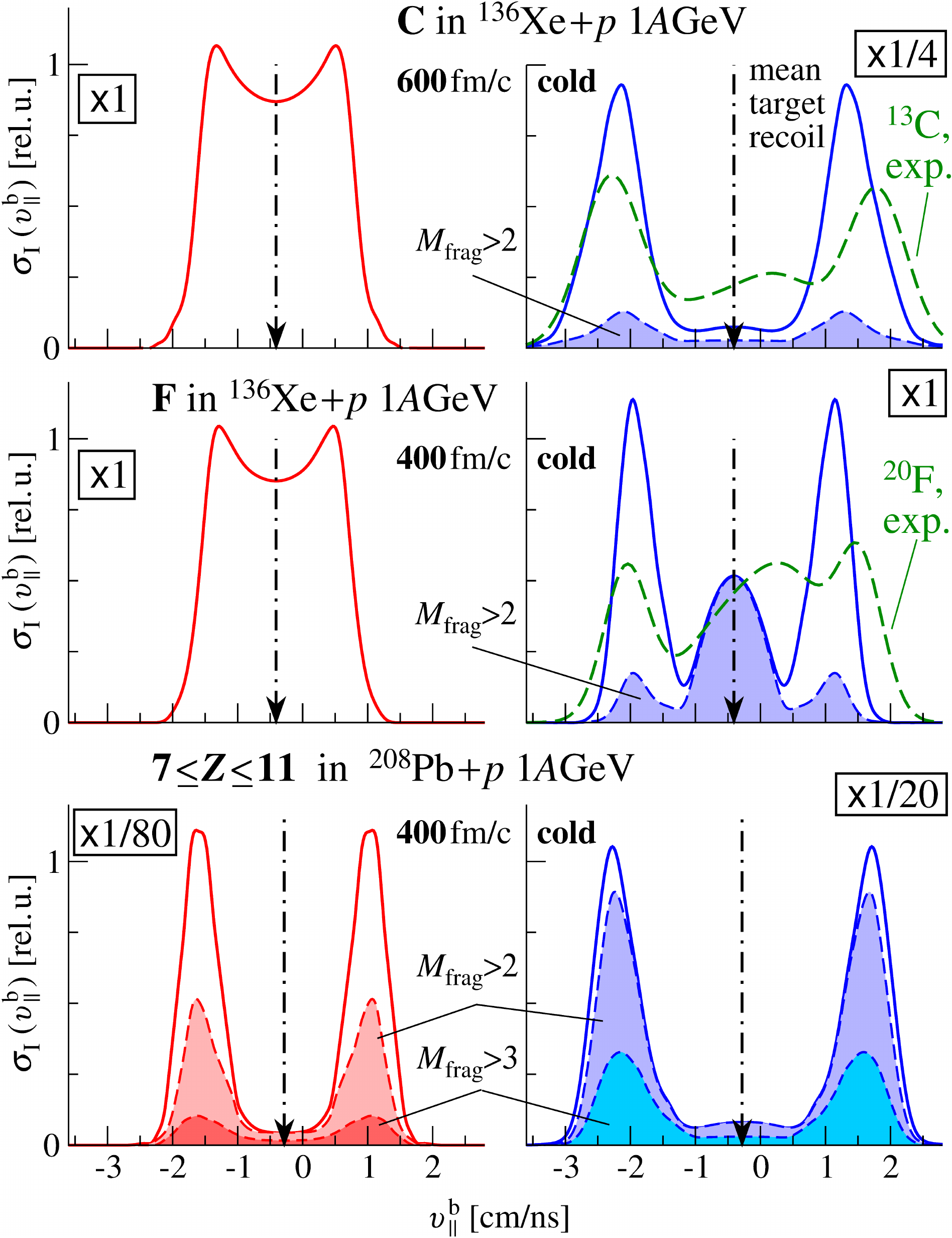}
\end{center}
\caption{
	Zero-angle invariant velocity distributions $\sigma_{\textrm{I}}(v_{||}^{\textrm{b}})/\sigma$  of carbon and fluorine isotopes calculated for the system $^{136}$Xe+$p$ at 1 $A$ GeV and of IMFs with $7\!\le\!Z\!\le\!11$  calculated for $^{208}$Pb+$p$ at 1 $A$ GeV.
	Left panels present distributions of hot fragments at 700 fm/c, while right panels present distributions of cold fragments after secondary decay.
	The integral of the distributions reflects the same number of events exploited for the two systems and, for better comparison, they are scaled by the factors indicated in the boxes. The spectra are shifted with respect to zero by the mean recoil velocity of the  $^{136}$Xe and $^{208}$Pb (indicated by arrows).
	The values of $M_{\textrm{frag}}$ indicate the multiplicity of fragments of $Z>4$ (including heavy residues) associated to the events, and label the corresponding contributions to the spectra.
	The simulation, restricted to central impact parameters covering 1\% of the total geometric cross section, are compared qualitatively to the measured spectra~\cite{Napolitani2011} of the isotopes with the largest production cross section, $^{13}$C and $^{20}$F.	
}
\label{fig_v_BLOB}
\end{figure}

	We conclude the analysis by recalling the initial inspiring experimental finding of Fig.~\ref{fig_v_distr}. 
	Due to the computational complexity, we could not collect enough statistics to reproduce the same kinematic observable of Fig.~\ref{fig_v_distr} for single isotopes, but we could produce a similar observable by collecting, for instance, the velocity distributions of all isotopes of carbon and fluorine for the system $^{136}$Xe+$p$ at 1 $A$ GeV, and all IMFs with $7\!\le\!Z\!\le\!11$ (interval chosen around oxygen and neon, which are elements frequently produced in multifragmentation) for the system $^{208}$Pb+$p$ at 1 $A$ GeV: this study is illustrated in Fig.~\ref{fig_v_BLOB} in the reference of the heavy nucleus before the collision; the shift with respect to zero corresponds therefore to the mean recoil of the target.
	The spectra could be symmetrised because the global reaction configuration studied with a central impact parameter is symmetric.

%
%
Carbon and fluorine in the $^{136}$Xe system change from a wide-hump distribution, for the hot IMFs, to a concave distribution for the cold IMFs. 
	When present, the contribution of events with fragment multiplicity larger than two are indicated.
For the $^{136}$Xe system it appears only in the cold system, leading to a wide convex portion of the spectrum, especially in the fluorine case: the convexity results from the variety of possible sizes and patterns involved in the splitting configurations, mainly when $M_{\textrm{frag}}>2$. 
	The resulting overall concave or two-humped wide spectra of the cold carbon and fluorine isotopes is produced by imparting different boosts to the fragments issued of binary events as a function of the partner size, producing wide humps from the folding of different Coulomb boosts, and by an additional contribution from asymmetric fission of the heavy residues, which selects a narrower Coulomb peak.
	From the analysis of Fig.~\ref{fig_Dalitz} we infer that, even when the multiplicity is larger than two, the kinematics of the splits should however manifest a binary-like character due to the size asymmetry among fragments: the kinematics reflects in this case the prominent Coulomb repulsion imparted by the largest fragment. 
	This effect becomes dominant in the $^{208}$Pb system, where concave wide spectra are also observed for larger fragment multiplicities.
	The calculation was limited to a small interval of impact parameters. 
The extension to the full range of impact parameters would, firstly, add or enhance the feeding of Coulomb peaks in the cold-IMF spectrum from asymmetric fission of heavy residues.
	Secondly, it would produce a folding over a span of recoil velocities for the target. Events where IMF are produced are related to a large range of central to semi-central impact parameters, and are mostly contributing to the centre of the distribution. Thus such folding would deform the central portion of the spectrum into an asymmetric shape with more extended tails for negative values of the velocity.
	In general, we observe that the more or less pronounced filling of the centre and the appearing of wide humps in the zero-angle spectra signs the presence of mechanisms possibly related to the sudden production of a few IMFs in a same short interval of time, as suggested in experimental observations~\cite{Napolitani04,Napolitani2011}.

	The calculated spectra for carbon and fluorine in $^{136}$Xe+$p$ (corresponding to central impact parameters covering 1\% of the total geometric cross section) are compared to the experimental spectra for the isotopes with the largest production cross section $^{13}$C and $^{20}$F, respectively.
	The comparison, very qualitative, shows that the peaks have comparable position and the centre of the spectra are filled also in the simulation; the data show however an even more pronounced contribution from violent mechanisms.

\section{Phenomenology}

	Even though a quantitative comparison with experimental data is beyond the purpose of this work, we observe that the model describes a large range of observable, from nuclide production to kinematics, which is globally consistent with the available experimental information.
	
The dynamical evolution leads to a chaotic population of the available phase space, which makes the final result quite similar to the
predictions of statistical multifragmentation models \cite {Raduta2006} (statistical investigations along this line can be found in refs~\cite{Souza2009}).
	The interest of the dynamical approach is that it can be used to extract further information on the phenomenology of the process at any time. 
	Moreover, kinematical effects connected to the expansion dynamics can only be described within a dynamical model. 

\subsection{Frustrated multifragmentation}

%
%
\begin{figure}[b!]\begin{center}
	\includegraphics[angle=0, width=.7\textwidth]{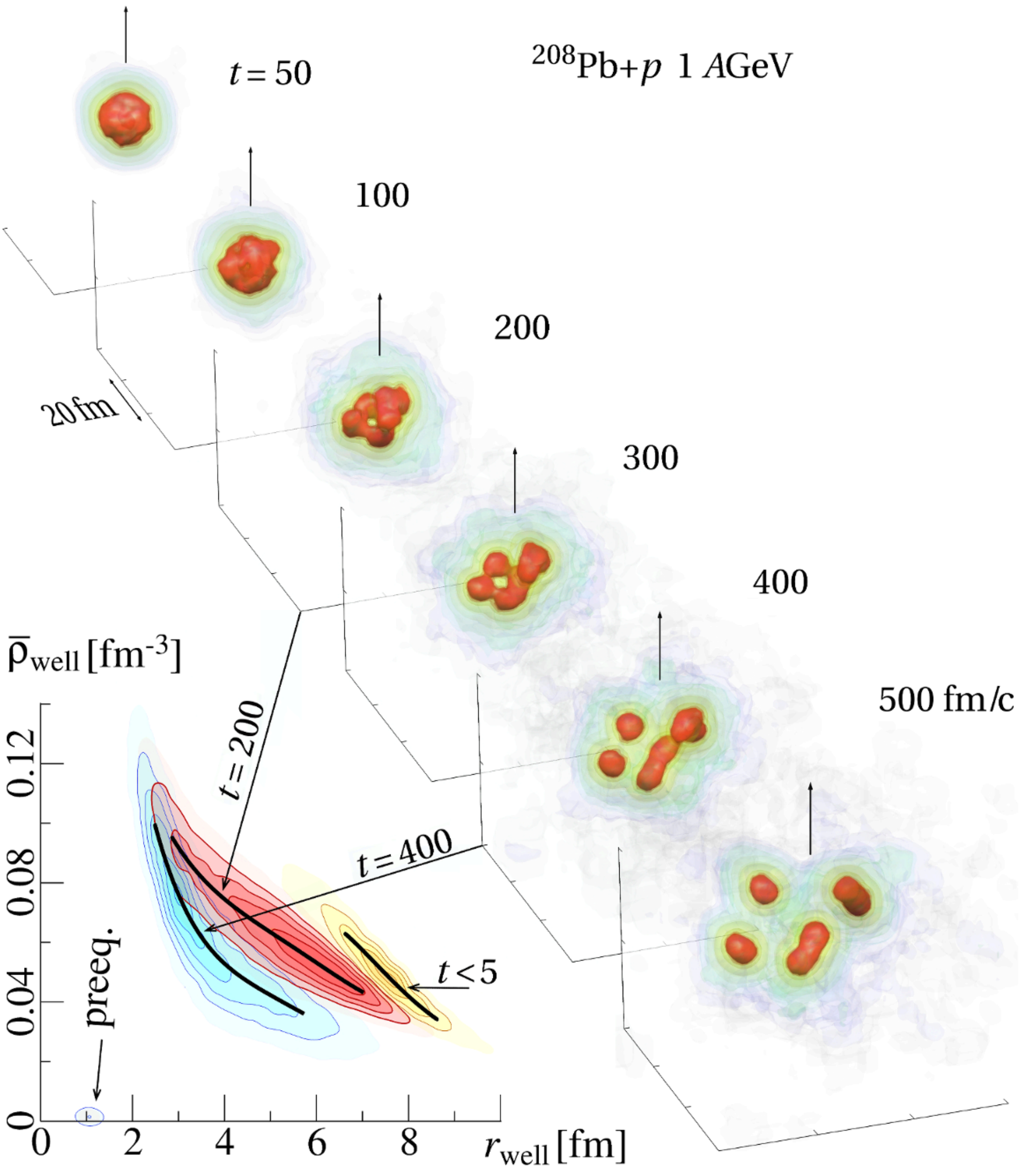}
\end{center}
\caption{
	Top. Time evolution of the density profile of the systems $p$~(1GeV)$+^{208}$Pb for one specific event selected among those giving multifragmentation.
	The system undergoes a spinodal behaviour, visible from 100 to 200~fm/c, when developing inhomogeneities of comparable size.
	Later on, the fragmentation mechanism is frustrated by the mean-field resilience, resulting into a rather asymmetric fragmentation.
	Bottom. Time evolution of the size of potential concavities associated to the evolution of the density profile at the beginning of the process ($t\!<\!5$~fm/c), during the phase of instability growth ($t\!=\!200$~fm/c), and when fragments appear ($t\!=\!400$~fm/c). See text for details. 
}

\label{fig_animation_frag}
\end{figure}

	Fig.~\ref{fig_animation_frag} gives an overview of the richness of the dynamic behaviour.
	In the first instants which follow the collision, low-density tails appear in correspondence with the emission of preequilibrium particles, proceeding from forward angles and later extending more isotropically  to all solid angles.

	After this time, the system starts expanding and the dynamical fluctuations handled by the BLOB treatment become a dominant mechanism in the process.
	With reducing bulk density, phase-space fluctuations grow in amplitude and potential ripples develop, becoming the nesting sites of fragments: inhomogeneities in the bulk density profile stand out at around 100 fm/c, but it takes them long time to eventually separate into fragments.
	The process exhibits a typical characteristic of the spinodal instability, i.e. the arising of blobs of similar size in the bulk.
	The inset of Fig.~\ref{fig_animation_frag} analyses for this same system the density averaged over those blobs as a function of their size at different times.
	The blobs are identified as any potential concavity found in the system and the size 
is their average radius (their shape is nearly spherical).
	At early times potential concavities coincide with the whole expanding system or with some large portions of it when particle flow develops. 
	At late times potential concavities have a large probability to coincide with the inhomogeneities arising in the density landscape: because their size reflects the leading instability mode~\cite{Napolitani_IWM2014}, they are all comparable in size, corresponding with larger probability to neon or oxygen nuclei~\cite{Chomaz2004,Borderie2008}.
	At intermediate times, sizes range from the whole system to the size of the spinodal undulations in the density landscape.
	This coexistence recalls phase-transition signals and corresponding results at Fermi energies, where suitable 
observables (such as the asymmetry between the charges of the two heaviest fragments produced in one collision event~\cite{Pichon2006} or the size of the largest fragment produced in one event~\cite{Bonnet2009}) have been proposed. 

	However, only in presence of a sufficiently large radial expansion these blobs can separate into fragments of comparable size and preserve the spinodal signal also in the exit channel.
	We can observe that this is definitely not the case: not all blobs succeed in separating in single fragments but they bond together in groups.
	The event of Fig.~\ref{fig_animation_frag} finally results in the fragmentation of the system into four asymmetric parts.
	Such a scenario seems to be general for this kind of spallation mechanisms and it is reflected in the low multiplicity of hot fragments analysed in Fig.~\ref{fig_multiplicity} and in the mass distribution of hot remnants of Fig.~\ref{fig_Adistr}. 
	In fact this latter, even when displaying a peak around the elements selected by the spinodal instability like oxygen and neon, presents a rather flat distribution which implies a large recombination of the spinodal inhomogeneities into larger fragments.
As it was already argued in ref.\cite{Colonna1997bis}, we can conclude on this phenomenology that the multifragmentation mechanism in spallation in the 1~$A$GeV energy regime is frustrated by the action of the mean field which tends to recompact the system as far as not enough energy is spent in the radial expansion.
%

\subsection{Exit-channel chaos and binary events}

%
%
\begin{figure}[t!]\begin{center}
	\includegraphics[angle=0, width=.53\textwidth]{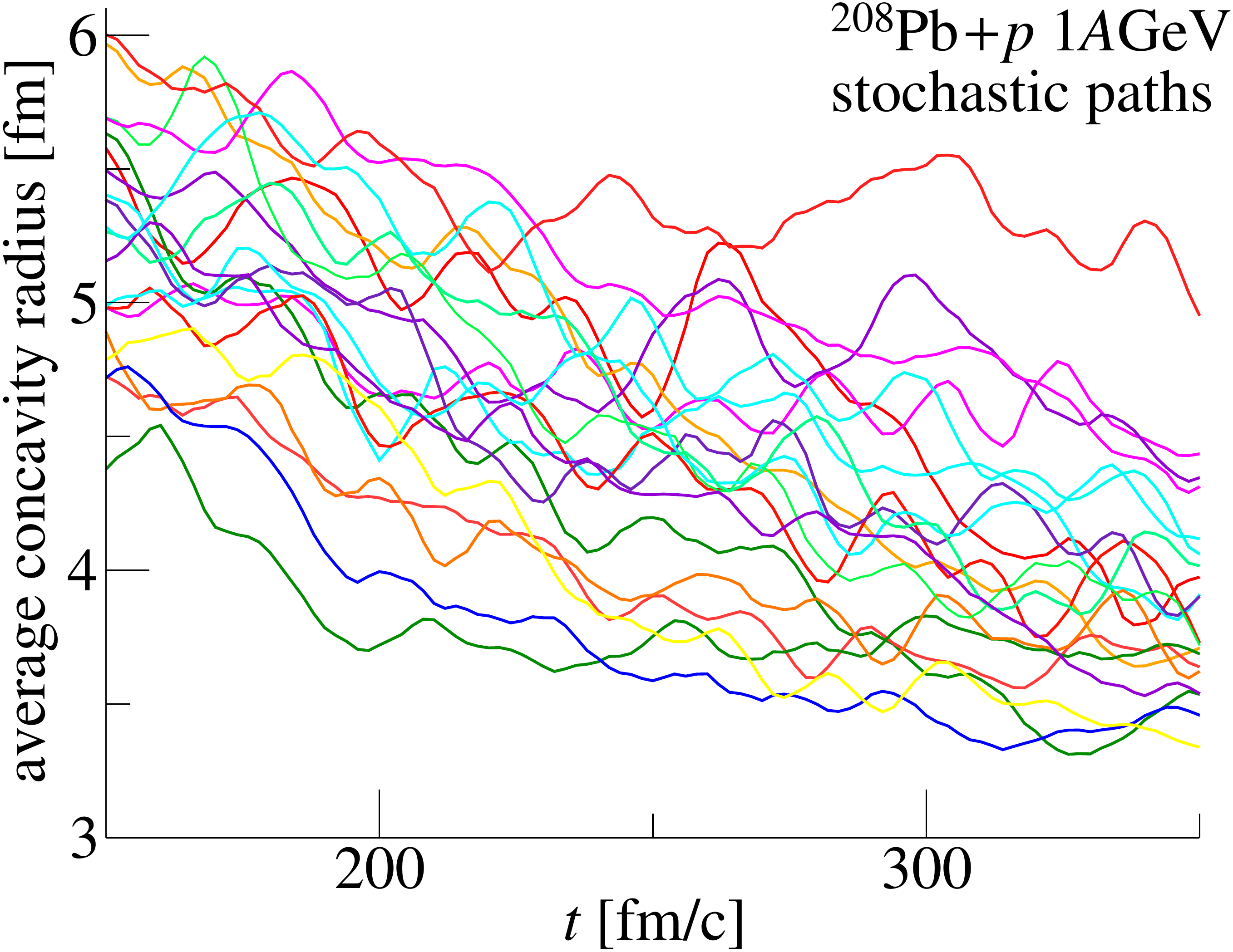}\\
\end{center}
\caption{
	A random selection of some reaction paths followed by the system $^{208}$Pb+$p$ at 1 $A$ GeV on the time-dependent potential landscape, represented by the average size of potential concavities as a function of time.
}
\label{fig_trajectories}
\end{figure}

	Phase-space fluctuations favour fragmentation.  Moreover,
	they also act, over many events, in expanding the bundle of dynamical reaction paths into a large chaotic pattern of bifurcations~\cite{Randrup1990}:
this leads to a variety of exit channels.
	Fig.~\ref{fig_trajectories} illustrates this effect by plotting the average radius of inhomogeneities found in the density landscape as a function of time for several stochastic evolutions of the systems $^{208}$Pb+$p$ at 1 $A$ GeV.
	Each trajectory could lead to a different exit channel, characterised by a different kinematics and larger or smaller fragment multiplicities and charge asymmetries.

\begin{figure}[b!]\begin{center}
	\includegraphics[angle=0, width=.8\textwidth]{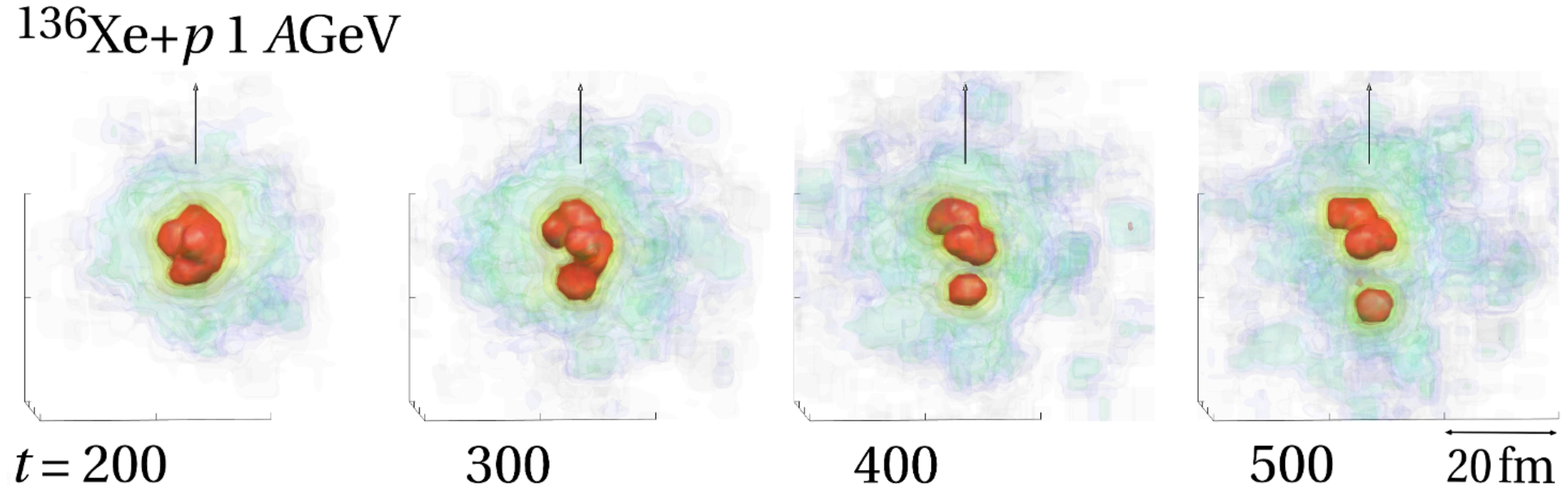}
\end{center}\caption
{
	Time evolution of the density profile for the systems $^{136}$Xe+$p$ at 1 $A$ GeV for one events resulting into an asymmetric binary split.
}
\label{fig_animation_fiss}
\end{figure}

	Particularly interesting are trajectories leading to only two fragments in the exit channel, as it can occur rather often in systems like $^{136}$Xe+$p$ at 1 $A$ GeV, as illustrated in Fig.~\ref{fig_animation_fiss}.
	Also in this case, a spinodal process may activate and immediately enter in competition with  the action of the mean field which tends to reverse the fragmentation pattern into a compact shape.
	Most of the times, this frustrating process results in one single compound nucleus.
	Very seldom the mean field succeeds only partially in coalescing the inhomogeneities of the density profile, and one or a group of those separates into a fragment and leaves the system.
	Depending on the stochastic configuration of the fragmenting system, the partial coalescence could recompact the inhomogeneities in many combinations resulting in different asymmetries.

%
%
%
%

	To characterise the final exit channel, we have combined the dynamical process and the additional secondary decay, which we simulated through a fission-evaporation afterburner.
	It turns out that two mechanisms may coexist in the spallation systems we examined.
	One is a multifragmentation process frustrated by the coalescence effect of the mean field, which constrains the fragment multiplicity to small values, rarely equal to three or four, more often equal to two: this work focused mostly on this process.
	The other mechanism, related to the production of heavy residues, is also a binary split, but it coincides with asymmetric fission from a compound nucleus in the secondary decay process~\cite{Sanders99}.
	For different causes, both processes contribute to the same IMF production when regarding the charge distribution and even when regarding the isotopic content.
	However, these two contributions are fundamentally different because of two reasons.
	First: They are separate in time; multifragmentation develops within the short time of the dynamical collision process, while the second occurs during the longer time of the fission decay process.
	Second: from a dynamical point of view  binary splits emerging from fission or from multifragmentation are different: fission of a compound nucleus is a trajectory in a deformation landscape which passes through the development of a neck; on the other hand, multifragmentation leads initially to a mottling topology in density space, possibly driven by spinodal instabilities, and then the expansion dynamics attempts to keep this topology, in competition with the antagonist tendency of reverting into a compact shape.
This would lead to binary channels, similar to fission, but obtained through re-aggregation processes. The latter could generally have
different kinematic features with respect to standard fission processes.  
	Experimentally, this difference would be reflected in the velocity spectra of the IMFs if the resolution is sufficiently high.

	The length of the transport calculation makes it prohibitive to track all contributions to the yields coming from the full distribution of excitation energies of the spallation system, and we had to restrict to the most violent collisions only.
	More quantitative simulations are left for further works.

\subsection{Spallation to ion-ion collision at Fermi energies}

	We find that unstable isoscalar modes can actually arise in such systems, like heavy nuclei bombarded by protons and deuterons in the 1 A GeV regime and, with a low but not negligible probability, become responsible for the fragmentation of the system.
	According to our theoretical approach, we find that unstable modes in spallation should exhibit quite a similar phenomenology as spinodal instability in dissipative central heavy-ion collisions when the incident energy corresponds to the threshold between fusion and multifragmentation~\cite{Napolitani2013}.

	In the case of dissipative ion-ion collisions, the excitation of the system is mostly determined by mechanical perturbations, while in the spallation process the excitation originates from an almost isotropic propagation of the energy deposited by the light projectile.
	Though these situations are different in some aspects, they both drive phase-space fluctuations of large amplitude and they may both activate the spinodal behaviour and the amplification of mechanically unstable modes.
	In both cases, the mean field may then have the effect of reverting the whole system, or part of it, into a compact shape, smearing, modifying, or completely erasing the fragment configuration.
	In particular, if the kinetic energy feeding the expansion dynamics is not sufficient to disintegrate the system but it is still larger than the system can hold, the fragment multiplicity reduces and the fragment configuration becomes asymmetric.
	As an extreme situation, the process may look like asymmetric fission, but the chronology, as well as the violence of the process will be incompatible with the conventional fission picture: this process would correspond to a binary channel obtained by re-aggregation, keeping some dynamical aspects.
	In particular, the physical description that we suggest for IMF production in spallation may solve apparent discrepancies between some experimental interpretations of inclusive and exclusive data, the former revealing multifragmentation-like kinetic energies, the latter revealing small fragment multiplicities compatible with compound-nucleus decays.
	These experimental results would actually be perfectly coherent.

	After having speculated about a bridge between spallation and ion-ion collision at Fermi energies, we investigate in detail this latter situation in the following chapter.

\chapter{Fluctuation and bifurcations in dissipative heavy-ion collisions
\label{ch_HIC}}

	In this chapter we focus on intermediate-mass cluster formation in heavy-ion collisions around and slightly above Fermi energies.
	We investigate the phenomenology of fragment production from central to peripheral impact parameters, which is the object of many recent experimental projects.
	This mechanism emerges in low-density portions of the system from the onset of mean-field instabilities which drive fluctuations and bifurcations in dynamical trajectories.
	To describe this situation, in-medium dissipation and phase-space fluctuations should be accounted for. 
	The interplay of these correlations with the one-body collective behaviour determines the properties (kinematics and fragment production), the variety of mechanisms of the exit channel (from fusion to neck formation and multifragmentation) and the corresponding thresholds.
	In particular, in the proximity of a threshold, fluctuations between two energetically favourable mechanisms stand out, so that when evolving from the same entrance channel, a variety of exit channels are accessible.
	A possible situation is the fusion-to-multifragmentation threshold in central collisions, where the system is shown to fluctuate between two energetically favourable mechanisms: either reverting to a compact shape or rather disintegrating into several fragments of similar size. 
	Another situation is the transition from binary mechanisms to neck fragmentation (ternary or even quaternary channels), in peripheral collisions.

	To account for competing contributions from one and two-body effects and handle large-amplitude phase-space fluctuations, we rely on the stochastic transport descriptions presented in \textsection~\ref{ch_instabilities}, which have already been tested on nuclear matter in \textsection~\ref{ch_inhomogeneities} and then addressed to open systems in the context of spallation reactions in \textsection~\ref{ch_spallation}.
	In this chapter we apply both SMF and BLOB approaches to heavy-ion collisions and we report on some significant results, showing that the models are close to the observation.

	The most relevant finding reported in this chapter is that phase-transition signatures are obtained as the result of a non-equilibrium description which includes the mechanism of spinodal instability.
	A second finding reported thereafter concerns the mechanism of frustrated fragmentation arising from a competition between spinodal instability and mean-field resilience: this process can describe the low-energy threshold of multifragmentation and its possible association to asymmetric splits in two or few fragments in close analogy to the phenomenology explored in \textsection~\ref{ch_spallation} for spallation.
	We also investigate the richness of the process of neck fragmentation, which may result in various unusual patterns to be further investigated in forthcoming theoretical and experimental works.

\textit{Main sources for this chapter}: 
this chapter exploits discussions and studies developed in refs.~\cite{Napolitani2013,Napolitani2012,Colonna2017,Napolitani2017} and profits from the experimental results of refs.~\cite{Moisan2012,Ademard2014}.

\section[Dynamical description of HIC at Fermi energy]{Dynamical description of heavy-ion\\ collisions at Fermi energy
\label{sec_FermiEnergy}}

	Heavy-ion collisions at Fermi energies are open systems which require a non--equilibrium dynamical description when the process should be followed from the first instants.
	The heated system produced in such collisions cannot be treated within an independent-particle picture and additional correlations should be taken into account: they rely on in-medium dissipation and phase-space fluctuations.
	At large beam energies, when exceeding few hundred MeV per nucleon, those beyond-mean-field contributions become prominent (see \textsection~\ref{sec_flow}), while (the rest of this chapter) the Fermi-energy domain is characterised by the interplay of nucleon-nucleon (N-N) collisions and one-body collective behaviour.
	As related to these contributions, the fragment production in dissipative heavy-ion collisions at Fermi energies is driven by instabilities which induce fluctuations of so large amplitude to produce bifurcations in the dynamical evolution.
 	In this case, one given projectile-target system evolving from a given entrance channel, defined by an impact parameter and an incident energy, may produce a variety of exit channels. 
	For instance, when the incident energy is increased while approaching Fermi energies, the system may oscillate between fusion and multifragmentation for small impact parameters, or between binary and ternary mechanisms for large impact parameters. 
	To describe such a chaotic behaviour, dominated by fluctuations and bifurcations, pure mean-field equations should be replaced by more adapted theories

	As done in \textsection~\ref{ch_inhomogeneities} in the context of nuclear matter, in order to exploit both mean-field and beyond-mean-field contributions, the one-body description can be extended in terms of BBGKY hierarchy to include N-N collisions and, as their natural corollaries, isoscalar and isovector fluctuations.
	N-N collisions affect flow and stopping~\cite{Lopez2014,Basrak2016} while fluctuations produce an ensemble of mean fields and a corresponding variety of exit channels.
	This approach to open systems has already been described in \textsection~\ref{ch_spallation} for spallation reactions, where the entrance channel had essentially the role of introducing an excitation energy in the system.
	In ion-ion collisions at Fermi energies, on the other hand, the entrance channel has also the role of strongly affecting the density distribution of the system with a strong dependence on the impact parameter, the target and projectile sizes and the isospin properties of the participants~\cite{Baran2005}. 

%
%
\begin{figure}[b!]\begin{center}
	\includegraphics[angle=0, width=1\textwidth]{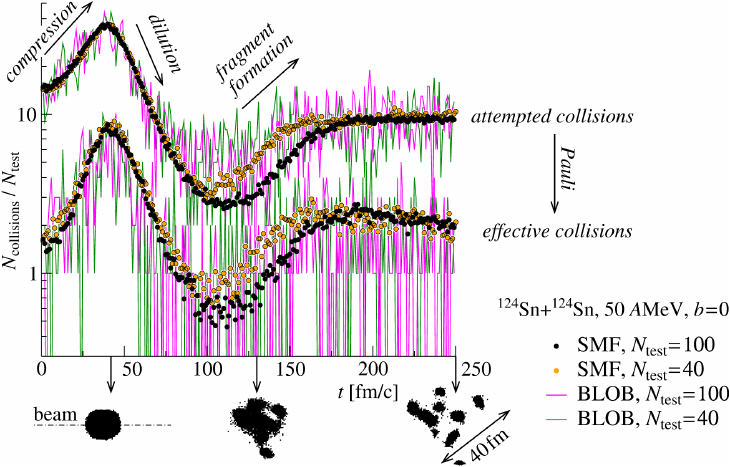}
\end{center}
\caption{
SMF and BLOB compared for the same system $^{124}$Sn$+^{124}$Sn at 50 $A$MeV and $b=0$. 
Evolution of the number of attempted and effective (not Pauli blocked) collisions per nucleon for one single dynamical trajectory (one event) reflecting the amplitude of the fluctuations.
Labels and density plots indicate different phases of the process.
SMF calculations show a dependence on $\Ntest$ and a small variance.
BLOB calculations show no dependence on $\Ntest$ and a huge variance.
}
\label{fig_collisions_SMF_BLOB}
\end{figure}

	Numerical solutions have been worked out for the dynamical description of ion-ion systems at Fermi energies in the framework of the Boltzmann-Langevin (BL) equation which is well suited to describe out-of-equilibrium processes.
	In the regime of small-amplitude fluctuations, a stochastic definition of the initial states is sufficient, restricting to mean-field (quantum) fluctuations of collective observables~\cite{Abe1996,Lacroix2012}, while Fermi energies are mostly related to large-amplitude fluctuations.
	For instance, when low densities are attained, the nuclear system is brought to explore regions of the phase diagram where it becomes unstable against density fluctuations, like the spinodal region. 
As demonstrated in \textsection~\ref{ch_inhomogeneities}, the action of the BL term results in agitating the density profile over several wave lengths and amplifying the unstable modes according to the specific dispersion relation, associated with the employed mean-field interaction.
	In this chapter we rely therefore on stochastic one-body approaches like SMF~\cite{Colonna1998} and BLOB, Eq.~(\ref{eq:BLOB}), that were introduced in \textsection~\ref{ch_instabilities} to let develop fluctuations spontaneously and continuously through the collision term.
	For this latter, in the present chapter, if not otherwise specified, all calculations are done by introducing a screened in-medium cross section $\sigma_{\textrm s}$ (from ref.~\cite{Danielewicz2002,Coupland2011}). 
	As in \textsection~\ref{ch_spallation} and in ref.~\cite{Baran2005}, for the simulation we use a soft equation of state with $k_{\inf}\!=\!200$ MeV; when not otherwise specified, the potential component of the symmetry energy of the EOS is given by a linear term as a function of the density (asy-stiff).

	Besides defining the entrance channel for target and projectile nuclei, the application of these models to heavy ion collisions requires some additional refinements.
	The extension of the wave packets makes necessary to pay special attention to scatterings close to the surface of the system, i.e. occurring across potential boundaries: confronting the shape of the wave packet to the shape of the surface, the blocking factors are increased in proportion to the spread of the nucleon packet outside of the boundary (similarly to what is done in some molecular-dynamics approaches~\cite{Aichelin1991,Bohnet1991}).

	Fig.~\ref{fig_collisions_SMF_BLOB} illustrates a study of the N-N collision statistics for the collision $^{124}$Sn$+^{124}$Sn at 50 $A$MeV, at a central impact parameter ($b=0$), studied with both models SMF and BLOB.
	Since the same prescription for the in-medium cross section is used, the two models give closely comparable average number of attempted and effective N-N collisions, but differences stand out in the variance which is strongly reduced in SMF due to the more approximated scheme of introducing fluctuations. 
	For the same reason, SMF also exhibits a dependence on the number of test particles per nucleons $\Ntest$, which is absent in BLOB.

	The average number of effective N-N collisions $\langle N_{\textrm{eff}}\rangle$ reflects the average density evolution of the system: a maximum is encountered in correspondence with the largest compression and a minimum marks the subsequent expansion mechanism.
	Later on, $\langle N_{\textrm{eff}}\rangle$ increases again indicating that the process of fragment formation sets in, and it finally levels off till, around $250$ fm/c, the system has stabilised into a well defined configuration.
	The velocity of this stage of fragment formation depends on the model, getting faster for implementations favouring larger fluctuation amplitudes: in fact, it is faster for BLOB than for SMF when the same number of test particles per nucleon is used, and it is faster when fewer test particles per nucleon are used within SMF.
	
\begin{figure}[t!]
	\begin{minipage}[b]{0.475\linewidth}
	\centering
		\includegraphics[angle=0, width=1\textwidth]{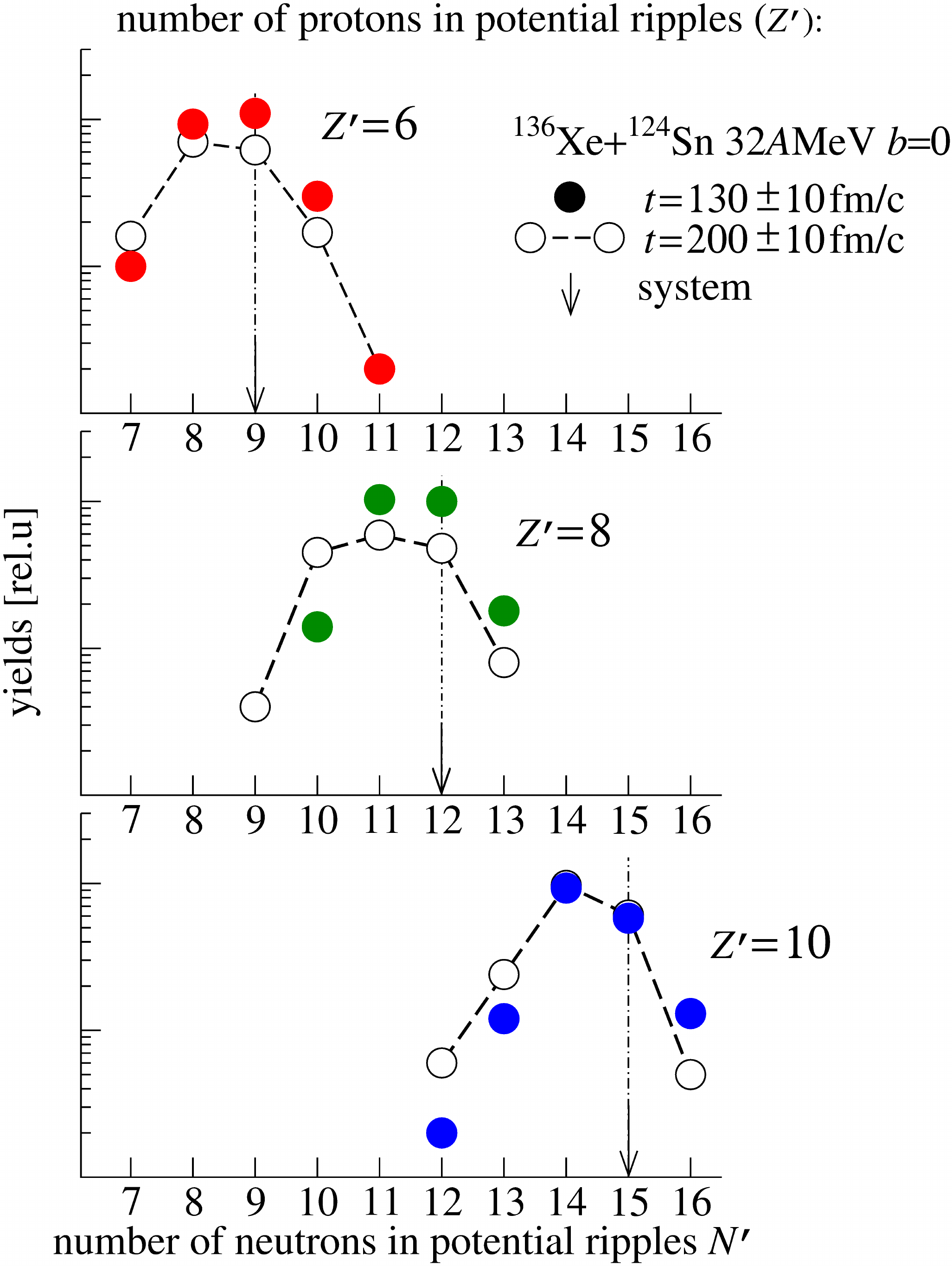}\\
		\vspace{8ex}
	\end{minipage}
	\hspace{0.05cm}
	\begin{minipage}[b]{0.475\linewidth}
	\centering
		\includegraphics[angle=0, width=1\textwidth]{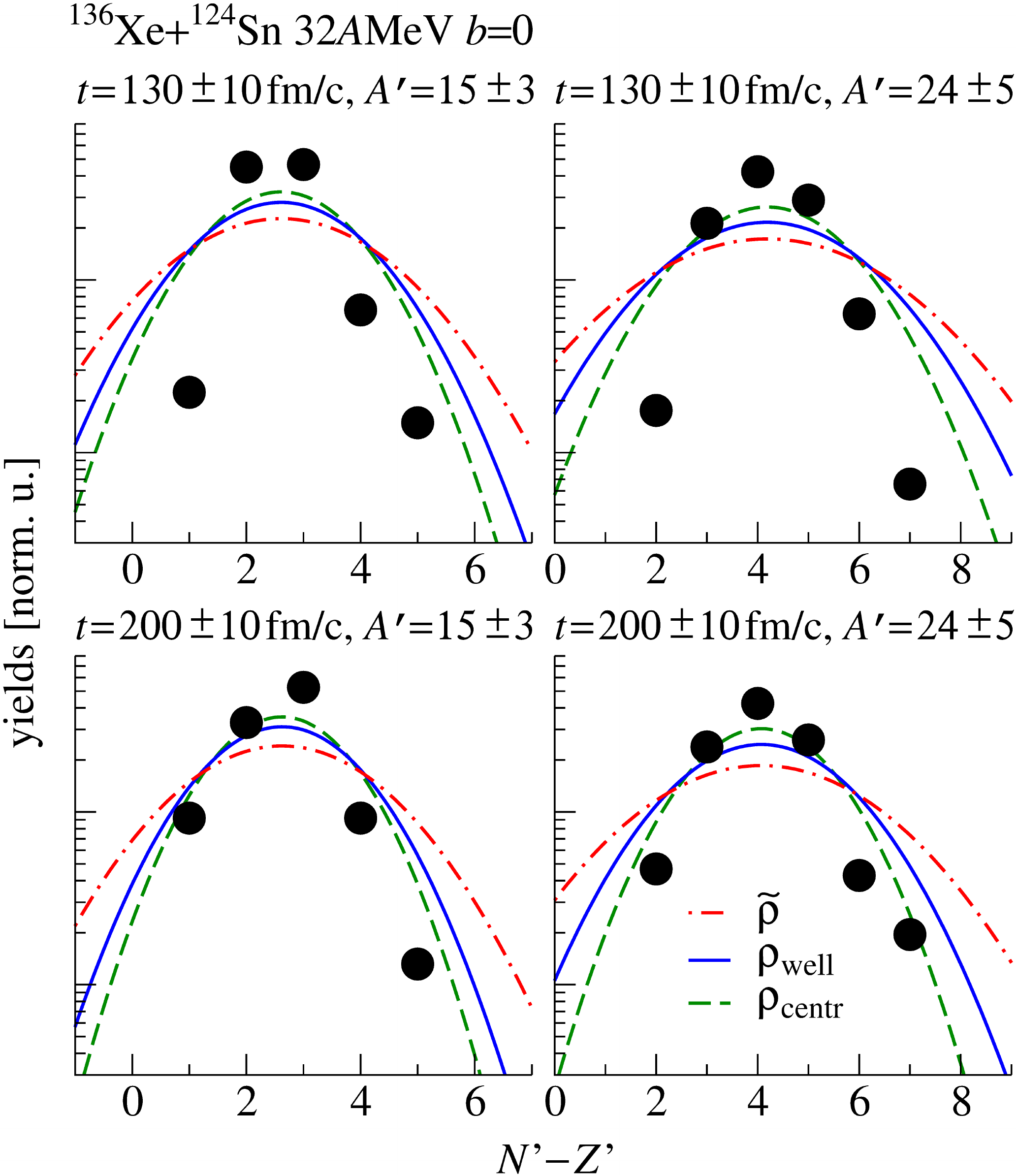}\\
		\includegraphics[angle=0, width=1\textwidth]{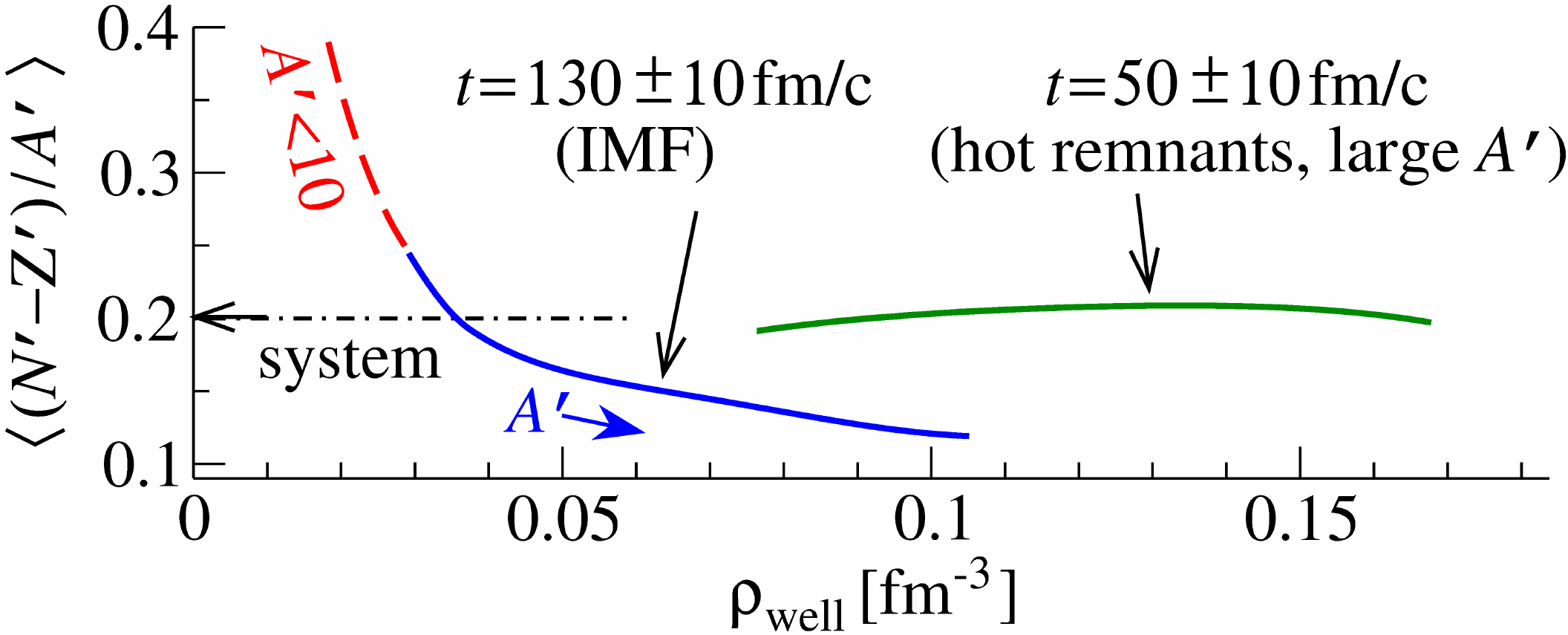}
	\end{minipage}
	\caption{
	Left. Isotopic distributions of forming clusters of Carbon, Oxygen and Neon at different times.
	Right, upper and middle rows. Distributions of isotopic variances in potential ripples containing $N'$ neutrons and $Z'$ protons for the most probable configurations for forming clusters in the regions around Carbon ($Z'=6$ and $A'=15$) and Neon ($Z'=10$ and $A'=24$) at different times.
	For comparison, analytic distributions are plotted, as a function of the density measured for different portions of the potential ripples (see text).
	Right, bottom. Average isospin content measured in potential ripples at different times as a function of the $\rho_{\textrm{well}}$.
	IMF mass $A'$ evolves according to a process of isospin distillation (see text).
}
\label{fig_isotopicdistributions}
\end{figure}

	Heavy-ion collisions are also suited systems to explore isospin effects related to isovector fluctuations.
	Fig.~\ref{fig_isotopicdistributions} translates the survey of \textsection~\ref{ch_inhomogeneities} to heavy-ion collisions, investigating the isospin content in potential ripples containing $N'$ neutrons and $Z'$ protons for the system $^{136}$Xe$+^{124}$Sn at 32 $A$MeV.
	Distributions of isotopic variances (right, two upper rows) are calculated for the most probable mass range around a forming Carbon and a forming Neon (defined from the isotopic spectra in the left panel).
	The distributions are studied in an early time span (before fragment formation around 130fm/c) and in a late time interval (during fragment formation around 200fm/c). 
	They are compared with the analytic distributions obtained at a temperature $T=5.5$MeV, as extracted from the calculation, and at the local density $\rho$.
	The isotopic variance can be studied as the probability of variation $\delta$ around the mean value of $N'-Z'$ and for a given $A'$ yielding the distribution 
\begin{equation}
	Y \approx \textrm{exp}[-(\delta^2/A')\,C_{\textrm{sym}}(\rho)/T]
	\label{eq_Y}
\end{equation}
(see Eq.~(\ref{eq:fluctuation_dissipation}) for the variance).
	The local density is evaluated either in the (denser) centroid of the potential ripples $\rho_{\textrm{centroid}}$, or averaged all over the volume of the emerging fragments $\widetilde{\rho}$, or, more significantly, corresponding to the matter contained in the volume of the potential ripples $\rho_{\textrm{well}}$.
	We deduce that, as expected from the calculation in stable nuclear matter discussed above, the isotopic width results underestimated with respect to the analytic prediction of Eq.~(\ref{eq:fluctuation_dissipation}). 
The difference is still acceptable due to the following two effects.
	First of all, fluctuations are built out of equilibrium: this implies that the collision rate is higher, generally leading to larger variances.
	Secondly, in open systems, particle evaporation may contribute in widening the isotopic spectra.
	The bottom row of Fig.~\ref{fig_isotopicdistributions}, right, investigates the average isospin content measured in potential ripples in the corresponding early and late time intervals as a function of the $\rho_{\textrm{well}}$.
	The IMF mass $A'$ grows with the density and the corresponding isospin content decreases signing a process of isospin distillation (see \textsection~\ref{sec_fusiontomultifragmentation}).
	In addition to distillation, heavy-ion collisions are also the place to trace isospin-transport processes driven by a different isotopic content in the target and in the projectile or by the developing of regions of different density during the collision; few more insights will be given in the following.
%
%

	Before undertaking a more detailed analysis, we use thereafter the BLOB model to simulate collisions of $^{130}$Xe nuclei, in order to have some general examples. 
	This system is chosen because it recalls widely investigated systems in the region from Sn to Xe~\cite{Kohley2014,DeFilippo2014,Ademard2014}; 
in order to rely on simpler entrance-channel properties, the system is chosen symmetric and along $\beta$ stability.

\subsection[Central collisions and the fusion-to-multifragmentation threshold]{Central collisions and the\\ fusion-to-multifragmentation threshold
\label{sec_fusiontomultifragmentation}}
\begin{figure}[t!]\begin{center}
	\includegraphics[angle=0, width=1\textwidth]{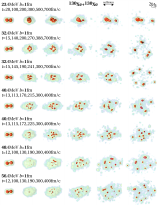}
\end{center}
\caption{
	$^{130}$Xe$+^{130}$Xe, examples of most probable mechanisms producing IMFs in central collisions as a function of the incident energy.
}
\label{fig_order_disorder_map}
\end{figure}

	Head-on collisions are investigated with BLOB in Fig.~\ref{fig_order_disorder_map}, where density distributions are analysed as a function of time for some events related to different incident energies.
	Each event samples the most probable mechanism leading to intermediate-mass fragments (IMFs) in the exit channel for the selected incident energy.
	The event at 32 $A$MeV shows the arising at early times (from 100 to 200 fm/c) of a pattern of several almost-equal-size inhomogeneities in the density landscape.
	These undulations reflect a spinodal behaviour~\cite{Chomaz2004,Borderie2008}, i.e. a condition of mechanical instability where the size of the emerging blobs reflects the leading instability mode as discussed in \textsection~\ref{ch_inhomogeneities} when studying the corresponding dispersion relation.

	At later times, if the radial expansion is not sufficient, the equal-size inhomogeneities may reaggregate producing a less symmetric pattern and a small multiplicity of fragments.
	In such chaotic process, the competition between these two antagonist tendencies, the disintegration into several pieces driven by spinodal instability on the one hand, and the action of the attractive nuclear force which tends to bond fragments together, imposes that different exit channels are favoured depending on the bombarding energy.
	The event at 23 $A$MeV shows for instance a highly frustrated fragmentation resulting into an almost complete re-aggregation, so that the final exit channel appears as a very asymmetric binary split.	

	In \textsection~\ref{ch_spallation} we argued that a similar mechanism also appears in spallation reactions induced by protons and deuterons in the 1 GeV range on heavy nuclei~\cite{Napolitani2015}.
	The spinodal signal, imparting fragment-size symmetry, becomes prominent in the event at 40 $A$MeV and fades at larger bombarding energies (i.e. the event at 56 $A$MeV).

	Within this same approach, the analysis of a statistics of exit channels connected to a given set of initial conditions could reveal phase-transition signals in central collisions: this issue is examined in details in \textsection~\ref{sec_central}

%
%
%
%
%
%
\begin{figure}[t!]\begin{center}
	\includegraphics[angle=0, width=1\textwidth]{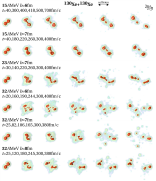}
\end{center}
\caption{
	$^{130}$Xe$+^{130}$Xe, examples of most probable mechanisms leading to IMFs in the exit channel for semiperipheral impact parameters ($b\approx$ 6 to 8~fm ) as a function of the incident energy.
}
\label{fig_neck_to_midrapidity_map}
\end{figure}
%
%
\subsection[Peripheral collisions and the binary-to-neck transition]{Peripheral collisions and the binary-to-neck\\ transition}
%
%
\begin{figure}[t!]\begin{center}
	\includegraphics[angle=0, width=.6\textwidth]{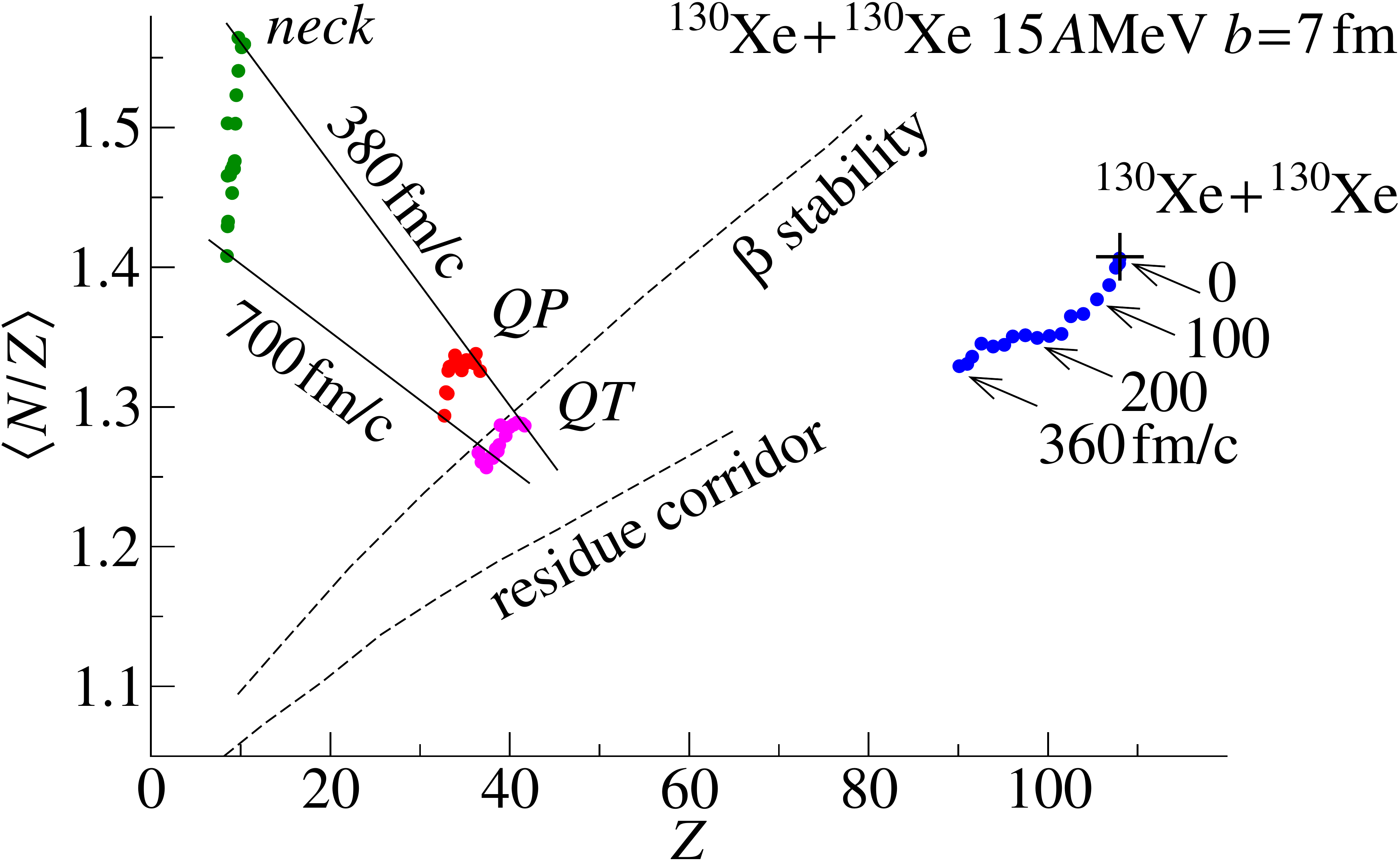}
\end{center}
\caption{
	Study of neck formation in BLOB for one single typical event in $^{130}$Xe$+^{130}$Xe at 15$A$MeV at $b=7$.
	The neutron enrichment ($N/Z$) is shown for different components of the system at different times.
	A similar study with SMF is discussed in~\cite{Lionti2005}.
}
\label{fig_neck_NZ}
\end{figure}
%
%
%
%
\begin{figure}[b!]\begin{center}
	\includegraphics[angle=0, width=.9\textwidth]{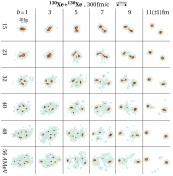}
\end{center}
\caption{
	Examples of most probable exit-channel configurations with IMFs at 300 fm/c in a incident-energy-versus-impact parameter table.
}
\label{fig_Fermi_map}
\end{figure}
%


	Fig.~\ref{fig_neck_to_midrapidity_map} investigates semi-peripheral collisions sampling the most probable mechanisms of IMF production for the selected incident energy, as calculated with BLOB.
	At 15 $A$MeV the typical mechanism where IMFs are produced is the formation and separation of a neck region in a rather long process.
	This mechanism was described as ruled by Rayleigh instability~\cite{Montoya1994}.
	At larger bombarding energies the neck gradually transforms into a diluted midrapidity region where more than one blob can form. 
	At 23 $A$MeV the process is still long and the separation of more than one IMF is too rare.
	At 32 $A$MeV neck fragmentation producing two or more IMFs becomes a favoured mechanism. 
	In less peripheral collisions (i.e. $b=7$ fm) IMFs arise close to the centre of the midrapidity region and they are repelled at large angles with respect to the collision axis, while in more peripheral collisions (i.e. $b=8$ fm) at the same bombarding energy two IMFs tend to form in the proximity of the quasiprojectile (QP) and the quasitarget (QT), respectively; in this case the IMFs may orbit around the QP or QT and be eventually pulled outside of the collision axis with forward angles.
	These exotic mechanisms were suggested in refs.~\cite{Baran2012,Rizzo2014,Colonna2015}. 
	The neck process has been widely investigated especially for its connection to the isospin migration mechanism~\cite{Baran2005,Lionti2005,DeFilippo2005bis,DiToro2006,Rizzo2008bis}: this process, driven by density gradients, induces neutron currents from the QP/QT regions towards the diluted midrapidity region where the neck forms.
	As illustrated in Fig.~\ref{fig_neck_NZ} for an event at 15 $A$MeV, the consequence is the prominent neutron enrichment of the neck fragment with respect to the QP and QT, which were chosen along $\beta$-stability; in this situation, the hot neck fragment neither succeeds in de-exciting towards the residue corridor, nor it can approach $\beta$-stability.
	More insights on semiperipheral and peripheral channels and their relation with isospin transport are given in \textsection~\ref{sec_neck} and \textsection~\ref{sec_peripheral}.


%
%
\begin{figure}[b!]\begin{center}
	\includegraphics[angle=0, width=.9\textwidth]{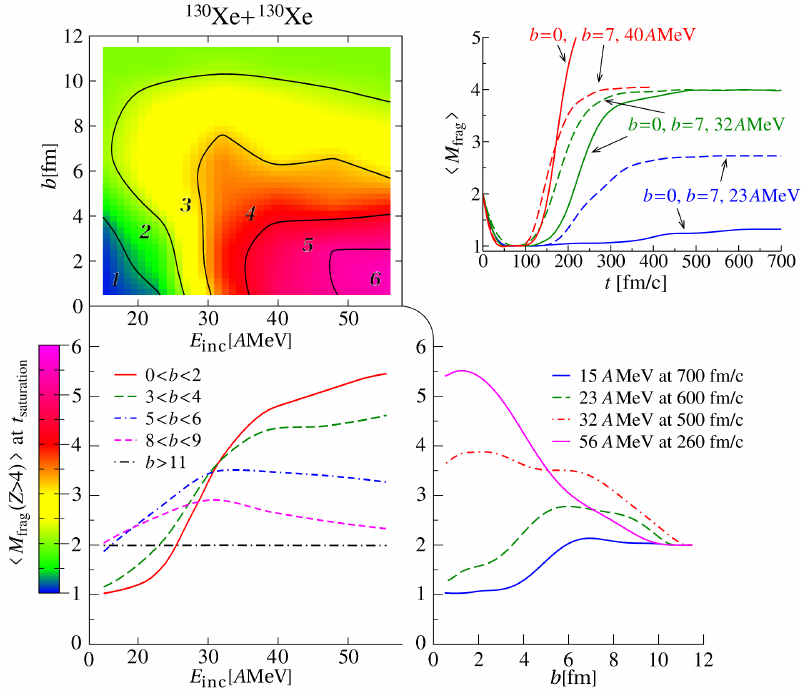}
\end{center}
\caption{
	Left. Mean multiplicity of fragments with $Z>4$ as a function of impact-parameter and incident energy for the system $^{130}$Xe$+^{130}$Xe at multiplicity-saturation time.
	Right. Time evolution of the mean multiplicity of fragments with $A>1$ in central and peripheral collisions.
}
\label{fig_Fermiland_bdistr}
\end{figure}
%
%
%

\subsection{Overview on fragment multiplicity and timing
\label{subsec_Fermiland}}

	Samples for the most probable fragment configurations at 300fm/c for the system $^{130}$Xe$+^{130}$Xe as a function of incident energy and impact parameter are collected in Fig.~\ref{fig_Fermi_map}, where all transitions between the different mechanisms discussed above can be followed.
	More quantitatively, the corresponding map of mean multiplicity $\langle M_{\textrm{frag}}(Z>4) \rangle$ of fragments with $Z>4$ as a function of the impact parameter and incident energy is shown in Fig.~\ref{fig_Fermiland_bdistr} (top left), from analysing a statistics of about 3500 events: projections on the two coordinates are also shown.
	To build the map of Fig.~\ref{fig_Fermiland_bdistr} the multiplicity of fragments was extracted at the time when its mean value stops growing.
	The right panel illustrates for some systems that such saturation time can be extracted from analysing the time evolution of the mean multiplicity $\langle M_{\textrm{frag}}\rangle$ of fragments with $A>1$.

	The maps of figs.~\ref{fig_Fermi_map} and \ref{fig_Fermiland_bdistr} indicate in particular the regions where transitions between different mechanisms can arise, and they adapt without significant changes to collisions at Fermi energies of nuclei in the region of Sn and Xe.
	Some of those transitions may be pointed out.
	The fusion cross section fades in favour of asymmetric binary splits above 20AMeV in central collisions and, when approaching 30AMeV, binary splits gradually change into the regular pattern of almost equal-size IMFs, indicating the onset of spinodal multifragmentation.
	Still for central collisions, such symmetric pattern persists till around 45AMeV, and further increasing of bombarding energy brings the system outside of the spinodal region.
	Below 20 $A$MeV, along the impact-parameter coordinate $b$, fusion changes into a binary mechanism.
	In the interval of about $5<b<9$fm, semiperipheral collisions lead to neck formation.
	A transition from ternary channels to events with more than one IMF at midrapidity appears when moving to larger bombarding energies (above about 30AMeV) and smaller impact parameters (below about 8fm).

\section[Multifragmentation threshold in central collisions]{Multifragmentation threshold and the \\onset of bifurcations in central collisions
\label{sec_central}}

	Experimental observables may exhibit a bimodal character in some suitable conditions~\cite{Pichon2006, Bonnet2009}: such signature, which in a statistical framework is a robust indication of a first-order phase-transition~\cite{Binder1984,Chomaz2000,Chomaz2003}, was also suggested to have a possible dynamical origin, related to the fragment-formation mechanism~\cite{LeFevre2009}, without necessarily requiring the reaching of thermodynamic equilibrium.  

	The aim of this section is to further investigate with both SMF and BLOB models, which have been explicitly conceived for describing spinodal instabilities, the dynamical trajectory of disassembling nuclear systems, seeking for features associated with phase transitions.
	We explore the possible occurrence of bifurcation patterns and bimodal behaviour in central heavy-ion reactions at beam energies around the multifragmentation threshold.
	Central collisions provide in fact the largest density perturbation along the compression-expansion path traced by the nuclear reaction.
	In this case, the system may access the spinodal region of the EOS, where a density rise is related to a pressure fall; there, phase-space fluctuations are even amplified, leading to phase separation~\cite{Chomaz2004,Borderie2008}.
	In particular, fluctuations act on both isoscalar and isovector degrees of freedom.
	As soon as a mottling pattern stands out at low density, i.e. at the boundary of the phase separation, a bundle of bifurcations into a variety of different dynamical paths may set in.
	These features are the fundamental ingredients for handling the formation of clusters and nuclear fragments at subsaturation densities in a one-body description.

%
%
%
\begin{figure}[b!]
\begin{center}
	\includegraphics[angle=0, width=.75\textwidth]{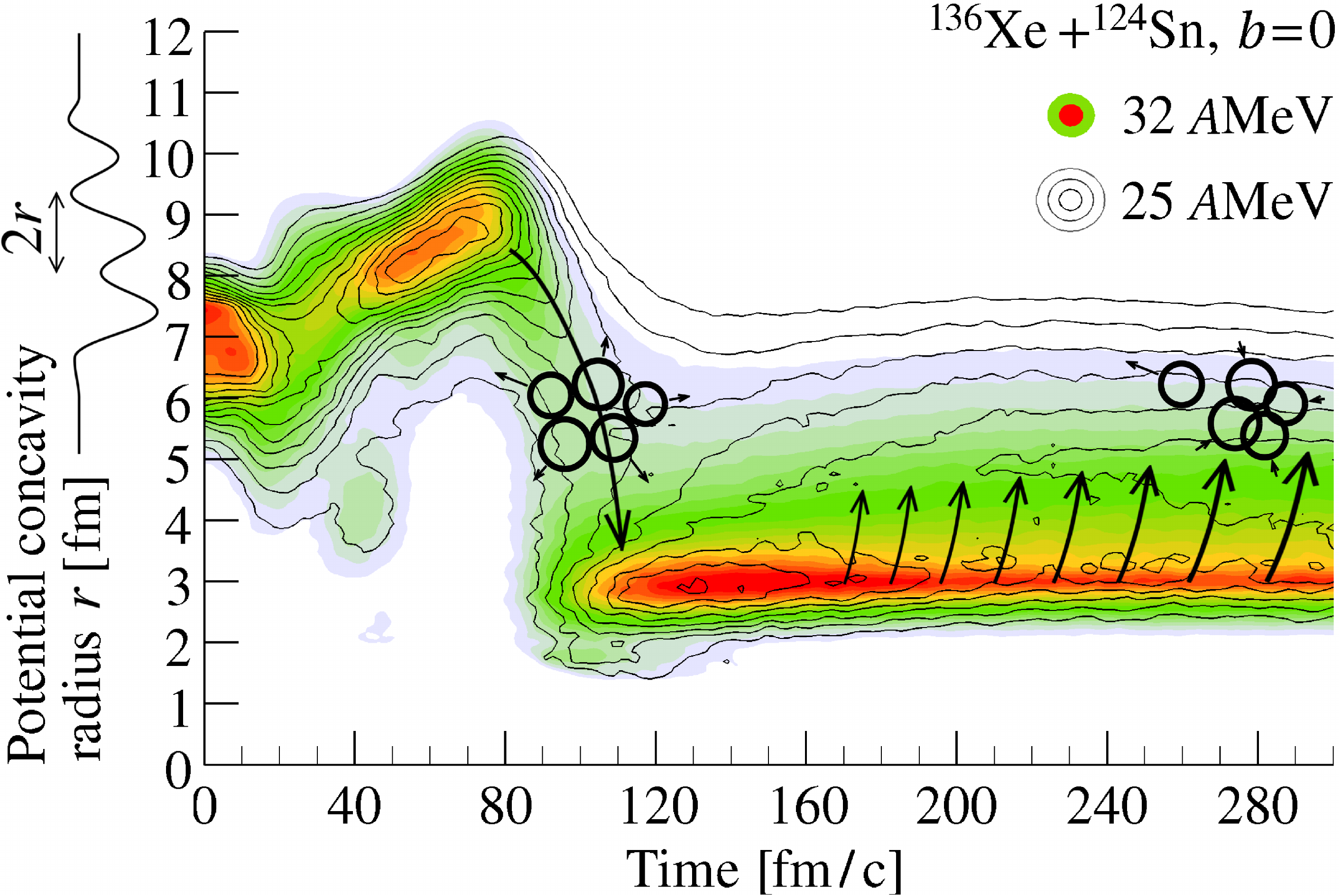}\hspace{1ex}
\end{center}\caption{
\label{fig_phase_trans}
Left.
	Temporal evolution of the size of potential ripples in $^{136}$Xe$+^{124}$Sn at 25 and 32$A$MeV in central collisions for an incident energy of $32\,A$MeV (filled contours) and $25\,A$MeV (contour lines).
Spinodal-like fragmentation occurring at around 80--100fm/c is followed by a process of recombination at later times.
}
\end{figure}
	Such process is explored in Fig.~\ref{fig_phase_trans} in the system $^{136}$Xe$+^{124}$Sn for central collisions at 25 and 32 $A$MeV; a probability map shows how ripples in the potential landscape evolve in size as a function of time (from ref.\cite{Napolitani2013}).
	Ripple radii probe the size of forming blobs of matter which are the fragment nesting sites; those latter may eventually separate into clusters and leave the system (prominent mechanism at $32\,A$MeV), or merge into larger fragments, or fuse back together forming a large residue (as clearly seen at $25\,A$MeV).
	Ripple radii evolve initially with the composite-system size, as marked by the upper branch in Fig.~\ref{fig_phase_trans}, and suddenly drops to smaller sizes which characterise all forming droplets of matter within a small variance: this latter defines the lowest branch in Fig.~\ref{fig_phase_trans}.
	From this structure we find that at around 100fm/c large sizes, corresponding to the whole system, coexists with small sizes, which are consistent with the leading wavelength of the dispersion relation, i.e. about the size of nuclei in the region of oxygen and neon~\cite{Colonna1994}.
	If in this latter situation all potential concavities could come apart into fragments, a pure signature of the spinodal decomposition would stand out as a set of equal-size fragments~\cite{Borderie2001}.
	However, it should be mentioned that not only one leading wavelength prevails, but also several multiple modes might develop with comparable growth rate in finite systems~\cite{Jacquot1996b}.
	In addition, when the radial expansion of the system is not sufficient to outweigh the mean-field resilience, the density landscape continues to evolve towards a more compact shape so that, if the system still succeeds in disassembling, potential concavities can merge giving rise to larger fragments and producing asymmetric fragment configurations.
	This process of recombination prevails at 25$A$MeV and weakens at larger incident energies.

	As anticipated in \textsection~\ref{sec_FermiEnergy}, the right-bottom panel of Fig.~\ref{fig_isotopicdistributions} shows the action of isovector fluctuations on the same system in a process of isospin distillation.
	As a function of the local density averaged on potential ripples, and for different sizes of those latter, the average isospin distribution for the corresponding sites is studied at different intervals of time.
	Different sets of potential-concavity sizes are indicated by extracting a corresponding mass $A'$.
	From an initial situation where the system is close to saturation density and the average isospin is defined by the target and projectile nuclei, densities drop to smaller values and the isospin distribution extends over a large range.
	In particular, the smaller is the local density, the larger is the isospin measured in corresponding sites as neutrons favour the most volatile phase.
	Such ordering have been intensively studied within the SMF approach~\cite{Chomaz2004}.
	However, this second isovector observable can only be exploited if isoscalar fluctuations have the large amplitude required to disassemble the system into clusters and fragments.

\begin{figure}[b!]\begin{center}
	\includegraphics[angle=0, width=.7\textwidth]{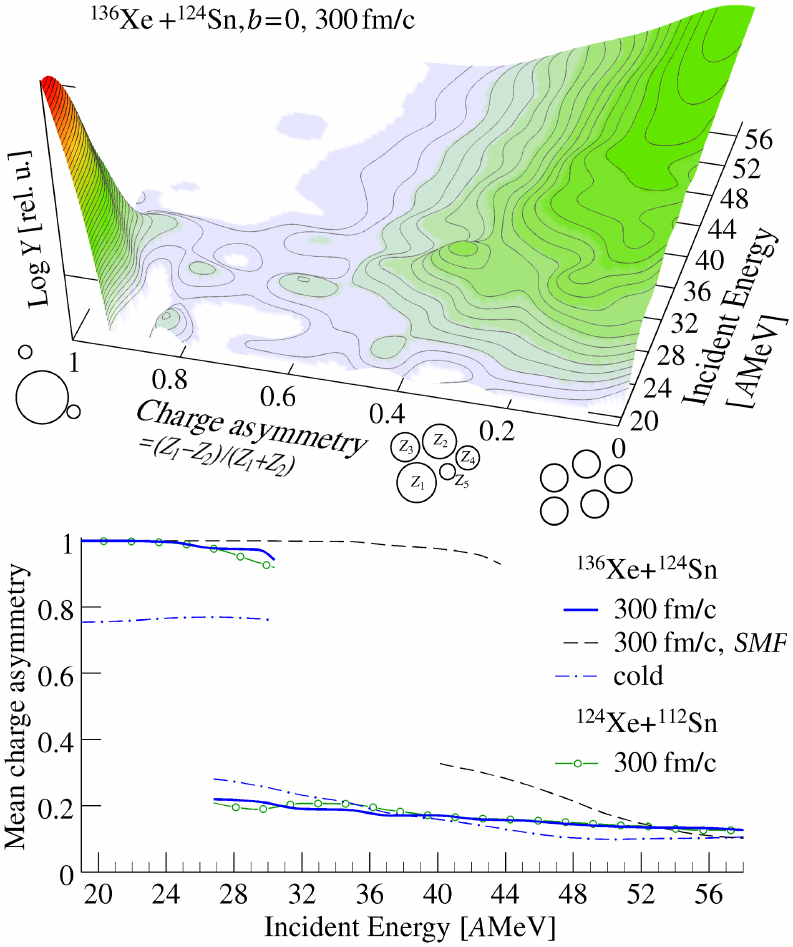}
\end{center}\caption
{
	Top. Survey of the reaction mechanism:
distribution of charge asymmetry $\alpha$ as a function of the incident energy 
in $^{136}$Xe$+^{124}$Sn ($b\!=\!0$) at $300$~fm/c.
	Bottom. Corresponding mean values.
Additional calculation for the system $^{124}$Xe$+^{112}$Sn,
	and simulation of the heaviest system within the SMF approach
and after secondary decay are also displayed.
	The cooling of the hot system (300fm/c) is undertaken by the use of the decay model Simon~\cite{Durand1992}.
}
\label{fig_Zasymm}
\end{figure}

	The analyses of Fig.~\ref{fig_phase_trans} and Fig.~\ref{fig_isotopicdistributions} have the advantage of applying to earlier times, before fragment formation or when fragments do not appear in the exit channel.
	It shows that, for incident energies above the fragmentation threshold, fluctuations feed the large width of the distribution of ripple radii.
	A further analysis, directly applied to the fragments in the exit channel, is presented in Fig.~\ref{fig_Zasymm}.
	It explores the charge asymmetry $\alpha = (Z_1-Z_2)/(Z_1+Z_2)$ between the largest fragment $Z_1$ and the second largest fragment $Z_2$, emitted in each event as a function of the incident energy.
	This observable illustrates that the reaction mechanism changes from incomplete fusion (charge asymmetry close to unity) to fragmentation (very small charge asymmetry) discontinuously, passing through a bombarding-energy range, around $28\,A$MeV, where  both mechanisms coexist.
	This latter range where the transition between incomplete fusion and fragmentation arises corresponds to the study of Fig.~\ref{fig_phase_trans}.
	At this transition energy we observe a wide, but discontinuous, bunch of trajectories and a bimodal behaviour of $\alpha$, indicating that either the system recompacts or it breaks up into several pieces of similar size.
	Intermediate charge-asymmetry values are never populated.   
	It should be noted that this variety of configurations is related to the same ``macroscopic'' initial conditions, i.e. the same beam energy and one unique value for the impact parameter.  
	These features, often discussed in the context of the thermodynamics of phase transitions and bimodal behaviour of the order parameter~\cite{Chomaz2000,Chomaz2003,Bonnet2009,Pichon2006}, are observed here as a result of the fragmentation dynamics, governed by the spinodal decomposition mechanism.
	As a consequence of the presence of fluctuations in the dynamical trajectories, the amount of light-particles early emitted may vary from one event to the other, leading to energy fluctuations in the fragmenting system.  
	It follows that this latter behaves as in a thermal bath and, at the transition energy, it oscillates between two configurations which are energetically favourable.    
	The same behaviour is also found in the corresponding SMF approach, but at larger incident energy.
	After secondary decay, the transition pattern is not washed out if events producing large fission residues are removed, and it could be searched in experimental data relative to the same systems~\cite{Moisan2012}.  
%
%

	The result of this study does not challenge the validity of thermodynamical analyses evidencing the occurrence of bimodal behaviour in nuclear fragmentation~\cite{Bonnet2009,Pichon2006}. 
	Indeed a large fraction of the available phase space is populated through the spinodal decomposition mechanism~\cite{Borderie2008}, thus legitimating the use of thermodynamical equilibrium concepts. 
	This finding provides a possible explanation, based on the occurrence of dynamical instabilities, of the origin of trajectory bifurcation and bimodality.
	We found this phenomenology in the bimodal behaviour of quantities related to fragmentation observables in the case of relatively low energies and head-on collisions.
	As bimodality has not been searched so far in this energy-centrality conditions, the present results are a suggestion for future experimental research.

	In the following, we apply both SMF and BLOB models to a highly constraining phenomenology: the low-energy threshold for multifragmentation which has been defined above.

	\subsubsection{Onset of fragmentation}
\begin{figure}[b!]\begin{center}
	\includegraphics[angle=0, width=.7\textwidth]{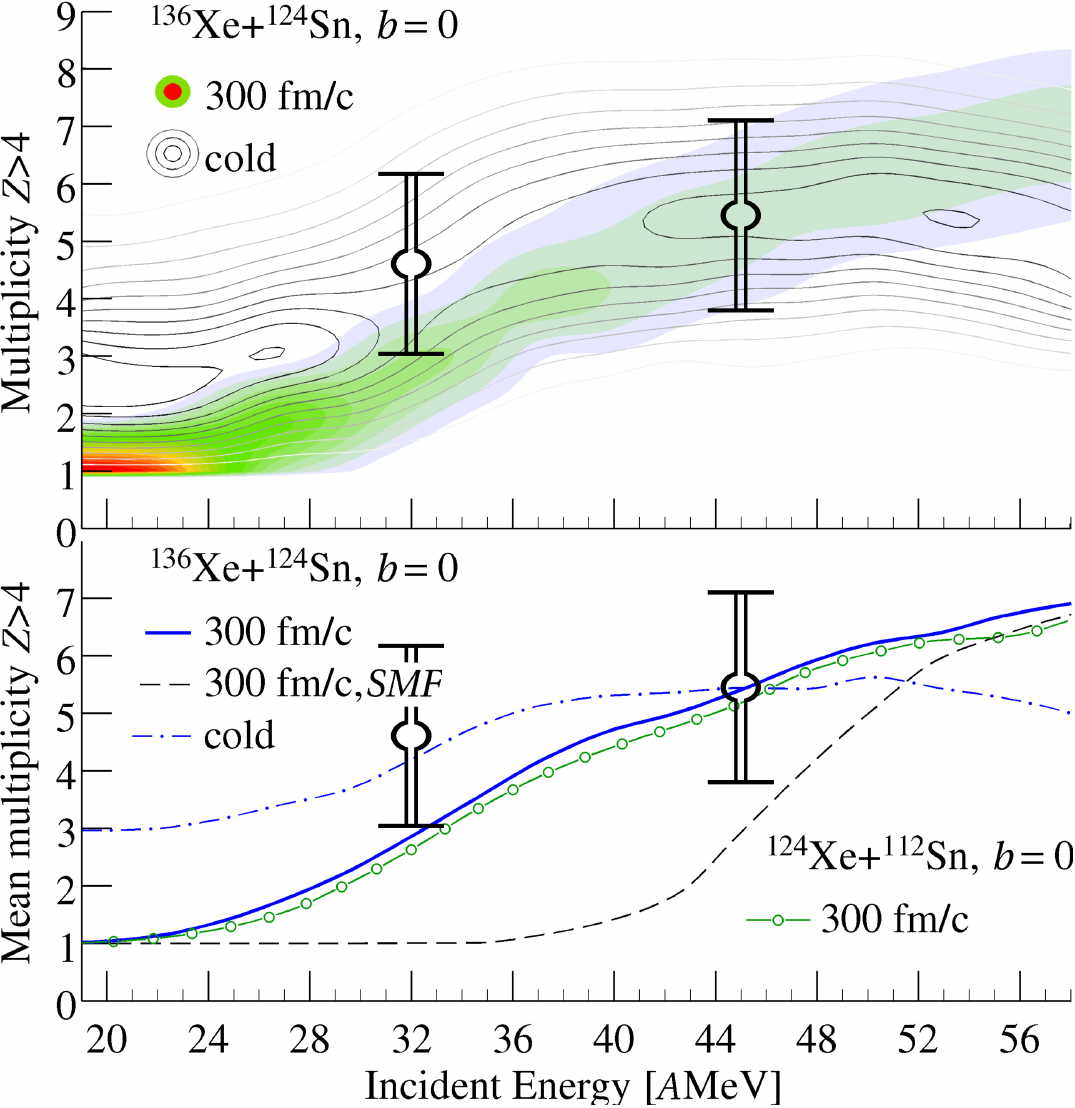}
\end{center}\caption
{
	Top. Multiplicity distribution of fragments with $Z\!>\!4$ 
as a function of the incident energy in $^{136}$Xe$+^{124}$Sn ($b\!=\!0$) 
at $300$ fm/c (filled contours) and after secondary decay (contour lines),
	Corresponding experimental data from Indra experiments~\cite{Moisan2012,Ademard2014} are added for comparison, giving average (symbols) and variance of a gaussian fitted to the multiplicity distributions (bars).
	Bottom. Corresponding mean values; 
additional calculation for the system 
$^{124}$Xe$+^{112}$Sn is also shown.
	The heaviest system is also simulated within the SMF approach.
}
\label{fig_M_full}
\end{figure}

	Having verified in \textsection~\ref{ch_inhomogeneities} that Langevin fluctuations show up with the expected properties, we study the low-energy threshold for multifragmentation by letting vary the incident energy over a large interval, from 19 to $58\,A$MeV.
	The corresponding evolution of the fragment multiplicity, calculated by counting all fragments (see \textsection~\ref{BLOB_opensystems} about fragment identification) with charge number $Z\!>\!4$ standing out at $300$ fm/c, is shown in Fig.~\ref{fig_M_full}.
	The general behaviour that BLOB traces is that a transition from incomplete fusion (just one nucleus observed) to multifragmentation occurs between 20 and $30\,A$MeV in $^{136}$Xe$+^{124}$Sn.
	The less neutron rich system $^{124}$Xe$+^{112}$Sn, also simulated, presents the same evolution, but retarded with a shift of about $2\,A$MeV to larger bombarding energies.
	Indeed the corresponding IMF multiplicity is smaller, in agreement with experimental findings~\cite{Moisan2012}.
	The same observable is also deduced at asymptotic time, by letting the system cool down after $300$ fm/c (where the excitation is around $3.5$MeV per nucleon) through a process of sequential evaporation, simulated in-flight in the swarm of light ejectiles and fragments (for this purpose, the model SIMON~\cite{Durand1992} was used).
	In the cold system the fragment multiplicity results increased by secondary decay up to around $45\,A$MeV and levels off for larger bombarding energies due to a more prominent decay into light fragments.
%
%
%

\begin{figure}[t!]
\begin{center}
	\includegraphics[angle=0, width=.7\textwidth]{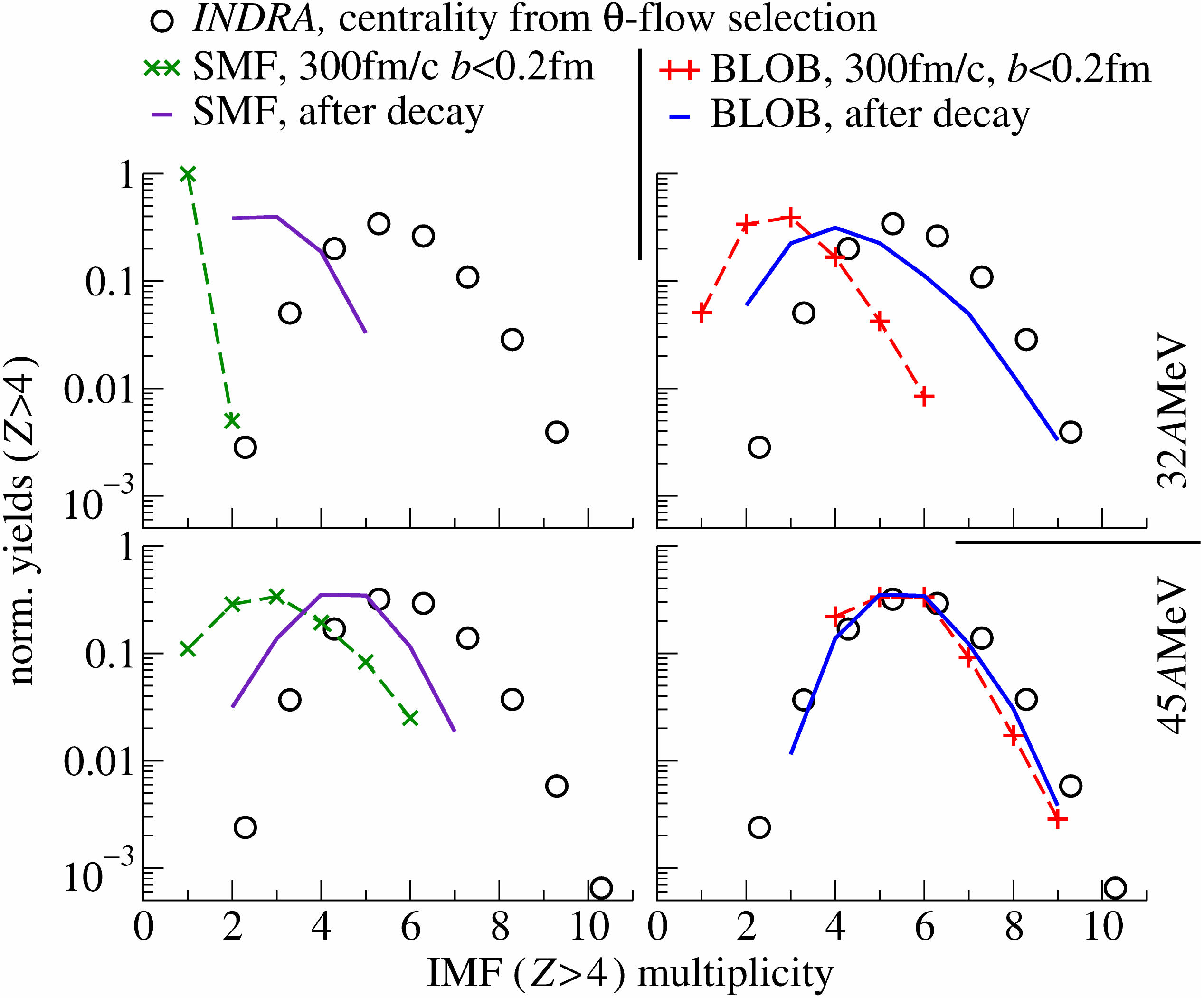}
\end{center}\caption{
\label{fig_M_spectra}
	BLOB and SMF simulations: IMF ($Z>4$) multiplicity distributions for the systems $^{136}$Xe$+^{124}$Sn at 32 and 45$A$MeV, for a selection of central collisions at 300fm/c and after secondary decay.
are compared to experimental data from Indra~\cite{Moisan2012,Ademard2014}.
}
\end{figure}

	The system evolves from a dominance of incomplete fusion around $20\,A$MeV to a full multifragmentation pattern. 
	This result is in close accordance with experimental observations addressed to the same system, where multifragmentation is already seen above $30\,A$MeV~\cite{Moisan2012}.
	On the other hand, if the BLOB approach is replaced by a SMF calculation with corresponding mean-field parameters, the transition moves above $40\,A$MeV.
	Two experimental distributions from INDRA~\cite{Moisan2012,Ademard2014} around the multifragmentation threshold (32$A$MeV) and in full multifragmentation regime (45$A$MeV) are shown in Fig.~\ref{fig_M_full} to indicate that the BLOB simulation is quantitatively consistent.
	The same distributions are shown in more details in Fig.~\ref{fig_M_spectra}, where the difference between SMF and BLOB is tested for the hot and the cold system. 
%
%

	We conclude that, beside a possible additional amplification from spinodal modes, the implementation of fluctuations in full phase space (as in BLOB) improves significantly the reaction dynamics, with respect to projecting fluctuations on the configuration space (as in SMF).
	Due to the enhanced formation of inhomogeneities, the BLOB approach is able to reproduce experimental observations related to the onset of multifragmentation better than previous mean-field-based models.
	Moreover, fluctuations have the effect of anticipating the formation of fragments.
	Indeed, the fraction of energy spent in light-particle emission reduces in favour of a larger contribution to the developing of inhomogeneities and of a more explosive dynamics, which drives the separation of those inhomogeneities into fragments.	
	This should reflect also in experimental observables like charge and velocity distributions~\cite{Napolitani2017}. 
	On the contrary, for small bombarding energies (below $30\,A$MeV), due to insufficient kinetic energy, the mean field often succeeds to coalesce all the inhomogeneities or part of them into a compact system.

\section{Semicentral to semiperipheral collisions
\label{sec_peripheral}}

%
%

	We described the phenomenology of central collisions as related to the transition from fusion to multifragmentation, triggered by the liquid-gas phase transition in two-component nuclear matter.
Isoscalar and isovector properties of the fragments were in this case associated to the interplay between spinodal decomposition and isospin distillation.
	In semicentral or semiperipheral collisions at Fermi energies the characteristic mechanism is a transition from fusion or multifragmentation to binary channels depensing on the incident energy, as a function of the characteristics of the nuclear interaction and of the involved isospin component (refs.~\cite{Colonna1998c,Amorini2009} give two didactic examples).
	In this transition an intermediate mechanism may appear where a neck is produced~\cite{Baran2012,Rizzo2014}.
	This latter, when it is not reabsorbed by the projectile- or target-like nucleus, separates as one individual fragment or disintegrates into few IMFs~\cite{Lionti2005,DiToro2006}.
	These processes are especially interesting because the isospin properties of the projectile, target and neck fragments are a probe of the symmetry energy through the mechanism of isospin transport.

	This mechanism is fundamental in many general frameworks, from the dynamics of nuclear collisions to the cooling of compact stars.
	It is reflected in the currents of neutrons $\boldsymbol{j}_n$ and protons $\boldsymbol{j}_p$, which are inversely proportional to the spacial gradients of the chemical potentials of the corresponding species: each species migrates towards the lowest value of its chemical potential~\cite{Baran2005b,Colonna2006,Ditoro2006bis,Rizzo2008bis} so that  
\begin{equation}
	\boldsymbol{j}_n \propto -\kappa\;\nabla\mu_n \;\;;\;\;\; 
	\boldsymbol{j}_p \propto -\kappa\;\nabla\mu_p \;.
\label{eq_grad_mu}
\end{equation}
	The chemical potential gradient can be separated into two contributions, one governed by 
a spacial gradient of isospin $I$, called ``diffusion'', the other by a spacial gradient of total density $\rho$, called ``migration'' or ``drift''. 
	Within a simple hydrodynamical picture (the detailed steps can be found in ref.~\cite{Baran2005}), the difference between the currents of neutrons and protons can be expressed by defining two transport coefficients, the isovector diffusion coefficient $D_{\textrm{iv}}^{I}$ and the isovector migration coefficient $D_{\textrm{iv}}^{\rho}$, which are strictly connected to the strength of the symmetry energy and of its derivative, respectively:
\begin{equation}
	\boldsymbol{j}_n -\boldsymbol{j}_p 
	= - D_{\textrm{iv}}^{\eta}\nabla I + D_{\textrm{iv}}^{\rho}\nabla\rho 
\propto
		\underbrace{S(\rho)\nabla I}_{\textrm{diffusion}} +
		\underbrace{\frac{\partial S(\rho)}{\partial\rho}I\nabla\rho}_{\textrm{migration}}
\;.
\label{eq_isospintransport}
\end{equation}
	Hence, in presence of asymmetry gradients (diffusive processes) we test essentially the strength of the symmetry energy 
while, when density gradients are encountered along the dynamical path, we observe `isospin migration' towards the low density regions, ruled by the derivative of the symmetry energy. 

	Alternatively, when the mechanism results binary, interesting effects of isospin diffusion can be studied~\cite{Tsang2004,Chen2005,Shetty2007,Galichet2009,Galichet2009b,Napolitani2010,Hudan2014}.

\subsection{Neck}
%
%
%
\begin{figure}[b!]\begin{center}
	\floatbox[{\capbeside\thisfloatsetup{capbesideposition={left,top},capbesidewidth=.35\textwidth}}]{figure}[\FBwidth]
	{\caption{
Bottom. Survey of average fragment multiplicity in peripheral collisions for BLOB and SMF simulations for the system $^{136}$Xe$+^{124}$Sn at 25, 32 and 45$A$MeV at 300fm/c.
The 32$A$MeV system is also shown after secondary decay.
Top. corresponding mass distributions at 32$A$MeV for the hot system.
	}
	\label{fig_peripheral_neck}}
	{\includegraphics[angle=0, width=.6\textwidth]{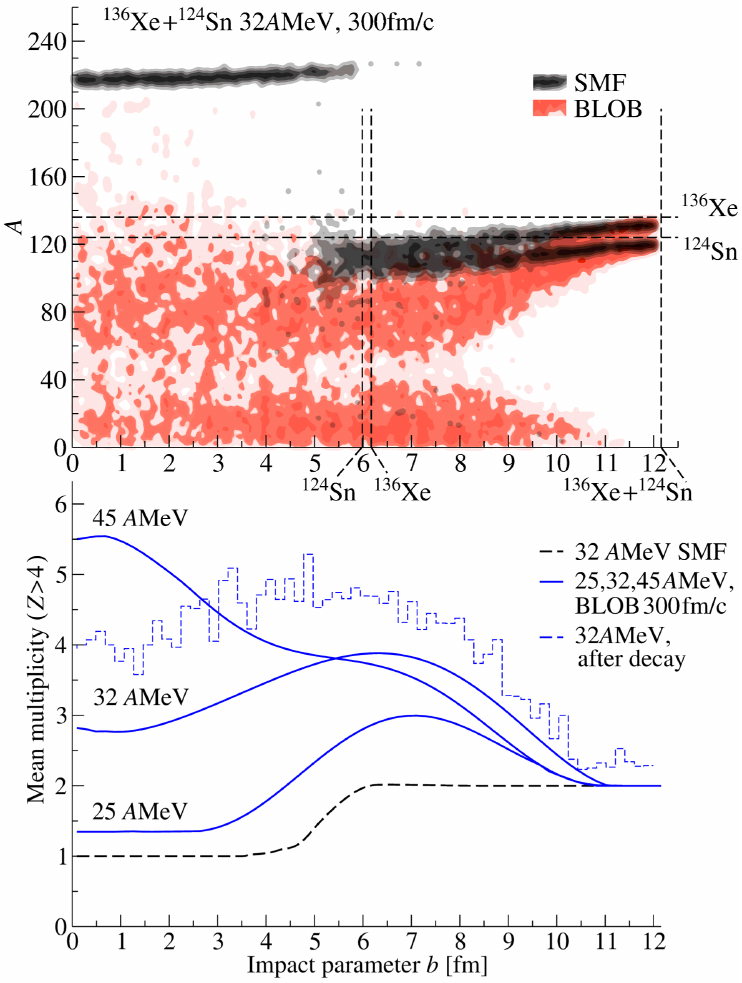}}
\end{center}
\end{figure}

	Fig.~\ref{fig_peripheral_neck} extends the survey of Fig.~\ref{fig_M_full} to all impact parameters for the same system $^{136}$Xe$+^{124}$Sn.
	The top panel of Fig.~\ref{fig_peripheral_neck}, tracks the mass of the IMFs registered at 300fm/c for the 32$A$MeV incident energy.
	While SMF jumps from fusion to binary splits, BLOB produces IMF at all impact parameters $b$, reaching a maximum of production above $b=6$~fm for fragments with $A<40$.

	The large density gradients induced by the neck formation offer favoured conditions for studying isospin migration, and many investigations have been carried out within the SMF approach~\cite{Baran2005,Rizzo2008bis}.
	The improvement that the BLOB approach introduces is not directly in the description of the isovector modes, but rather in the more complete description of the isoscalar modes.
	Those latter allow for extending the mechanism of neck fragmentation down to low incident energies (below 15$A$MeV) and accessing isospin migration when the process of ternary splits is the most relevant for semiperipheral impact parameters.
	This is illustrated in the bottom panel of Fig.~\ref{fig_peripheral_neck}, where ternary and quaternary splits (related to neck formation) at 300fm/c are shown to dominate the semiperipheral impact parameters below 32$A$MeV.
	A maximum of IMF multiplicity (quaternary splits) is reached at 32$A$MeV; on the contrary, at this same energy, the rate of ternary breakups in SMF is still negligible.

\subsection{Isospin transport}

%
%
\begin{figure}[b!]
\begin{center}
	\includegraphics[angle=0, width=.53\textwidth]{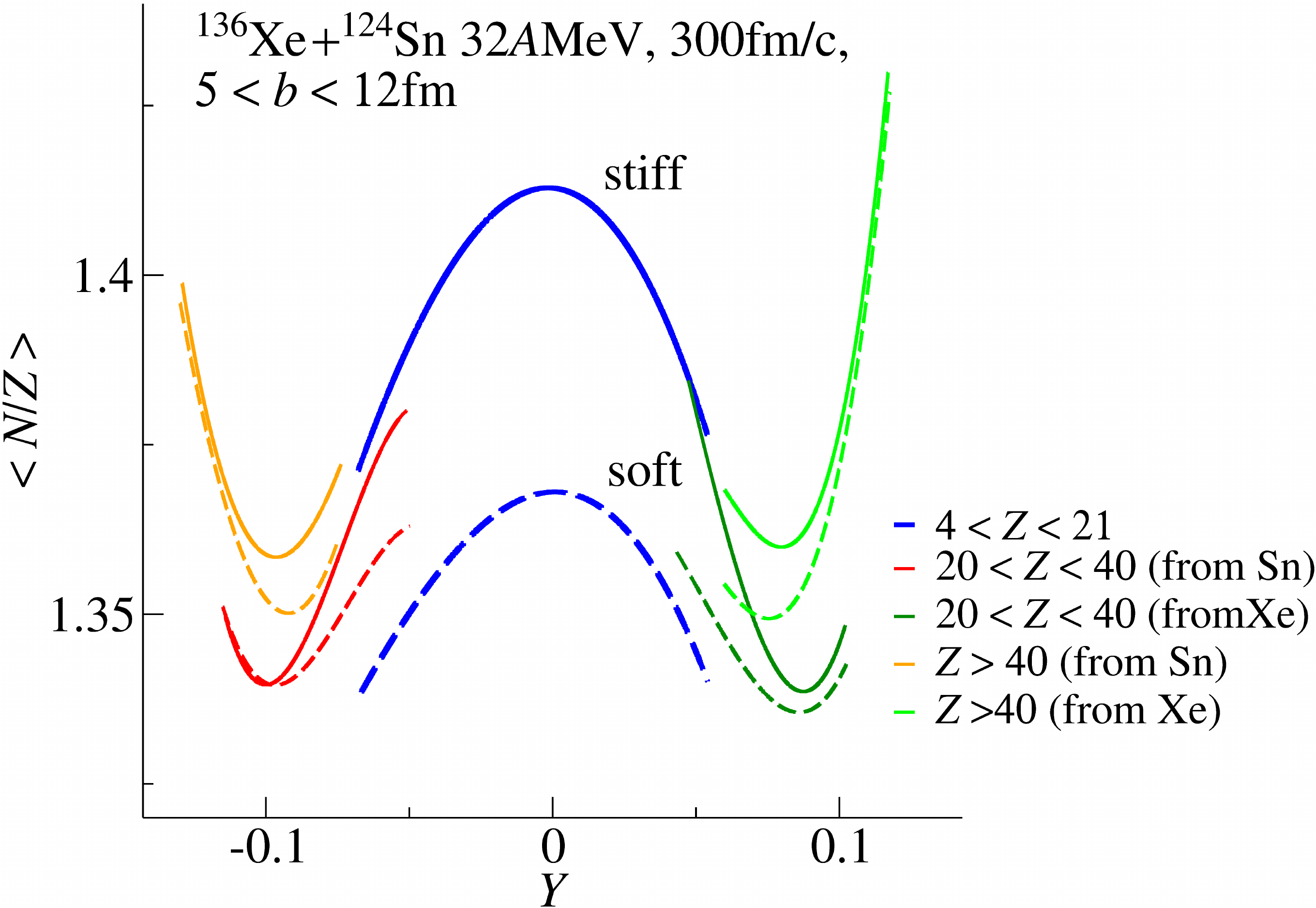}\hspace{1ex}
	\includegraphics[angle=0, width=.44\textwidth]{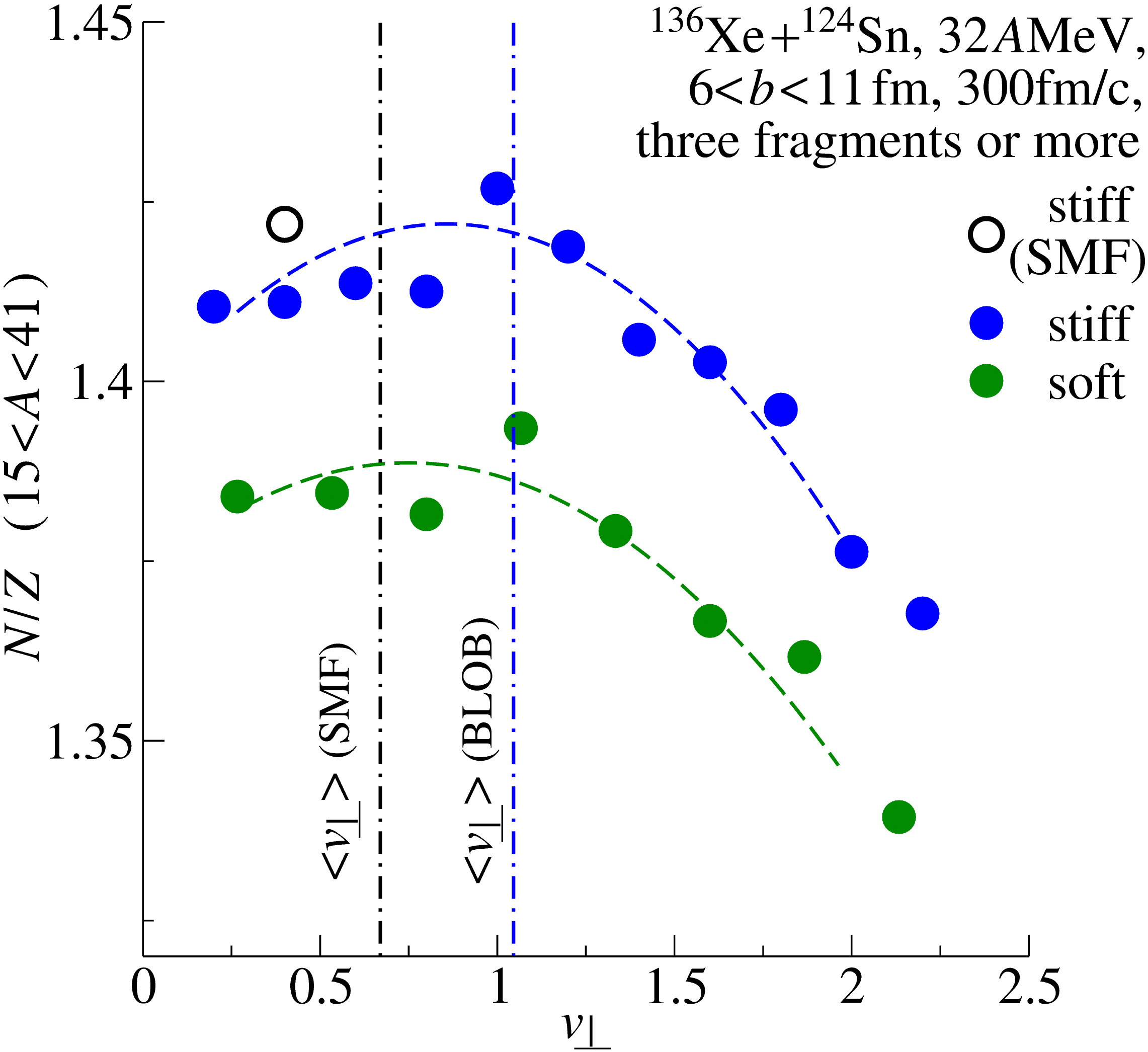}
\end{center}\caption{
\label{fig_Neck_NZ}
Left. 
	Migration and diffusion in $^{136}$Xe$+^{124}$Sn at 32$A$MeV for $b>5$ at 300fm/c.
BLOB simulation. Study of isotopic content of the midrapidity region 
$^{136}$Xe$+^{124}$Sn at 32~$A$MeV selecting different element intervals, for a stiff and a
soft form of the symmetry energy potential component.
Right. 
	Isotopic content of neck fragments at 300fm/c, identified in the region of maximum production, 
as a function of the transverse velocity component with respect to the quasi-projectile--quasi-target axis for a stiff and a soft form of the symmetry energy potential component. 
	Dashed lines are fits; vertical lines indicate mean transverse velocity components for SMF and BLOB.
}
\end{figure}
%

	The region of incident energy and impact parameters where fragments are predominantly produced in the midrapidity region (i.e. at intermediate rapidities with respect to the projectile and the target) is further studied in the left panel of Fig.~\ref{fig_Neck_NZ}, where the average isotopic content $<N/Z>$ of the IMFs is studied as a function of rapidity without any selection of mechanism (i.e. of fragment multiplicity): the different action of dissipation (in heavy residues) or migration (in lighter fragments) on the isotopic content signs different time lengths available for the process of isospin transport according to the form of the nuclear interaction.
	For this system, the midrapidity region is more affected by isospin transport which results connected to the process of migration towards the neck.

	Such situation is specifically addressed in the right panel of Fig.~\ref{fig_Neck_NZ}, by selecting ternary and quaternary events at midrapidity: the isotopic content $N/Z$ of the IMFs is studied as a function of the transverse velocity component with respect to the quasi-projectile--quasi-target axis for a stiff and a soft form of the symmetry-energy potential component.
	The dramatically reduced production of IMF in SMF, compared to BLOB, is related to a much smaller mean transverse velocity component, as indicated by the vertical lines in Fig.~\ref{fig_Neck_NZ}.
	The isovector features are expected from the properties of the employed interaction.
	A stiff form of the symmetry-energy potential component gives a less repulsive dynamics at low density enhancing the isospin migration towards the diluted neck region, resulting into a larger isospin content.
	Smaller transverse velocity components are related to fragment configurations more aligned along the quasi-projectile--quasi-target axis, exploiting a longer interval of time available for the process of isospin migration, and finally favour more neutron rich fragments. 
	The neutron enrichment reduces for less aligned configurations and larger transverse velocities.
	This study is consistent with the experimental investigation reported in ref.~\cite{DeFilippo2005,DeFilippo2005bis,DeFilippo2009,DeFilippo2014}.

\subsection{Quaternary splits
\label{subsec_quaternary}}
%
%
\begin{figure}[b!]
\begin{center}
	\includegraphics[angle=0, width=.49\textwidth]{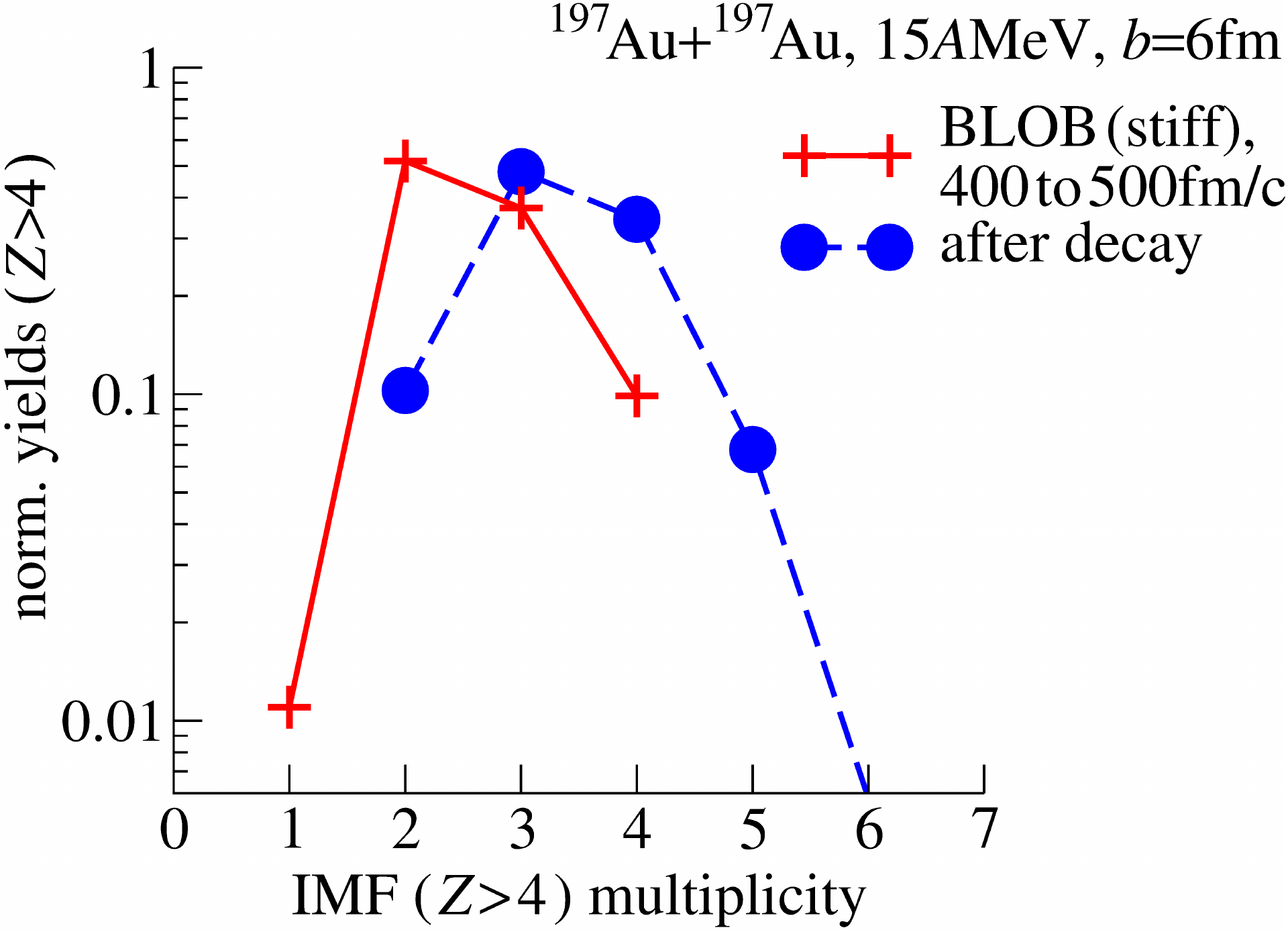}
	\includegraphics[angle=0, width=.49\textwidth]{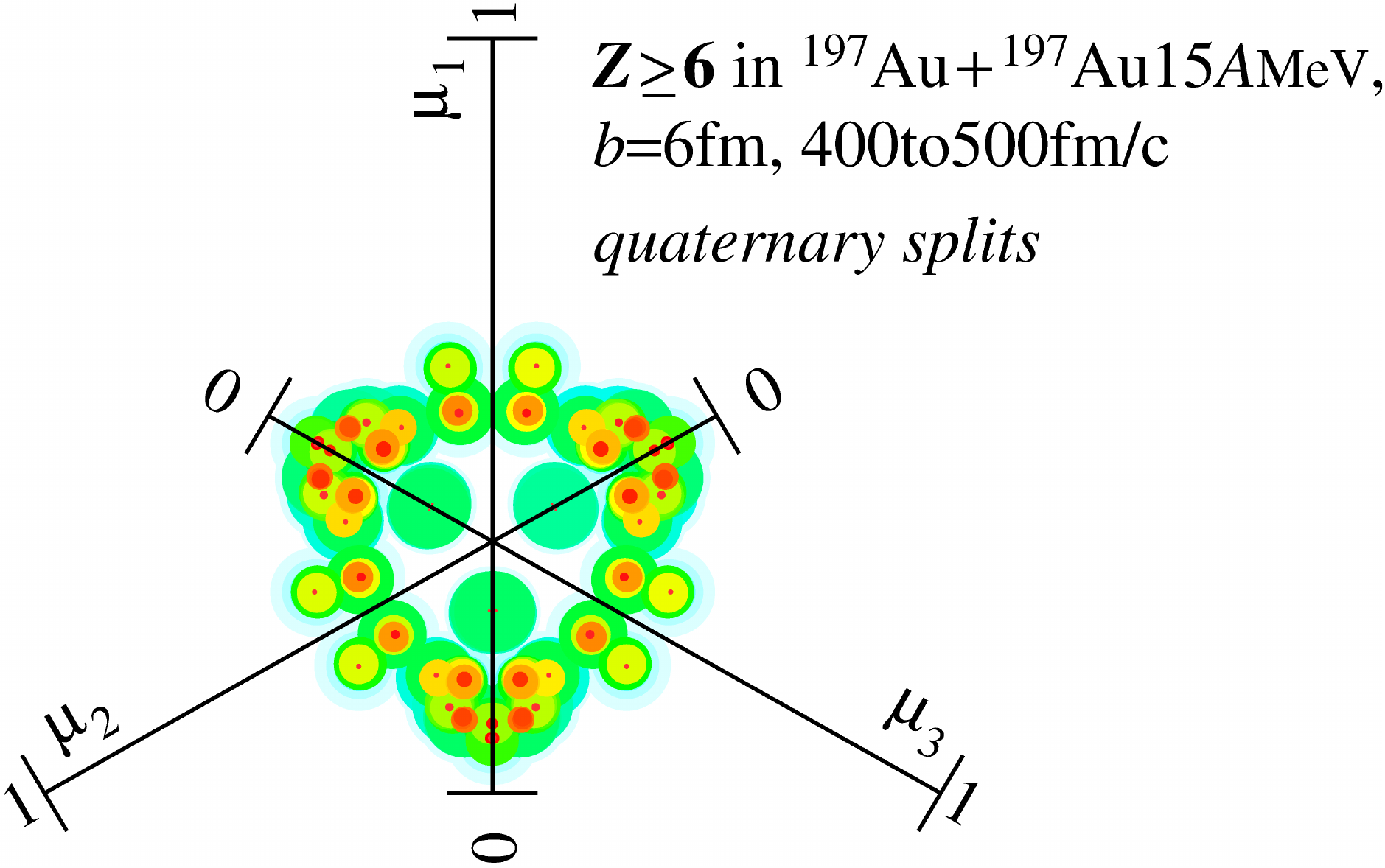}
\end{center}\caption{
\label{fig_quaternaryAuAu}
Left. Multiplicity of fragments in the system $^{197}$Au$+^{197}$Au at 15 $A$MeV 
calculated with BLOB for hot systems in a time window ranging from 400 and 500fm/c
(the stopping time is set after the last split).
Right. Dalitz plots illustrating the corresponding fragment configuration for selections of quaternary splits.
These results can be closely compared to the experimental study of ref.~\cite{SkwiraChalot2008}.
}
\end{figure}

	Fig.~\ref{fig_quaternaryAuAu} is the study of a system, $^{197}$Au$+^{197}$Au at 15 $A$MeV (measured with CHIMERA, LNS, Catania) which was found to lead to quaternary splits in a recent experiment~\cite{SkwiraChalot2008}.
	It was found that fragments were produced in an aligned configuration and with a kinematics as fast as the neck mechanism which, at these energy characterises ternary events.

	This mechanism is not accessible within an SMF approach, so that refined analysis methods are necessary to extrapolate the fragmented configurations from density distributions~\cite{Rizzo2014,Colonna2015}.
	Neck fragmentation at 15 $A$MeV is on the other hand a favoured channel in BLOB simulations, as shown in Fig.~\ref{fig_quaternaryAuAu}, left: ternary and quaternary splits have large yields already in the excited system and they become dominant after secondary decay, showing an almost equal probability for either ternary or quaternary splits. 
	The Dalitz plot, evaluated for quaternary splits in the hot system is shown in Fig.~\ref{fig_quaternaryAuAu}, right, recalling very closely the experimental results reported in ref.~\cite{SkwiraChalot2008}.
	Also as deduced from the experiment, ternary or quaternary splits are related to the same conditions and show up for the same range of impact parameters.




\section[Violent size-asymmetric collisions at Fermi energies]{Violent size-asymmetric collisions at\\ Fermi energies
\label{sec_neck}}

	We address in this section situations which are at the edge of the mechanisms we analysed above.
	At rather large incident energy, peripheral collisions evolve towards more complex mechanisms with more than one fragment at midrapidity~\cite{Baran2012}.
	When, in additions, peripheral collisions involve nuclei of asymmetric size, 
the mechanism may become largely incompatible with equilibrium conditions.
	These studies are exclusively possible within a stochastic dynamical approach, as they go beyond situations where comparisons can be made with statistical approaches at equilibrium.
	At the same time, hydrodynamic and kinematic conditions for the neck mechanisms are simply suppressed because light projectiles do not persist as a quasi-projectile so that the outcome of the collision is a heavy target remnant and a set of intermediate-mass clusters.
	Nevertheless, configurations which may be confused with neck fragmentation are observed experimentally~\cite{Nyibule2016}, like several aligned fragments.

	As an example we simulate the system $^{48}$Ca$+^{124}$Sn at 45$A$MeV (measured with CHIMERA, LNS, Catania) for impact parameters ranging from 0 to $7$~fm (with the usual prescription of this chapter and an asy-stiff form of the symmetry energy).
	Multifragmentation-like events, in competition with fusion characterise central collisions with impact parameter $b\le2$fm. 
	Beyond $b=3~fm$, collisions display similar behaviours quite independently on the impact parameter: they can reach an IMF multiplicity of three to four fragments (target-remnant included) which saturates at around 300 to 340~fm/c. 
	The largest production of small IMF is for $b$=3 to 4~fm; in this case, a time window extending from 200~fm/c to 340~fm/c was used for the identification of fragments.

\subsection{nuclear ``jets''}
%
%
\begin{figure}[b!]
\begin{center}
	\includegraphics[angle=0, width=1\textwidth]{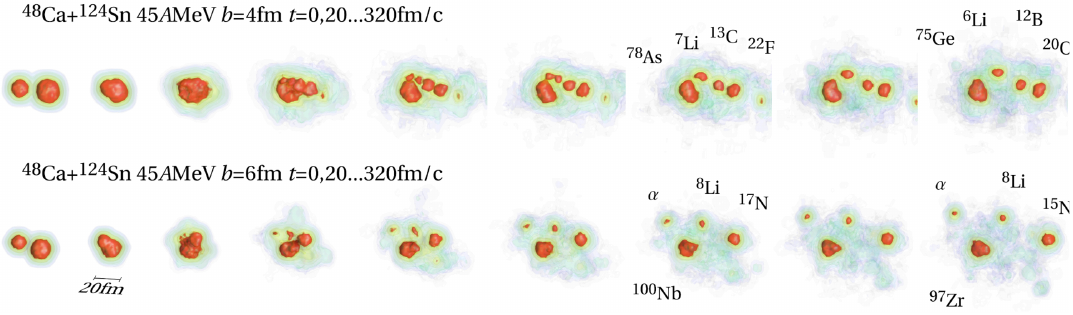}
\end{center}\caption{
\label{fig_jets_animation}
Study of two sample event where four fragments are found in the exit channel.
top. $b=4$fm. The process passes through the formation of an arc (as a remnant of $^{48}$Ca) which is successively strained and broken in four pieces. Two fragments merge back together.
bottom. $b=6$fm. The process is closer to a multifragmentation event. Some light fragments come from the target as well. 
}
\end{figure}
%
%
%
\begin{figure}[b!]
\begin{center}
	\includegraphics[angle=0, width=.7\textwidth]{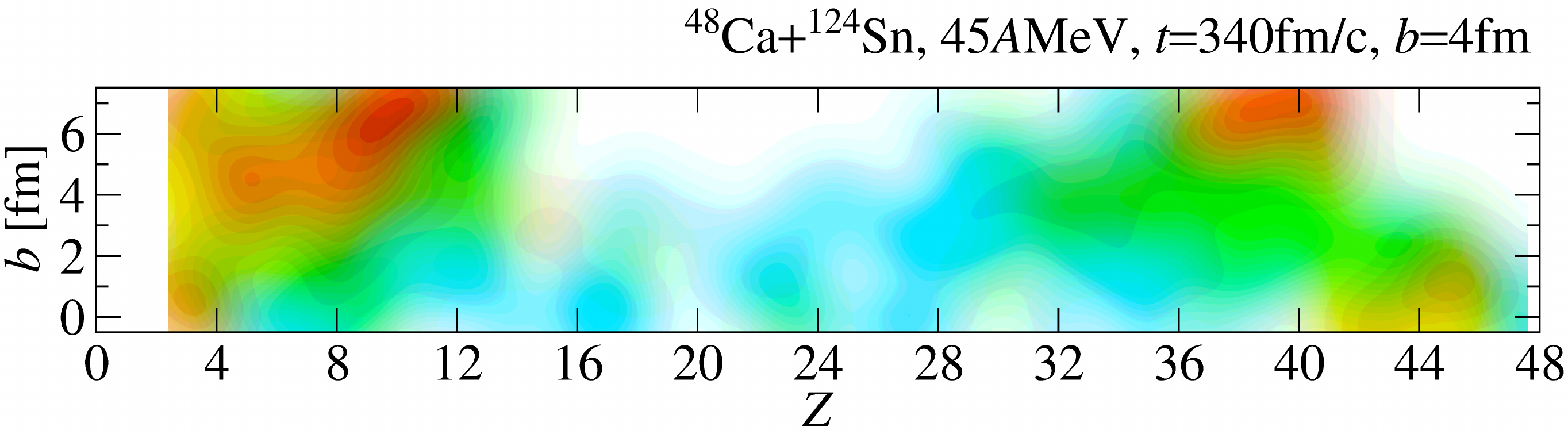}\\
	\includegraphics[angle=0, width=.9\textwidth]{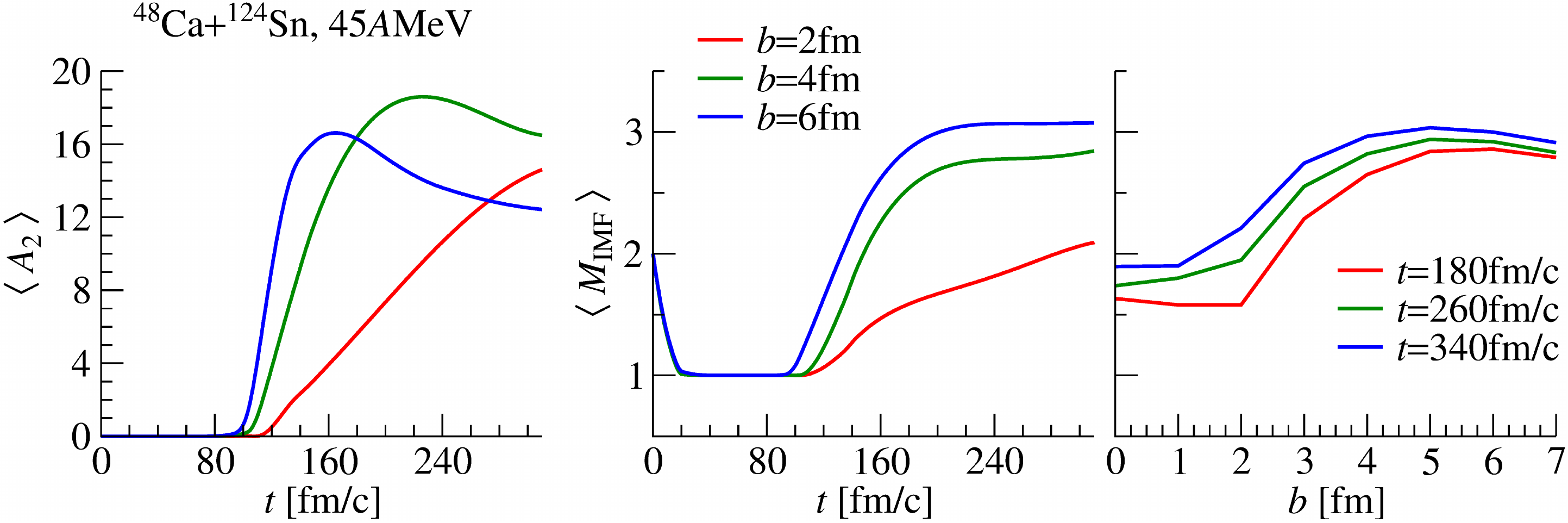}\\
\end{center}\caption{
\label{fig_production_jet}
Top. Production ($Z$ of any produced fragment) as a function of $b$.
Nuclides are produced also below the most probable spinodal IMF size.
Bottom. 
Evolution of the second largest fragment in the system, corresponding to the largest fragment from the disassemby of $^{48}$Ca) as a function of time for three impact parameters (left),
evolution of the average multiplicity of fragments (including the target remnant) as a function of time for three impact parameters (centre) and as a function of impact parameter for three different times (right).
Contact time occurs at around 20fm/c.
}
\end{figure}

%
%
\begin{figure}[b!]
\begin{center}
	\includegraphics[angle=0, width=.5\textwidth]{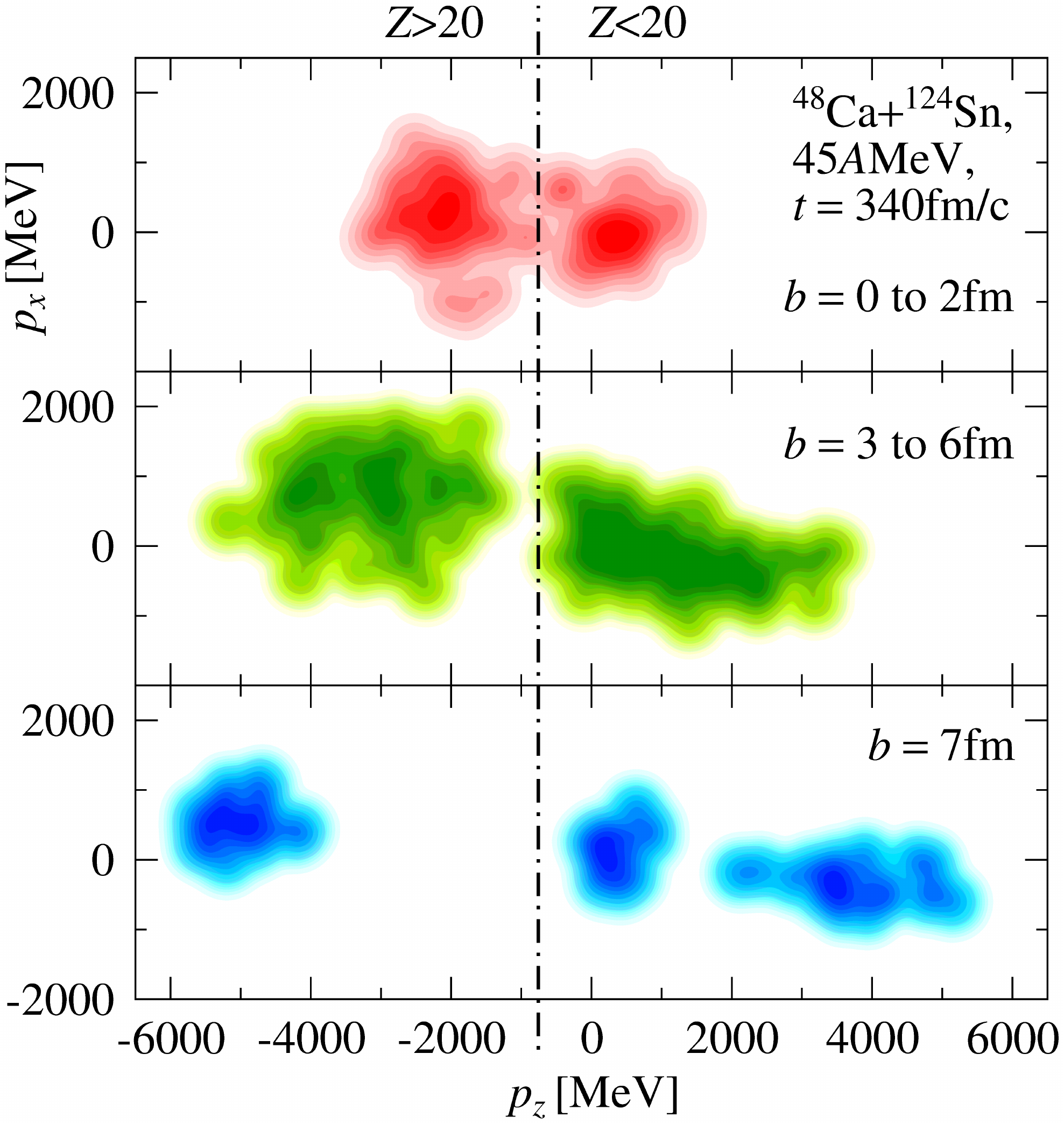}\\
	\includegraphics[angle=0, width=.85\textwidth]{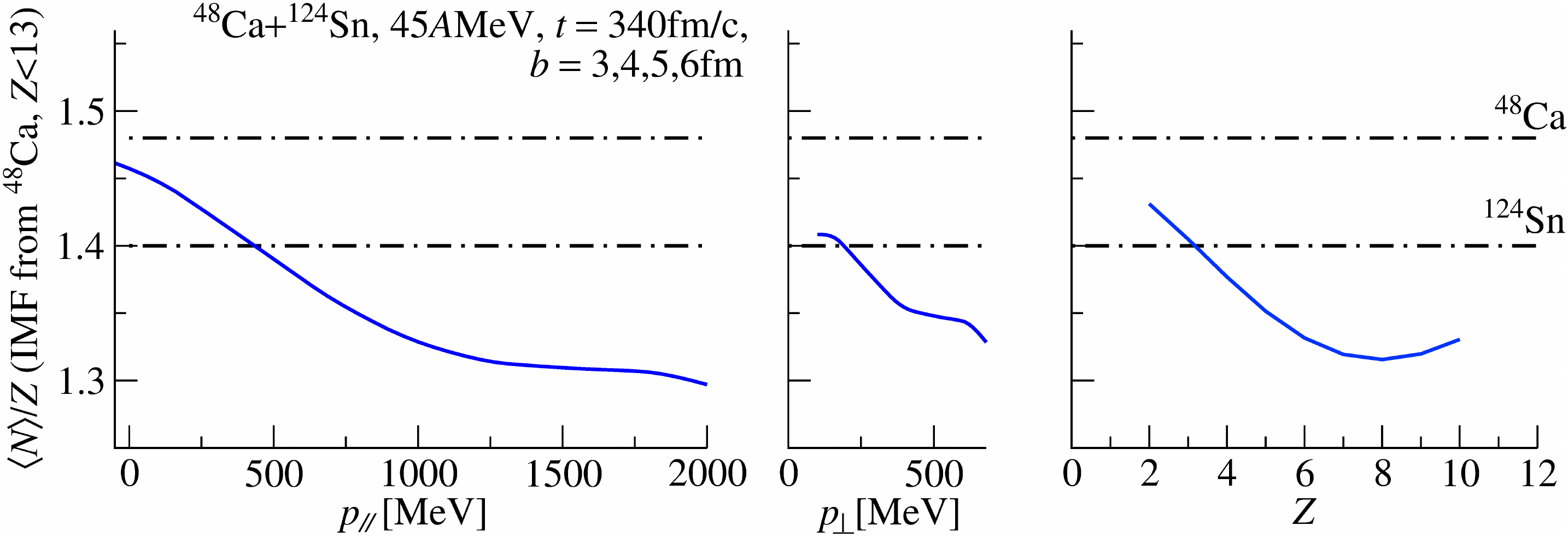}
\end{center}\caption{
\label{fig_kinematics_NZ_jet}
Top. Projections of momentum vectors on the reaction plane along the beam axis ($p_z$) and along the
impact-parameter distance ($p_x$) for all produced fragments. Colours indicate impact-parameter selections.
Bottom. Study of isotopic component of forward emitted IMF selected for $3\le Z\le 6$ from the fragmentation of $^{48}$Ca as a function of $p_{||}$ (left), of $p_{\perp}$ (centre) and of $Z$ (right).
Due to the total disintegration of the projectile and the impossibility of defining a target-projectile axis, $p_{||}$ is taken along the beam direction $p_z$.
}
\end{figure}

	In asymmetric semiperipheral collisions at Fermi energy, four fragments can be obtained in the exit channel with non-negligible probability.
	Two significant simulated examples illustrate such mechanism in Fig.~\ref{fig_jets_animation}.
	The corresponding exit channel might be confused with an aligned neck fragmentation but it is rather the effect of the total disintegration of the light projectile. 
	The ensemble of IMF is organised along a large arc which keeps them almost aligned. 
	They are however not aligned with the heavy target due to the rotation effect imparted in the collision phase.

	Fragments emerge as potential concavities very early, immediately after the compression phase like in central collisions and not from the breaking of a neck (which, in this case, is rather an arc). 
	In reality the process is the opposite of the splitting of a neck.
	Fragments form in a very short time span along an arc from the residue of the light projectile
according to a size hierarchy.
	Immediately after formation they might partially re-coalesce together.
	Compared to central collisions, IMF have rather small size, smaller that the typical Oxygen-Neon range in spinodal fragmentation.
	These features, related to the IMF production are studied in Fig.~\ref{fig_production_jet}.

	Very different from the other mechanisms previously discussed for heavy-ion collisions at Fermi energies, these exit channels recall features of the hydrodynamics of liquid jets, like Plateau-Rayleigh instabilities developing along columnar structures.
	It is possible that, seemingly, ``nuclear jets'' might develop similar hydrodynamic behaviour from the action of binary nucleon-nucleon collision, which might have the effect of restructuring the jet into a sequence of droplets.

	These exit channels may recall features of the hydrodynamics of liquid jets~\cite{Eggers2008}.
	A difference should however be mentioned.
	Jets of classical liquids, like the very early descriptions of Fr\"olich and Griffin~\cite{Frolich1973,Griffin1976,Brosa1990} for nuclear collisions are governed by surface energy, and in particular they undergo Plateau-Rayleigh instabilities developing along columnar structures~\cite{Brosa1990}, till fractioning the jet into a sequence of droplets~\cite{Gopan2014}.
	However, the calculations discussed in this section (to be investigated further) present a very fast dynamics, fragments seem to form very early and even emerge as separate fragments right at the exit of the target nucleus.
	Thanks to this rapid dynamics they can reflect distillation features in their dependence of isospin content with size.
	Consistently with this ensemble of characteristics, the fastest process of fragment formation is a volume instability, triggered by binary nucleon-nucleon collisions at low-density.

	Those configurations are incompatible with the quaternary splits observed in 
$^{197}$Au$+^{197}$Au in \textsection~\ref{subsec_quaternary} or in $^{130}$Xe$+^{130}$Xe in \textsection~\ref{subsec_Fermiland} also when analysing isospin effects.
	In those symmetric collisions quaternary splits show the typical features of neck formation, where isospin effects depend on how long fragments are involved in the migration process: differences in the isospin component of the IMF may come from being exposed for different intervals of time, in different events, to comparable density gradients (\textsection~\ref{sec_peripheral}). 
	On the other hand, in quaternary splits occurring in $^{48}$Ca$+^{124}$Sn, isospin effects are also showing up, but they are correlated to different density gradients in the same event in the same very short interval of time.
	Features related to the isospin content and kinematics of IMF are studied in Fig.~\ref{fig_kinematics_NZ_jet}.

\section{Fluctuations and transparency at intermediate energy
\label{sec_flow}}
%
%
\begin{figure}[b!]\begin{center}
	\includegraphics[angle=0, width=.8\columnwidth]{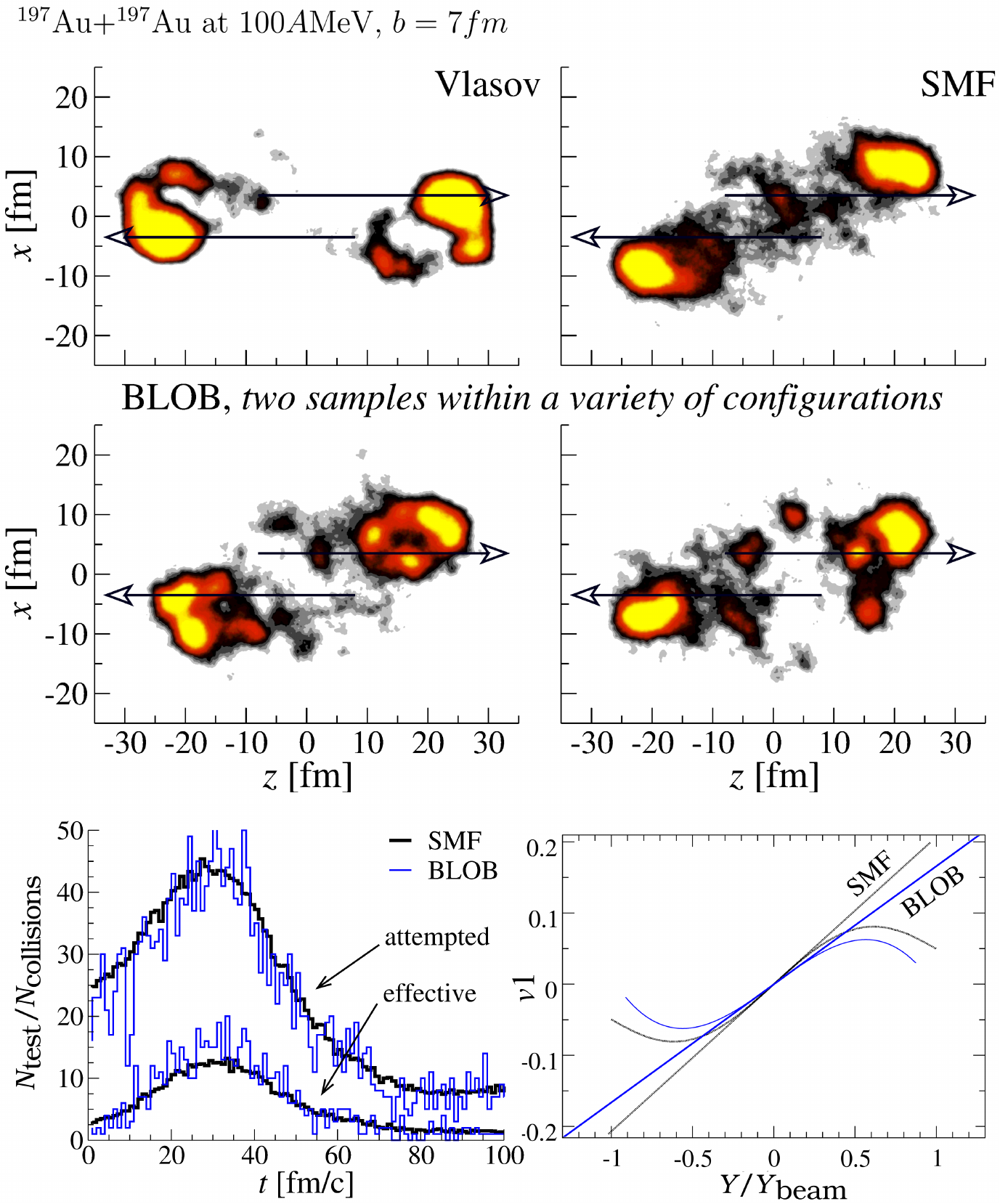}
\end{center}\caption
{
	Simulation of the collision $^{197}$Au$+^{197}$Au at $100\,A$MeV for an impact parameter of 7fm, studied at the time $t=140$ fm/c and simulated through three approaches: Vlasov (top, left), SMF (top, right) and BLOB (middle row).
	The arrows indicate the direction of the target and projectile; their origin indicate the centres of target and projectile at the initial time $t=0$ for the simulation.
	Bottom, left: attempted and effective number of collisions per nucleon calculated with SMF and BLOB in an interval of time of 1~fm/c, as a function of time.
	Bottom, right: directed flow as a function of reduced rapidity for SMF and BLOB. The corresponding slope at zero reduced rapidity is indicated.
}
\label{fig_flow}
\end{figure}

	This section, far from entering in the vast domain of intermediate energies, discusses an aspect which is strictly related to the inclusion of fluctuations in a transport model adapted to this energy range: the connection between fragment production and hydrodynamic properties like the characterisation of the flow~\cite{Reisdorf1997,ReisdorfRitter1997,Russotto2014}.


	At intermediate energy the inclusion of fluctuations has two antagonist effects: on the one hand, it enhances the fragmentation of the system, on the other hand it reduces the directed flow. 
	This effect can be studied in the comparisons of Fig.\ref{fig_flow}, for the collision $^{197}$Au$+^{197}$Au at 100 $A$MeV for an impact parameter of 7fm.
	The simulation is performed with three approaches, Vlasov, SMF and BLOB, using identical parameters for the mean field and for the two-body collision term~\footnote{For the simulation a soft equation of state with $k_{\inf}\!=\!240$ MeV was used, with a linear density dependence for the potential component of the symmetry energy (asy-stiff) and a constant isotropic cross section of 40 mb.}, i.e. for a comparable number of attempted and effective nucleon-nucleon collisions per time interval.
	The SMF approach describes the outward deflection of the trajectory imparted by the directed flow, which is absent in the Vlasov description.
	The BLOB approach exhibits a reduced directed flow with respect to SMF, because it competes with a more explosive dynamics.
	This latter, due to the Langevin fluctuations, results in a large variety of very different fragment configurations; two of those are shown, one where the fragmentation of the quasi-target and the quasi-projectile is observed (middle row, left), the other where the emitting source is situated at midrapidity (middle row, right).
	This example spots the main difference between the SMF approach and the BLOB approach when applied to intermediate energies.

	It may be observed that, quite independently on the metrics of the scattering wave packets, i.e. the extension of the wave packets in phase space, the fluctuation amplitude would no more be correct if the occupancy variance in phase-space cells is not compatible with satisfying the Pauli principle: the consequence would be the loss of consistency with the dispersion relation and some consequences on the fragmentation pattern.
	On the other hand, the metrics affects the collective effects and disregarding it, i.e. letting wave packets take too large widths in configuration space, would eventually result in a more transparent dynamics where the flow is reduced.
	A quantitative study of the flow is illustrated in the bottom row of Fig.\ref{fig_flow}.
	For a simulation where attempted and effective number of collisions per nucleon have the same mean value in SMF and BLOB (bottom, left, similar to the study of Fig.~\ref{fig_collisions_SMF_BLOB}) the larger variance which characterises BLOB and which is related to fragment formation is reflected in a smaller slope for the directed flow as a function of reduced rapidity (bottom, right).

\chapter{Conclusions}

%
%
\begin{figure}[b!]\begin{center}
	\includegraphics[angle=0, width=.6\columnwidth]{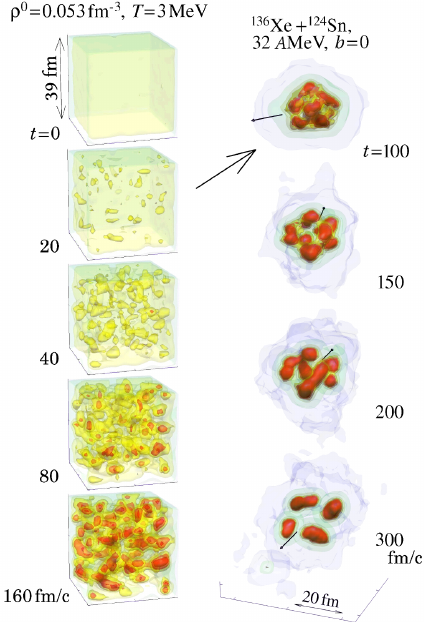}
\end{center}\caption
{
	Evolution of density landscape in configuration space in nuclear matter in a periodic box at $\rho^0\!=\!0.053$~fm$^{-3}$ (left) and in a hot open nuclear system formed in a head-on heavy ion collision of $^{136}$Xe+$^{124}$Sn at 32 $A$MeV (right); arrows indicate the beam direction. Both simulations employ the BLOB approach with the same mean-field properties, as defined in the text.
	The arrow proposes an analogy between nuclear matter and the open system in correspondence with the arising of the spinodal signal.
}
\label{fig_mottling}
\end{figure}

We reviewed some significant mechanisms among the vast phenomenology of violent nuclear reactions, with a focus on those processes which led the system to separate into fragments.
	The purpose is handling at least a qualitative and possibly sometimes a quantitative description within a transport theory which connects as many observables as possible.


	We followed the approach of the Boltzmann-Langevin equation and presented crucial steps to validate corresponding transport models applied to a fermionic system, especially when in presence of mechanical instabilities and isovector fluctuations.
	In practice, even though technically demanding, we could build a conceptually simple solution of the Boltzmann-Langevin equation for the evolution of the one-body distribution function in full phase space through a renormalisation of the two-body collision contribution: the Fermi statistics is in fact preserved for a long time and a correct isoscalar fluctuation amplitude is obtained independently of the ingredients of the numerical implementation (like the number of test particles, for instance).
	This feature is fundamental for satisfying our primary requirement of describing fragment formation in a nuclear system preserving the specific characteristics of a one-body theory, as far as the description of mean-field (spinodal) instabilities is concerned and, at the same time, solving the long-standing difficulty of introducing fluctuations of correct amplitude in a Boltzmann formalisation.
	Also the fluctuation variance of isovector observables are better treated than in conventional semiclassical approaches, even though the expected variance is still underestimated.
	We could conceive a full treatment of phase-space fluctuations which are continuously and spontaneously revived in time from collisional correlations, without any external action on the model.

	
	This approach can easily connect nuclear matter to heavy-ion collisions in the same framework, so that we could connect the phenomenology of fragment formation in heavy ion collisions (and very specific observables for instance, like the fragment size, isotopic composition, or multiplicity) to the dispersion relation for unstable modes which in nuclear matter connects the amplitude of the fluctuations and their growth to the a given form of the nuclear interaction.
	As an example, Fig.~\ref{fig_mottling} illustrates the correspondence between a portion of nuclear matter (simulated for $T=3$~MeV and $\rho^0\!=\!0.053$~fm$^{-3}$ for an interaction defined as in Eq.~(\ref{eq:pot}) 
and a hot system formed in the collision $^{136}$Xe$+^{124}$Sn at 32 $A$MeV for a central impact parameter $b=0$.
	In particular, we observe some analogy between the early time when inhomogeneities emerge in nuclear matter (20fm/c) and when fragments start forming in an open system (around 100fm/c) right after accessing low-density spinodal conditions (around 80fm/c).
	In both systems, a spinodal signal stands out by exhibiting equal-size inhomogeneities in configuration space within a similar time scale~\cite{Napolitani2013}, and it is smeared out by fragment recombination later on.
	Al even later times, the evolution is different, in the box calculation clusters continue interacting with each other while in the open system they split apart.
	The model, tested on the transition from fusion to spinodal fragmentation in central collisions at Fermi energies, on the transition from neck fragmentation to quaternary splits, on the transition from compound nucleus to fission-by-re-aggregation and to multifragmentation in relativistic spectators, reveals to be closer to the observation than previous attempts to include a Langevin term in Boltzmann theories.
	A transport approach constructed by requiring to satisfy the dispersion relation proves to be suited for the description of nuclear multifragmentation.
	Experimental investigations of heavy-ion collisions at Fermi energies already pointed out that the range of masses given by the dispersion relation of Fig.~\ref{fig_growth_rate} is actually favoured in multifragmentation mechanisms; the kinematics of the process was also found to be rather explosive.
	The spinodal mechanism was therefore proposed as a suited description~\cite{Borderie2001,Desesquelles2002}; 
	BLOB has already been adapted successfully to nuclear collisions and tested over various systems which experience spinodal instability~\cite{Napolitani2013,Napolitani2015,Colonna2017}.
	In particular, the timing of the process was found to be rather short and simulations of heavy-ion collisions at Fermi energies employing the BLOB approach described the disassembly of the system as progressing from around 100fm/c, when the lowest density conditions are experienced in the system, to about 130fm/c when inhomogeneityies stand out in the configuration space; those latter would however separate into fragments at later times.


	The dynamical approach presented in this work does not require any thermodynamic hypothesis (equilibration for instance), so that the characteristic thermodynamic features of multifragmentation, like the occurrence of a nuclear liquid-gas phase transition, are obtained as a result of the transport dynamics.
	We have identified the occurrence of bifurcations and bimodal behaviour in dynamical trajectories, linked to the fragment formation mechanism, so that at the transition energy, the system may either recompact or split into several pieces of similar sizes.
	This finding makes the dynamical description and alternative statistical approaches for multifragmentation mutually consistent.


	The processes discussed in this work, related to phase-space fluctuations and mechanical instabilities, involve a rich phenomenology of phase transitions and thresholds between very different reaction mechanisms.
	They may therefore also present some similarities in their outcome with other rather different processes and, in some situations, combine with them.
	For instance, the onset of instabilities of Rayleigh type is a common process in macroscopic hydrodynamic systems like classical fluids with a non-negligible surface tension~\cite{Ashgriz1990}, but it has also been proposed as a possible additional scenario for nuclear multifragmentation in heavy-ion collisions~\cite{Moretto1992,lukasik1997}, in some specific situations.
	Such process occurs in systems where a dilute core expands into a denser shell (Rayleigh-Taylor instability), or it acts on very deformed systems involving cohesional forces which respond to external perturbations (Plateau-Rayleigh instability).
	The system in such hydrodynamic scenario develops hole nucleation, evolving into a sponge-like or a filamented configurations which then relax into compact droplets.
	The mechanism is faster than ordinary fission and density variations of bound matter along the process do not need to be significant.
	On the other hand, the spinodal process described all throughout this report can only occur if the system traverses a specific region of the equation of state, characterised by negative incompressibility where nucleation progresses from a dilute phase letting blobs of larger density gradually emerge.
	From a microscopic point of view, it is rather associated to the nuclear liquid-gas phase transition and it requires a time comparable to the equilibration time of the system. 
	We point out that, in simulations of heavy-ion collision, the BLOB approach is actually able to describe the interplay between spinodal processes and the above mentioned hydrodynamic effects~\cite{Napolitani2016}.


	In the prospective of future developments, we point out that heavy-ion collisions and nuclear matter also involve processes of nuclear cluster formation, from light charge particles to heavier nuclear molecules, but those products emerge from an even different mechanisms~\cite{Typel2014}, which would require the explicit inclusion of additional correlations in the hierarchy of Eq.~(\ref{eq:BBGKY}).
	Light charged particles related to nuclear clustering have too small size, exceeding the ultraviolet cutoff of the dispersion relation, so that they can not belong to the unstable multipole modes which characterise spinodal fragmentation. 
	Solutions for an explicit treatment of cluster formation are proposed in refs.~\cite{Danielewicz1991,Kuhrts2001,Ono2016}.
	Connections between nuclear clustering and (spinodal) multifragmentation might be proposed, considering that multifragmentation might act on defining nuclear sources with rather complex shape from which clustered structures might eventually emerge.


	Last but not least, we may mention that the BLOB model, frequently referred to throughout this report, has been translated into a comprehensive numerical framework which has proven to be rather predictive and which is at the moment one of the most advanced simulation tools available for nuclear reactions at Fermi energies.

	Finally, we stress that the results presented here may be relevant not only for nuclear fragmentation studies, but in general for the dynamical description of quantum many-body systems.

%
%
%
%
%
\let\v\oldv

\end{document}